\documentclass{aa}  

\usepackage{graphicx}
\usepackage{txfonts}
\usepackage{subfigure}
\usepackage{longtable}
\usepackage{color}
\usepackage{icomma}
\usepackage{bm}
\usepackage{marginnote}
\usepackage{textcomp}
\usepackage{amstext}
\usepackage[normalem]{ulem}

\usepackage{xcolor}

\begin{document} 

%\onehalfspacing

   \title{Star-disk interaction in the T Tauri star V2129 Oph: An evolving accretion-ejection structure\thanks{
      Based on observations obtained at the Canada-France-Hawaii Telescope (CFHT), which is operated by the National Research Council (NRC) of Canada, the Institut National des Sciences de l'Univers of the Centre National de la Recherche Scientifique (CNRS) of France, and the University of Hawaii. Based on observations obtained with SPIRou, an international project led by the Institut de Recherche en Astrophysique et Planétologie, Toulouse, France.
   Based on observations collected at the European Southern Observatory under ESO program 2101.C-5015(A). Based on observations obtained with the robotic telescope REM, operated in Chile by the Italian institute INAF.}   
   }

   \author{A. P. Sousa 
          \inst{1}
           \and
          J. Bouvier\inst{1}
           \and
          S. H. P. Alencar\inst{2}          
           \and
          J.-F. Donati\inst{3}
           \and
          E. Alecian\inst{1}
           \and
          J. Roquette\inst{4}
           \and
          K. Perraut\inst{1}
           \and
          C. Dougados\inst{1}
           \and
          A. Carmona\inst{1}
           \and
          S. Covino\inst{5}
           \and
          D. Fugazza\inst{5}
           \and
          E. Molinari\inst{6}
           \and
          C. Moutou\inst{3}
           \and
          A. Santerne\inst{7}
           \and
          K. Grankin\inst{8}
           \and
          É. Artigau\inst{9}
           \and
          X. Delfosse\inst{1}
           \and
          G. Hebrard\inst{10}
           \and
          the SPIRou consortium
          } %

    \institute{Univ. Grenoble Alpes, CNRS, IPAG, 38000 Grenoble, France\\
               \email{alana.sousa@univ-grenoble-alpes.fr}
      \and
      Departamento de F\'isica-Icex-UFMG Ant\^onio Carlos, 6627, 31270-901. Belo Horizonte, MG, Brazil
      \and
      Univ. de Toulouse, CNRS, IRAP, 14 avenue Belin, 31400 Toulouse, France
      \and
      Department of Physics and Astronomy, University of Exeter, Physics Building, Stocker Road, Exeter, EX4 4QL, UK
      \and
      INAF - Osservatorio Astronomico di Brera - Merate, Italy
      \and
      INAF - Osservatorio Astronomico di Cagliari, Italy
      \and
      Laboratoire d'Astrophysique de Marseille, Université d'Aix-Marseille; CNRS, UMR 7326, 38 rue F. Joliot-Curie, 13388, Marseille Cedex 13, France
      \and 
      Crimean Astrophysical Observatory, 298409, Nauchny, Crimea
      \and
      Université de Montréal, Département de Physique, IREX, Montréal, QC, H3C 3J7, Canada
      \and
      Institut d'astrophysique de Paris, UMR7095 CNRS, Universit\'e Pierre \& Marie Curie, 98bis boulevard Arago, 75014 Paris, France    
    }
 
   \date{Received May 26, 2015; accepted September 04, 2015}

% \abstract{}{}{}{}{} 
% 5 {} token are mandatory
 
  \abstract
   % context heading (optional)
   {Classical T Tauri stars are young low-mass systems still accreting material from their disks. These systems are dynamic on timescales of hours to years. The observed variability can help us infer the physical processes that occur in the circumstellar environment.}
   % aims heading (mandatory)
   {In this work, we aim at understanding the dynamics of the magnetic interaction between the star and the inner accretion disk in young stellar objects. We present the case of the young stellar system V2129 Oph, which is a well-known T Tauri star with a K5 spectral type that is located in the $\rho$ Oph star formation region at a distance of $130\pm1\,\mathrm{pc}$.}
   % methods heading (mandatory)
   {We performed a time series analysis of this star using high-resolution spectroscopic data at optical wavelengths from CFHT/ESPaDOnS and ESO/HARPS and at infrared wavelengths from CFHT/SPIRou. We also obtained simultaneous photometry from REM and ASAS-SN. The new data sets allowed us to characterize the  accretion-ejection structure in this system and to investigate its evolution over a timescale of a decade via comparisons to previous observational campaigns.} 
   % results heading (mandatory)
   {We measure radial velocity variations and recover a stellar rotation period of 6.53 days. However, we do not recover the stellar rotation period in the variability of various circumstellar lines, such as H$\alpha$ and H$\beta$ in the optical or HeI $10830\,\mathring{\mathrm{A}}$ and Pa$\beta$ in the infrared.  
   Instead, we show that the optical and infrared line profile variations are consistent with a magnetospheric accretion scenario that shows variability with a period of about $6.0\,\mathrm{days}$, shorter than the stellar rotation period. Additionally, we find a period of $8.5\,\mathrm{days}$ in $\mathrm{H}\alpha$ and $\mathrm{H}\beta$ lines, probably due to a structure located beyond the corotation radius, at a distance of $\sim 0.09\,\mathrm{au}$. 
   We investigate whether this could be accounted for by a wind component,  twisted or multiple accretion funnel flows, or an external disturbance in the inner disk.} 
   %conclusion (optional)
   {We conclude that the dynamics of the accretion-ejection process can vary significantly on a timescale of just a few years in this source, presumably reflecting the evolving magnetic field topology at the stellar surface.}


 \titlerunning{Star-disk interaction in V2129 Oph}
 \authorrunning{A. Sousa et al.}
   \maketitle
   
\newcommand \ang{$\mathring{\mathrm{A}}$}
\newcommand \angn{\AA\, }
\newcommand \ha{$\mathrm{H}\alpha$}
\newcommand \hb{$\mathrm{H}\beta$}
\newcommand \hem{$\mathrm{He}$\,I\ 5876\,$\mathring{\mathrm{A}}$ }
\newcommand \he{$\mathrm{He}$\,I\ 10830$\mathring{\mathrm{A}}$ }
\newcommand \He{$\mathrm{He}$\,I }
\newcommand \naI{$\mathrm{Na}$\,I }
\newcommand \caII{$\mathrm{Ca}$\,II\ 8542\,$\mathring{\mathrm{A}}$ }
\newcommand \KI{$\mathrm{K}$\,I\ 7698.96\,$\mathring{\mathrm{A}}$ }
\newcommand \dn{Na\,{\sc i} D\,}
\newcommand \fc{$F_{\mbox{\small c}}$}
\newcommand \kms{$\mathrm{km s}^{-1}$ }
\newcommand \oi{[O\,{\sc i}]\,}
\newcommand \sii{[S\,{\sc ii}]}
\newcommand \ms{$\mathrm{M_{\odot}yr^{-1}}$} 
   
%________________________________________________________________
\section{Introduction}
Classical T Tauri stars (CTTSs) are young low-mass ($<2\,\mathrm{M_{\odot}}$) stellar objects surrounded by a circumstellar disk from which they still accrete. Accretion channeled by the stellar magnetic field is the most accepted theory for describing the observed characteristics of CTTSs \citep[e.g.,][]{2016ARA&A..54..135H,2007prpl.conf..479B}. In this scenario, the central star has a magnetic field that is strong enough to disrupt the circumstellar disk at a few radii from the star. The ionized gas in the inner disk follows the magnetic field and hits the star at high latitudes, forming hot spots \cite[e.g.,][]{camenzind1990magnetized, koenigl1991disk}. Several studies have shown that the magnetic field at the surface of young stars can be significantly more complicated than a simple dipole, including multipole components \citep{2007MNRAS.380.1297D,2008MNRAS.386.1234D,2019MNRAS.483L...1D,2009MNRAS.398..189H,2012ApJ...755...97G}. However, at $\sim0.1 \,\mathrm {au}$ from the star (i.e., about $11\,\mathrm{R_\ast}$ for a typical CTTS with a radius of $2\,\mathrm{R}_\sun$), the dipole component generally dominates the interaction with the inner disk. Thus, the star-disk interaction is expected to be mediated by the dipole component of the star’s magnetic field. 

The stellar rotation axis and the large-scale stellar magnetic field axis are often found to be tilted \citep{2020MNRAS.497.2142M,2011MNRAS.412.2454D,2011AN....332.1027G}. The tilt amplitude can influence the accretion regime, producing either stable accretion occurring through two main funnels, one in each stellar hemisphere, or unstable accretion that takes place through several funnels that appear and disappear on short timescales \citep[i.e., shorter than the stellar rotation period;][]{2016MNRAS.459.2354B,2009MNRAS.398..701K}. Other properties of the system, however, can also help determine the accretion regime, such as the stellar magnetic field strength and the mass accretion rate (Kulkarni \& Romanova 2008). Systems in unstable accretion usually present higher mass accretion rates than stable systems \citep{2013MNRAS.431.2673K}. Systems in a stable accretion present quasi-periodic photometric and spectroscopic variability, which is not expected in systems in the unstable accretion regime.

Accreting T Tauri stars present many characteristics that can be identified through spectroscopy, spectropolarimetry, and photometry, for example, cold spots, at the photospheric level, due to magnetic activity and hot spots, at the chromospheric level, associated with the gas accretion process \citep[e.g.,][]{2016ARA&A..54..135H,2007A&A...463.1017B}. These stars present intense and broad emission lines \citep[e.g.,][]{2003ApJ...582.1109W}, which are variable from night to night, such as $\mathrm{H}\alpha$, $\mathrm{H}\beta$, and $\mathrm{He}$\,I\ 5876$\mathring{\mathrm{A}}$ in the optical \citep[e.g.,][]{2018A&A...620A.195A,2016A&A...586A..47S,2012MNRAS.427.1344C,2011MNRAS.411.2383K,2009A&A...504..461F,1998AJ....116..455M} and $\mathrm{He}$\,I\ 10830$\mathring{\mathrm{A}}$, Br$\gamma$, and Pa$\beta$ in the infrared \citep[][]{2008ApJ...687.1117F,2007ApJ...657..897K,2001A&A...365...90F}. They also show continuum emission excess in the ultraviolet, due to accretion shocks  \citep{2004AJ....128.1294C,2008ApJ...681..594H,2014A&A...570A..82V}, and in the infrared, because of the presence of a dusty circumstellar disk \citep[e.g.,][]{2009AJ....138.1116S,2012A&A...540A..83T,2019A&A...629A..67S}. T Tauri stars are also optical and infrared photometric variables
\citep[e.g.,][]{2017A&A...603A.106R,2014AJ....147...82C,2014AJ....147...83S,2010A&A...519A..88A,2003A&A...409..169B} and X-ray emitters \citep[e.g.,][]{2005ApJS..160..401P,2016MNRAS.457.3836G}, which is the result of the intense coronal activity of these objects \citep{1999ARA&A..37..363F}. X-rays can also be produced in the accretion shock \citep{2006ApJ...649..914S,2011A&A...530A...1A}, which are less energetic than those produced in magnetic reconnections.

The magnetospheric accretion region extends over a few stellar radii around the central star, a compact volume not easily accessible with direct imaging techniques such as interferometry \citep{2020A&A...636A.108B, 2020Natur.584..547G}. Through the investigation of the photometric and spectroscopic variability in different wavelength ranges, we can infer physical processes at work in the inner circumstellar environment \citep[e.g.,][]{2020A&A...642A..99P,2020A&A...643A..99B,2018A&A...620A.195A,2016A&A...586A..47S,2014AJ....147...82C,1995AJ....109.2800J}. In \cite{2016A&A...586A..47S},  the accretion process of a sample of CTTSs belonging to the young star forming region NGC 2264, observed with the Convection Rotation and planetary Transits (CoRoT) satellite and/or from the FLAMES at the Very Large Telescope (VLT) spectrograph, was analyzed. It was shown that in $30\%$ of the CoRoT sample the morphology of the light curve changed in three years, switching from periodic to aperiodic accretion regimes and vice versa. They also analyzed the variability of the $\mathrm{H}\alpha$ and HeI lines of CTTSs in NGC 2264, showing that most of the targets were not periodic in these lines. The blue and red wings of the $\mathrm{H}\alpha$ line were, in general, dominated by different regions, such as the magnetosphere and/or disk wind. 
In the infrared, the $\mathrm{He}$\,I\ 10830$\mathring{\mathrm{A}}$ line commonly occurs in emission in CTTSs and often exhibits both redshifted and blueshifted absorption components that trace accretion and outflows from the system \citep[e.g.,][]{2007ApJ...657..897K}. 

In this work, we investigate the  CTTS V2129 Oph (SR9, DoAr 34, AS 207, ROX 29), a well-known young star ($1-2\,\mathrm{Myr}$) with a K5 spectral type \citep{2006A&A...460..695T} and a mass of $1.35\pm0.15\,\mathrm{M_\odot}$ \citep{2007MNRAS.380.1297D}, located in the $\rho$ Oph star forming region at a distance of $130\pm1\,\mathrm{pc}$ \citep{2018AJ....156...58B}. This is part of a binary system, with a separation of 0.59" \citep{2006ApJ...636..932M}. The companion is a faint low-mass object,  with a K-band flux ratio of $\sim0.09$ \citep{2006ApJ...636..932M,2011A&A...530A...1A}, and it is not expected to contribute significantly to the integrated optical or near-infrared spectra of the source.

V2129 Oph accretes from its circumstellar disk at a moderate rate of $(1.5\pm0.6)\times10^{-9}\,$\ms\  \citep{2012A&A...541A.116A}, and we view the system at a inclination of about $\sim60^o$ from our line of sight \citep{2011MNRAS.412.2454D}. From photometric spot modulation, \cite{2008A&A...479..827G} used  data obtained from 1986 to 2003 to derive a period of $6.55\,\mathrm{days}$ over the 15 seasons covered by their full data set. Analyzing each season separately, they found the period to vary between $6.34$ and $6.59\,\mathrm{days}$ over the seasons, a long-term variation that presumably reflects latitudinal differential rotation at the stellar surface. In this work, we adopt (and confirm) the rotational period of $6.53\,$days that was initially derived using photometric data obtained from 1986 to 1993 \citep{1998AJ....116.1419S} and later confirmed spectroscopically from spot-modulated radial velocity variations  \citep{2012A&A...541A.116A}. 

Previous monitoring campaigns of this star led to the derivation of its magnetic properties by \cite{2007MNRAS.380.1297D,2011MNRAS.412.2454D}. They measured an increase in the magnitude of the dipole and the octupole magnetic field components over a timescale of a few years. A change in the magnetic field intensity can significantly impact the star’s accretion properties. In this work, we analyze the accretion-ejection process in V2129 Oph through the variability of the circumstellar emission lines and compare our results with previous works to investigate the dynamics of the system on a timescale of a decade.

The paper is organized as follows. In Sect. \ref{sec:data}, we describe the data sets used in this work. The optical and infrared photometric analyses are presented in Sects. \ref{sec:photo} and  \ref{sec:asassn}, respectively. Section \ref{sec:optical} presents the results obtained with the CFHT/ESPaDOnS (Canada-France-Hawaii Telescope Echelle SpectroPolarimetric Device for the Observation of Stars) and ESO/HARPS (European Southern Observatory High Accuracy Radial velocity Planet Searcher) spectrographs at optical wavelengths. In Sect. \ref{sec:infrared} we perform the analysis of the infrared data obtained with the spectrograph CFHT/SPIRou (CFHT SPectropolarimètre InfraROUge). In Sect. \ref{sec:accrate}, we derive the mass accretion rate onto the star. In Sect. \ref{sec:discussion}, we discuss the structure of the star-disk interaction region that can yield the observed variability and compare it to previous results. In Sect. \ref{sec:concl}, we draw our conclusions.

\section{Observations}\label{sec:data}

We observed V2129 Oph during a campaign that included observations from the optical to the near infrared. We obtained data at optical wavelengths with two spectrographs: CFHT/ESPaDOnS (ten observations at a resolving power of $65\,000$) and ESO/HARPS (nine observations at a resolving power of $115\,000$). The CFHT/ESPaDOnS data were reduced using the pipeline Libre-ESpRIT \citep{1997MNRAS.291..658D}.  The HARPS data were automatically reduced by the HARPS data reduction software \citep{2003Msngr.114...20M}. We obtained infrared observations with CFHT/SPIRou (nine observations at a resolving power of $75\,000$). Telluric corrected SPIRou spectra were produced using the data reduction system APERO, version 0.6.131 (Cook et al., in prep.).

We also analyzed photometric observations obtained with the Rapid Eye Mount (REM) telescope. REM is a 60 cm diameter fast reacting telescope located at ESO La Silla.
The telescope hosts two instruments: REMIR, an infrared imaging camera, and ROS2, a visible imager with four simultaneous passbands. The two cameras can also simultaneously observe   the same field of view, of 10x10 arcmin, thanks to a dichroic placed before telescope focus. Thus, five images are obtained at the same time: $g,r,i,z,J,H,K$. The Observatory is operated for the Italian institute INAF by the REM team. Here we only used the $JHK$ images of REM as the optical ones do not contain enough reference stars to calibrate the image. The images were preprocessed by the telescope team, including precise World Coordinate System (WCS) information. Photometry extraction was performed using the AstrOmatic package\footnote{\url{https://www.astromatic.net/}}. For each filter, the images were aligned and used to create a master frame by stacking all the images using the tool SWARP  \citep{2002Bertin}. We used the Source Extractor \citep[SExtractor;][]{1996Bertin} to perform source detection in the master frame.

\begin{figure} 
 \centering
 \includegraphics[width=0.4\textwidth]{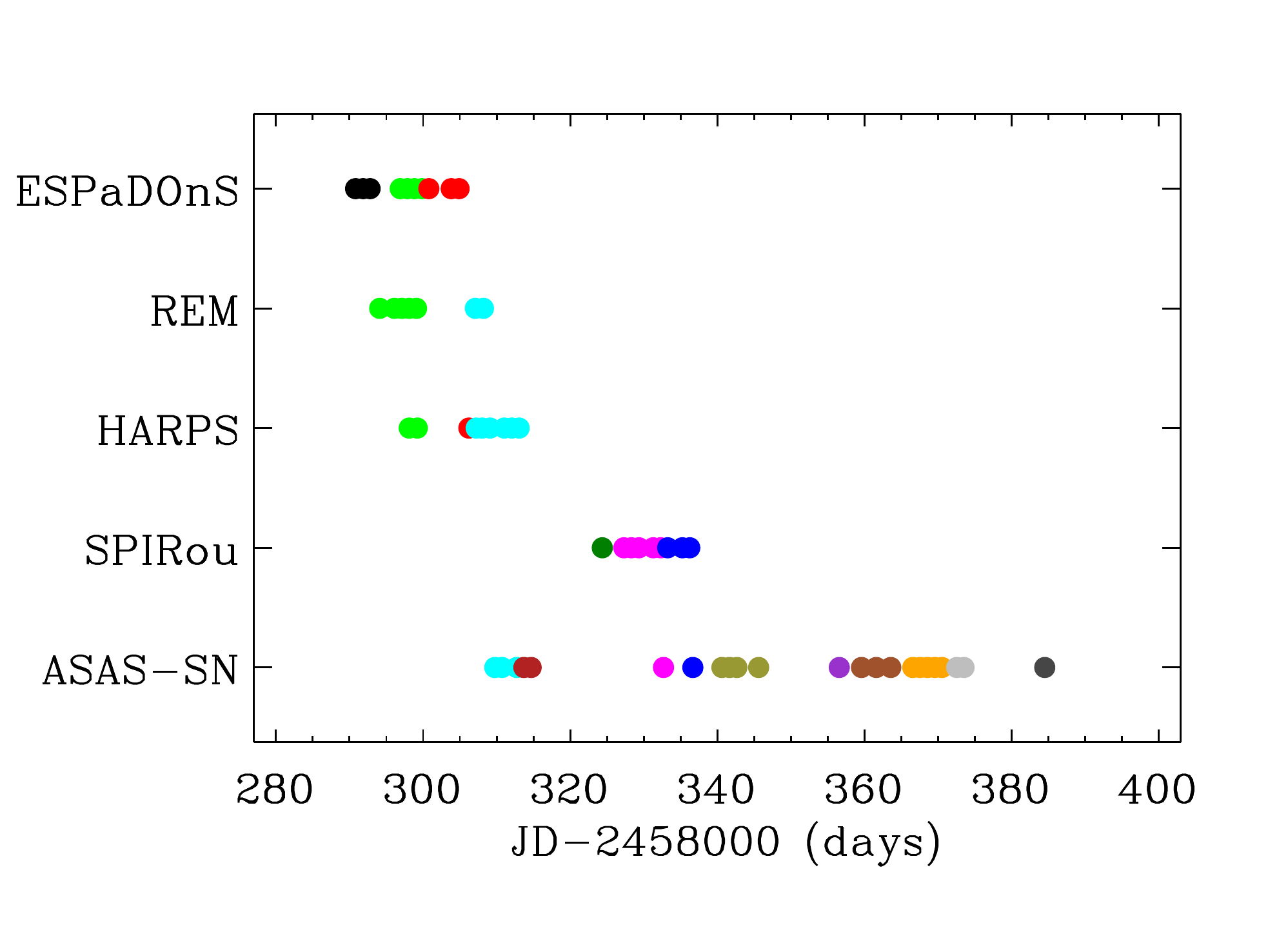}
\caption{\label{fig:chron} Chronological scheme of the observing runs from Table 1 and the ASAS-SN data covering the span of our observations. Colors represent different rotation cycles. 
}
\end{figure}

We present a summary of the observations in Table \ref{tab:obs}, and in Fig. \ref{fig:chron}
 we show a chronological scheme of the observations used in this work. 
To measure the rotation phases (E), we used the same ephemeris that was adopted by \cite{2012A&A...541A.116A} and \cite{2007MNRAS.380.1297D}: 

\begin{equation}\label{eq:eph}
 HJD=2\,453\,540.0+P_{rot}E,
\end{equation}
where $P_\mathrm{rot}=6.53\,\mathrm{days}$ is the stellar rotational period. However, due to the uncertainty in the period measurement and the long time span between the two data sets, the phases computed here have no correspondence to the phases in \cite{2012A&A...541A.116A}.

{\tiny
\begin{table}[htb!]
 \centering
 \tiny
 \addtolength{\tabcolsep}{-5pt}  
 \caption{Observations of V2129 Oph used in this work.}
 \label{tab:obs}
 \begin{tabular}{llll}
  \hline
  \hline
  Telescope & Instrument & Dates & Bands \\
  \hline
  REM               &  REMIR          & 2018-Jun-25 to 2018-Jul-9     & \textit{JHK}            \\  
  CFHT              & ESPaDOnS       & 2018-Jun-20 to 2018-Jul-4      & $3700-10000\,\mathring{\mathrm{A}}$       \\
  ESO               & HARPS          & 2018-Jun-28 to 2018-Jul-13     & $3780-6910\,\mathring{\mathrm{A}}$   \\
  CFHT              & SPIRou         & 2018-Jul-25 to 2018-Aug-6      & $9650-25000\,\mathring{\mathrm{A}}$  \\
 \hline
 \end{tabular} 
\end{table}
}

\section{Infrared photometry } \label{sec:photo}

\begin{figure} 
 \centering
 \subfigure[]{\includegraphics[scale=0.40]{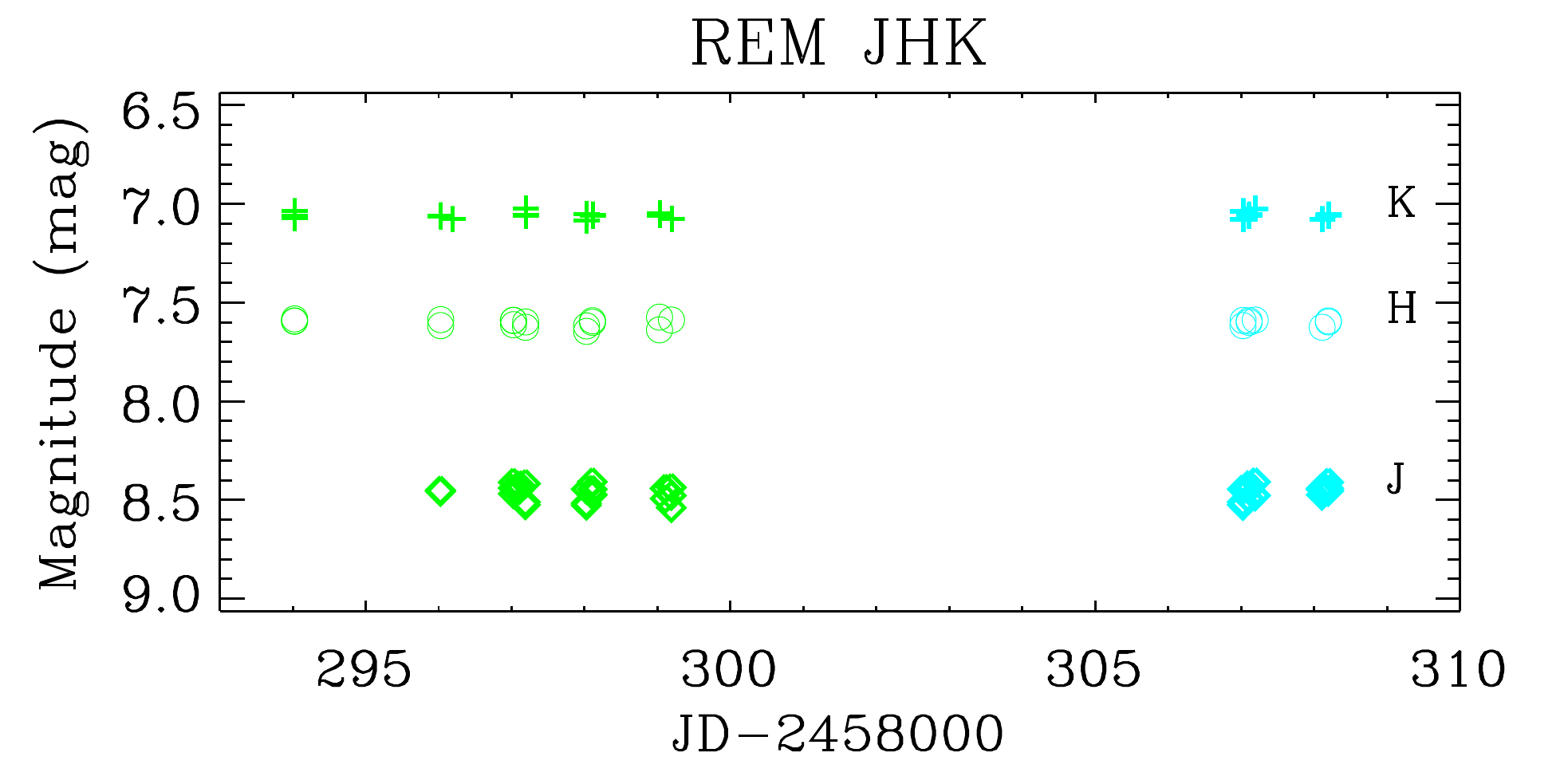}}
 \subfigure[]{\includegraphics[scale=0.40]{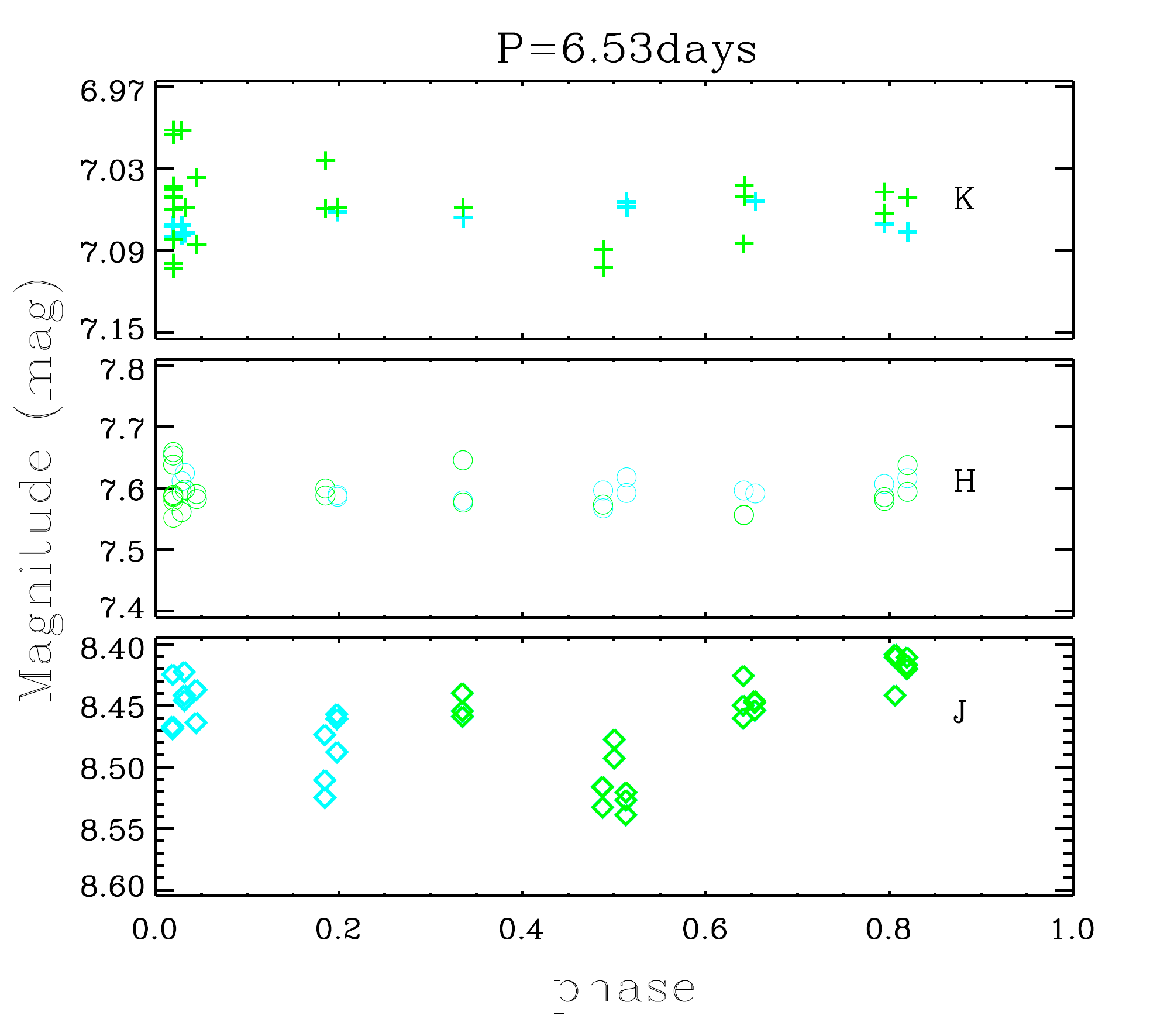}}
\caption{\label{fig:Rem_jhk} Photometric $JHK$ light curves obtained with the REM telescope as a function of observing date (a) and phase (b). We calibrated the magnitudes using the 2MASS $JHK$ magnitudes of a reference star (see text). The statistical uncertainties on the magnitudes are smaller than the symbols.} 
\end{figure}

V2129 Oph was monitored in the $J$-, $H$- and $K$-bands using the REM telescope. With the small field of view of the REMIR, only two reference stars were visible along with V2129 Oph on each observed night, and we adopted the brightest one, 2MASSJ16274987-2425402, as the reference star for differential photometry. The light curve for V2129 Oph is shown in Fig. \ref{fig:Rem_jhk}, and the mean magnitude of each night is listed in Table \ref{tab:mags}.

We used the modified Lomb-Scargle periodogram from  \cite{1986ApJ...302..757H} to investigate the periodicity in the $JHK$ light curves. Unfortunately, we could not identify any periodic signal in the light curves, which is likely due to the small number of observations obtained over only seven nights between June 25 and July 9, 2018. In Fig. \ref{fig:Rem_jhk}, we tentatively fold the $JHK$ light curves in phase with the  stellar rotation period of $6.53\,\mathrm{days}$ for V2129 Oph. 

\begin{table}[htb!]
\tiny
\caption{\label{tab:mags} Infrared photometric mean magnitudes of each night obtained with REM.}
\begin{center}
\begin{tabular}{lllllll}
  \hline\hline 
 JD\tablefootmark{a} & $J$ &  $H$ &  $K$   \\
\hline
 8294.1  &      -                        &      -                        &      7.014  $\pm$      0.003  \\
 8296.1  &      8.451  $\pm$      0.005  &      7.585   $\pm$     0.003  &      7.053  $\pm$      0.003  \\
 8297.1  &      8.515  $\pm$      0.005  &      7.646   $\pm$     0.003  &      7.094  $\pm$      0.003  \\
 8298.1  &      8.447  $\pm$      0.005  &      7.594   $\pm$     0.003  &      7.051  $\pm$      0.003  \\
 8299.1  &      8.418  $\pm$      0.006  &      7.577   $\pm$     0.003  &      7.059  $\pm$      0.004  \\
 8307.1  &      8.446  $\pm$      0.005  &      7.583   $\pm$     0.003  &      7.060  $\pm$      0.003  \\
 8308.2  &      8.486  $\pm$      0.005  &      7.613   $\pm$     0.003  &      7.073  $\pm$      0.003  \\
\hline
\end{tabular}
\end{center}
\tablefoot{
 \tablefoottext{a}{JD-$2\,450\,000$}
 }
 \end{table}
 
\section{Optical photometry} \label{sec:asassn}

\begin{figure} 
 \centering
\includegraphics[width=0.4\textwidth]{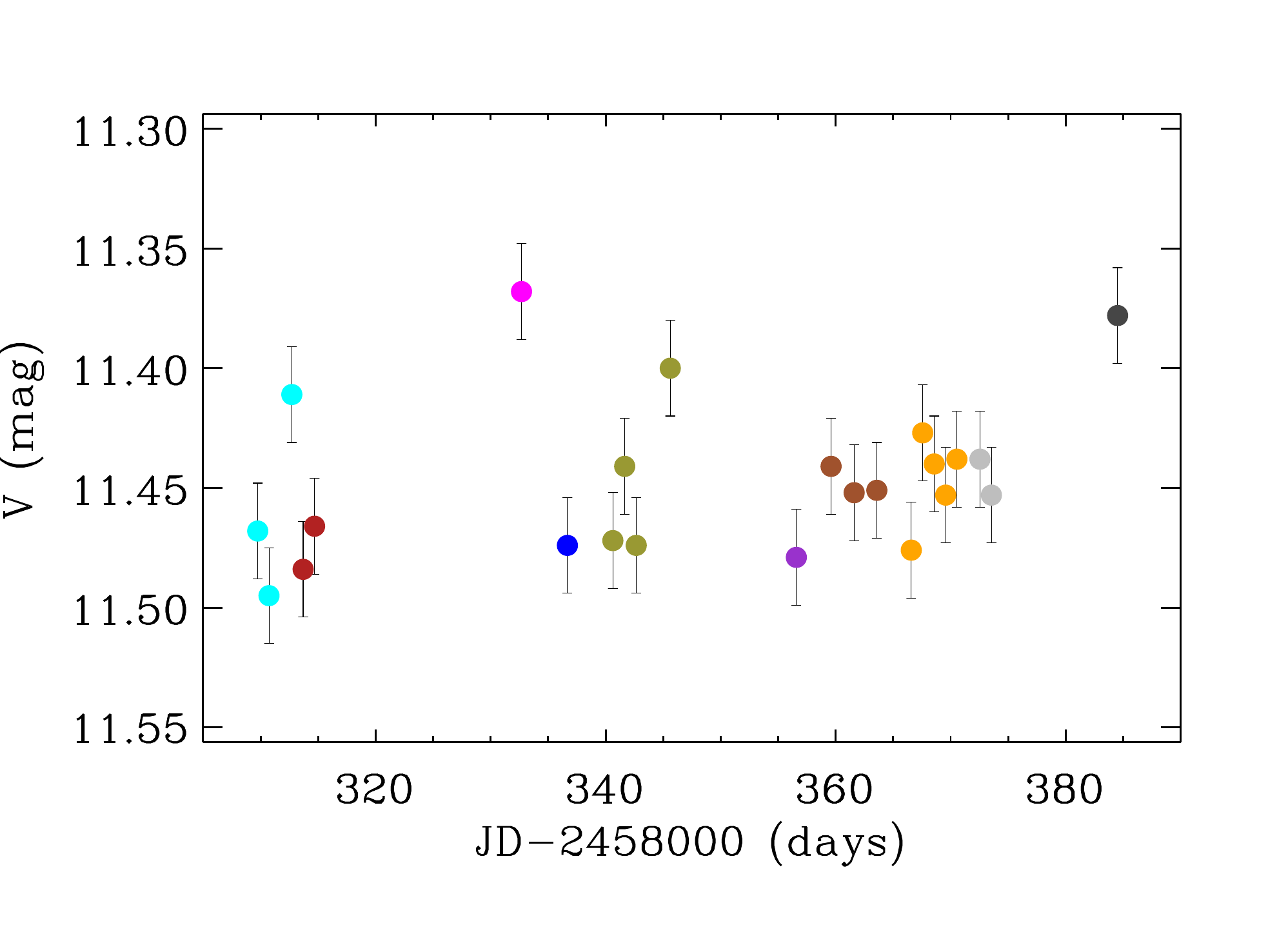}
\includegraphics[width=0.4\textwidth]{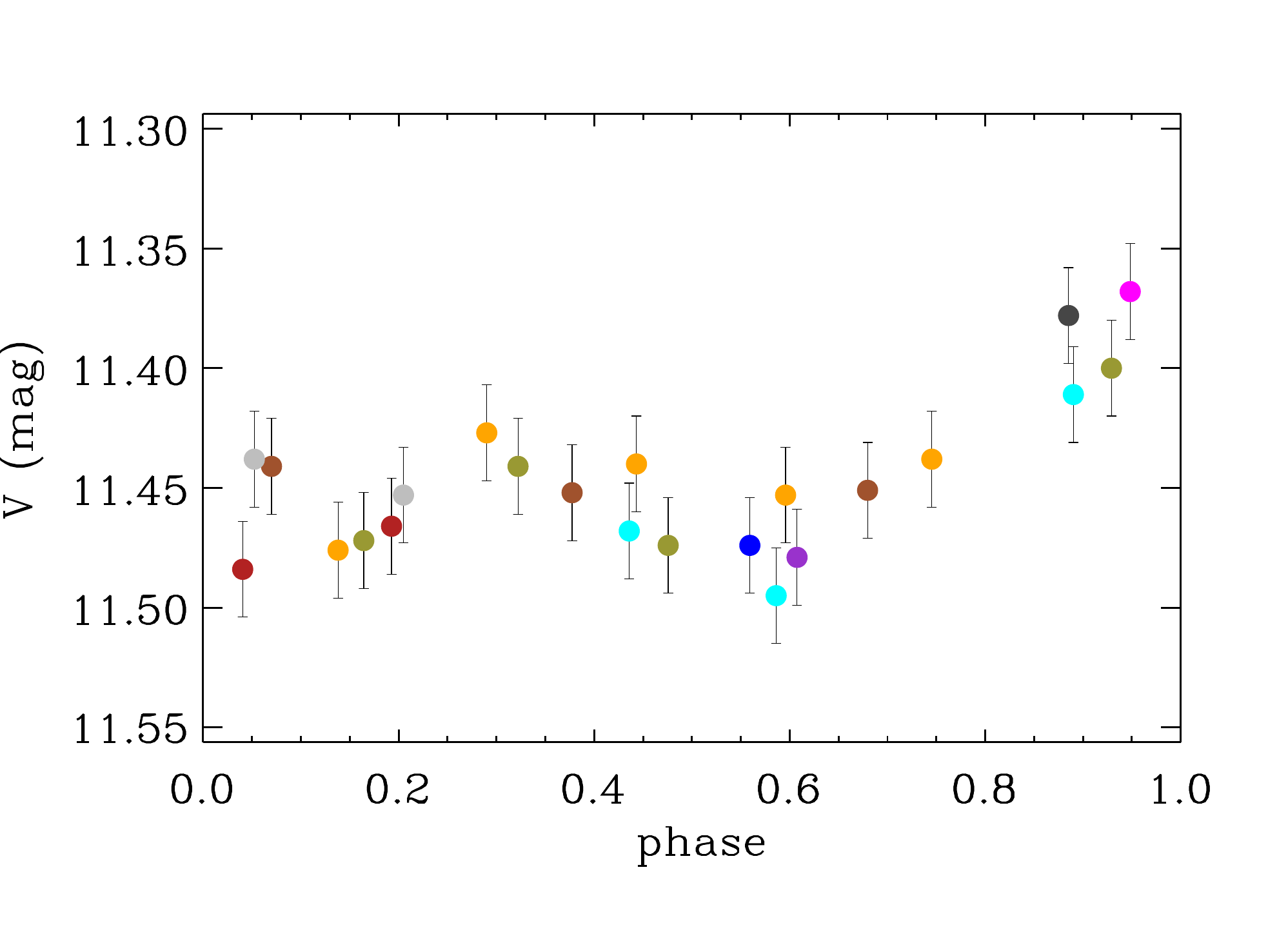}
\caption{\label{fig:asassn} V-band light curve of V2129 Oph from ASAS-SN photometry. {\it Top:} Direct light curve. {\it Bottom:} Light curve folded in phase with the ephemeris given in Eq. (1).  The colors represent different rotation cycles.} 
\end{figure}

We used the optical light curve provided by the ASAS-SN survey \citep{Shappee14, Kochanek17} to further explore the photometric behavior of V2129 Oph. We retrieved from the ASAS-SN database the source's V-band light curve from Modified Julian Date (MJD) 8309 to MJD 8385, which partly covers the time span of our spectroscopic observations (see Table  \ref{tab:parameters}). The V-band light curve is shown in Fig. \ref{fig:asassn}. The amplitude of variability reaches about 0.15 mag in the V-band. The light curve folded in phase using the ephemeris of Eq. (1) displays a smooth modulation of the brightness level at the rotational period of $6.53\,\mathrm{days}$. We note the existence of two photometric minima along each rotational cycle, one located around phase 0.15 and the other around phase 0.55, which suggests that the star's brightness is modulated by two cold spots located at nearly opposite longitudes. A polar spot extending at lower latitudes at phases 0.15 and 0.55 can, however, produce similar light curve characteristics. The cold spot configuration at the stellar surface of T Tauri stars is often found to be quite complex, consisting of several groups of spots  \citep[e.g.,][]{2019MNRAS.489.5556Y}. Therefore, we can interpret the optical light curve variability as being due to two main spot groups at opposite longitudes or to an elongated polar spot with  latitudinal extensions located at opposite azimuths at the time of our observations.  

The ASAS-SN monitoring of this source extends over a much longer timescale than shown here, from Julian date (JD) 6800 to JD 8385. Here we used only the last observing season, which is concurrent with the other data sets presented in this paper. The longer-term ASAS-SN light curve (not shown) indicates that the amplitude of variability can reach up to 0.3 mag in the V-band. It also shows that the maximum brightness level can lie at V=11.30, that is to say, 0.05 mag brighter than what it was at the time of our campaign. Indeed, the V-band light curve obtained in 2009 by \cite{2012A&A...541A.116A} exhibits an amplitude of 0.4 mag and a maximum brightness level of V=10.7. It also displays a single photometric minimum, which suggests a single cold spot dominating the rotational modulation of the stellar brightness at that time. At the time of our observation, both the lower amplitude of variability and the fainter maximum brightness level were consistent with the star's brightness being modulated by two sets of spots located at nearly opposite longitudes. In this configuration, one set of spots is always in view, and this effectively reduces both the photometric amplitude and the maximum brightness level. 

Finally, we note that \cite{2018AJ....156...71C} analyzed the Kepler K2 light curve of V2129 Oph obtained in 2014 and found it to be stochastic, with no discernible period, even though the light curve is continuous over 80 days. This suggests that the variability pattern of the source may drastically change over a timescale of a few years.   

\begin{table*}[htb!]
\tiny
\addtolength{\tabcolsep}{-4.8pt}  
\caption{\label{tab:parameters} Observational parameters measured in the V2129 Oph optical and infrared spectra.}
\begin{center}
\begin{tabular}{lllllllllllllllllll}
  \hline\hline 
Inst. & JD\tablefootmark{a} &$\phi$ & $\mathrm{v_{rad}}$  & $\mathrm{veil}$\tablefootmark{b} &  $\mathrm{v_{rad\ HeI}}$  & $\mathrm{EW}_{\mathrm{H}\alpha}$\tablefootmark{c} &$\mathrm{EW}_{\mathrm{H}\beta}$\tablefootmark{c}  & $\mathrm{EW_{HeI}}$\tablefootmark{c} & $\mathrm{EW}_{\mathrm{NaI}}$\tablefootmark{c} &$\mathrm{EW}_{\mathrm{CaII}}$\tablefootmark{c}  & $\mathrm{EW}_{\mathrm{HeI}\ }$\tablefootmark{c} & $\mathrm{EW}_{\mathrm{Pa\beta}}$\tablefootmark{c} & $\mathrm{EW}_{\mathrm{Pa\gamma}}$\tablefootmark{c} &
$\mathrm{\dot{M}_{H\alpha}}$\tablefootmark{d} & $\mathrm{\dot{M}_{HeI}}$\tablefootmark{d} & $\mathrm{\dot{M}_{Pa\beta}}$ \tablefootmark{d}\\

& &   & Phot. & & $(5876\mathrm{\mathring{A}})$  &  & & $(5876\mathrm{\mathring{A}})$  & Doublet  &  & $(10830\mathrm{\mathring{A}})$ &  &  &  &  $(5876\mathrm{\mathring{A}})$ \\

 & &   & ($\mathrm{km}\ \mathrm{s}^{-1})$  &  & ($\mathrm{km}\ \mathrm{s}^{-1})$ &$(\mathring{\mathrm{A}})$ &  $(\mathring{\mathrm{A}})$ &$(\mathring{\mathrm{A}})$ & $(\mathring{\mathrm{A}})$ & $(\mathring{\mathrm{A}})$ & $(\mathring{\mathrm{A}})$ & $(\mathring{\mathrm{A}})$ &$(\mathring{\mathrm{A}})$ &$(\mathrm{M_\odot\ yr^{-1}})$ &  $(\mathrm{M_\odot\ yr^{-1}})$ &  $(\mathrm{M_\odot\ yr^{-1}})$ \\
\hline
ESPaDOnS  &   8290.82  &  0.54  &  -6.8  $\pm$  0.2  &  0.12  $\pm$  0.03  &  1.7  $\pm$  0.2  &  15.92  &  3.41  &  0.35  &  1.62  &  0.50  &    -     &  -     &  -     &  2.17  &  1.55  &  -     \\
ESPaDOnS  &   8291.84  &  0.69  &  -7.2  $\pm$  0.2  &  0.17  $\pm$  0.03  &  2.4  $\pm$  0.1  &  18.92  &  4.59  &  0.38  &  1.92  &  0.97  &    -     &  -     &  -     &  2.64  &  1.68  &  -     \\
ESPaDOnS  &   8292.79  &  0.84  &  -6.6  $\pm$  0.3  &  0.17  $\pm$  0.03  &  0.9  $\pm$  0.1  &  16.13  &  3.59  &  0.34  &  1.70  &  0.74  &    -     &  -     &  -     &  2.20  &  1.50  &  -     \\
ESPaDOnS  &   8296.89  &  1.47  &  -5.7  $\pm$  0.2  &  0.08  $\pm$  0.04  &  1.5  $\pm$  0.2  &  10.90  &  1.47  &  0.31  &  1.33  &  0.31  &    -     &  -     &  -     &  1.41  &  1.32  &  -     \\
ESPaDOnS  &   8297.87  &  1.62  &  -7.4  $\pm$  0.2  &  0.11  $\pm$  0.03  &  3.3  $\pm$  0.2  &  11.06  &  1.30  &  0.29  &  0.52  &  0.11  &    -     &  -     &  -     &  1.44  &  1.23  &  -     \\
HARPS     &   8298.11  &  1.65  &  -8.9  $\pm$  0.3  &  0.08  $\pm$  0.03  &  5.6  $\pm$  0.2  &   9.87  &  2.63  &  0.40  &  0.67  &  -     &    -     &  -     &  -     &  1.26  &  1.78  &  -     \\
ESPaDOnS  &   8298.81  &  1.76  &  -7.7  $\pm$  0.2  &  0.08  $\pm$  0.03  &  3.8  $\pm$  0.2  &  10.43  &  2.03  &  0.32  &  1.37  &  0.27  &    -     &  -     &  -     &  1.34  &  1.39  &  -     \\
HARPS     &   8299.21  &  1.82  &  -7.0  $\pm$  0.1  &  0.05  $\pm$  0.02  &  5.2  $\pm$  0.3  &  10.61  &  2.04  &  0.22  &  1.21  &  -     &    -     &  -     &  -     &  1.37  &  0.92  &  -     \\
ESPaDOnS  &   8299.85  &  1.92  &  -6.6  $\pm$  0.2  &  0.05  $\pm$  0.03  &  4.9  $\pm$  0.2  &  11.97  &  1.90  &  0.18  &  1.84  &  0.34  &    -     &  -     &  -     &  1.57  &  0.71  &  -     \\
ESPaDOnS  &   8300.78  &  2.06  &  -6.3  $\pm$  0.2  &  0.05  $\pm$  0.03  &  1.9  $\pm$  0.3  &  13.07  &  2.39  &  0.15  &  1.87  &  0.35  &    -     &  -     &  -     &  1.74  &  0.56  &  -     \\
ESPaDOnS  &   8303.82  &  2.53  &  -7.2  $\pm$  0.3  &  0.06  $\pm$  0.03  &  1.5  $\pm$  0.2  &   7.00  &  0.49  &  0.18  &  0.92  &  0.02  &    -     &  -     &  -     &  0.86  &  0.70  &  -     \\
ESPaDOnS  &   8304.89  &  2.69  &  -7.7  $\pm$  0.2  &  0.05  $\pm$  0.03  &  3.2  $\pm$  0.2  &   8.15  &  0.88  &  0.14  &  1.36  &  0.17  &    -     &  -     &  -     &  1.02  &  0.55  &  -     \\
HARPS     &   8306.22  &  2.90  &  -6.6  $\pm$  0.2  &  0.05  $\pm$  0.02  &  3.7  $\pm$  0.3  &  10.41  &  2.83  &  0.24  &  1.19  &  -     &    -     &  -     &  -     &  1.34  &  0.98  &  -     \\
HARPS     &   8307.22  &  3.05  &  -6.2  $\pm$  0.1  &  0.05  $\pm$  0.02  &  0.2  $\pm$  0.3  &   9.71  &  2.70  &  0.27  &  1.17  &  -     &    -     &  -     &  -     &  1.24  &  1.15  &  -     \\
HARPS     &   8308.04  &  3.17  &  -8.7  $\pm$  0.4  &  0.07  $\pm$  0.02  &  0.1  $\pm$  0.4  &  13.23  &  3.18  &  0.30  &  0.66  &  -     &    -     &  -     &  -     &  1.76  &  1.27  &  -     \\
HARPS     &   8309.10  &  3.34  &  -6.4  $\pm$  0.1  &  0.04  $\pm$  0.02  &  2.5  $\pm$  0.3  &  15.96  &  3.27  &  0.33  &  1.55  &  -     &    -     &  -     &  -     &  2.17  &  1.45  &  -     \\
HARPS     &   8311.03  &  3.63  &  -8.1  $\pm$  0.2  &  0.06  $\pm$  0.02  &  6.5  $\pm$  0.3  &  12.09  &  2.87  &  0.30  &  1.32  &  -     &    -     &  -     &  -     &  1.59  &  1.30  &  -     \\
HARPS     &   8312.06  &  3.79  &  -7.0  $\pm$  0.1  &  0.05  $\pm$  0.02  &  4.9  $\pm$  0.3  &   8.01  &  1.70  &  0.24  &  1.25  &  -     &    -     &  -     &  -     &  1.00  &  0.97  &  -     \\
HARPS     &   8313.06  &  3.94  &  -6.6  $\pm$  0.1  &  0.05  $\pm$  0.02  &  2.4  $\pm$  0.4  &   6.06  &  1.62  &  0.23  &  1.00  &  -     &    -     &  -     &  -     &  0.73  &  0.97  &  -     \\
SPIRou    &   8324.36  &  5.67  &  -7.4   $\pm$  0.1   & -   &  -  &   -     &  -     &  -     &  -     &  -         &   0.42  &  0.66  &  1.42  &  -     &  -     &  0.89  \\
SPIRou    &   8327.30  &  6.12  &  -7.66  $\pm$  0.02  & -   &  -  &   -     &  -     &  -     &  -     &  -         &   2.19  &  0.79  &  1.49  &  -     &  -     &  1.07  \\
SPIRou    &   8328.30  &  6.28  &  -6.80  $\pm$  0.08  & -   &  -  &   -     &  -     &  -     &  -     &  -         &   1.85  &  0.89  &  1.43  &  -     &  -     &  1.22  \\
SPIRou    &   8329.34  &  6.44  &  -6.20  $\pm$  0.07  & -   &  -  &   -     &  -     &  -     &  -     &  -         &   0.37  &  0.70  &  1.47  &  -     &  -     &  0.95  \\
SPIRou    &   8331.28  &  6.73  &  -6.78  $\pm$  0.07  & -   &  -  &   -     &  -     &  -     &  -     &  -         &   1.57  &  0.90  &  1.51  &  -     &  -     &  1.23  \\
SPIRou    &   8332.30  &  6.89  &  -6.11  $\pm$  0.02  & -   &  -  &   -     &  -     &  -     &  -     &  -         &   1.54  &  0.82  &  1.52  &  -     &  -     &  1.12  \\
SPIRou    &   8333.25  &  7.04  &  -6.40  $\pm$  0.09  & -   &  -  &   -     &  -     &  -     &  -     &  -         &   3.04  &  1.49  &  1.67  &  -     &  -     &  2.10  \\
SPIRou    &   8335.26  &  7.34  &  -6.23  $\pm$  0.09  & -   &  -  &   -     &  -     &  -     &  -     &  -         &  -0.53  &  1.23  &  1.54  &  -     &  -     &  1.72  \\
SPIRou    &   8336.25  &  7.50  &  -7.3   $\pm$  0.2   & -   &  -  &   -     &  -     &  -     &  -     &  -         &  -0.95  &  0.56  &  1.59  &  -     &  -     &  0.74  \\
\hline
\end{tabular}
\end{center}
 \tablefoot{The table is ordered according to the observation date. 
 \tablefoottext{a}{JD-$2\,450\,000$}
 \tablefoottext{b}{The veiling obtained with the SPIRou data at different wavelengths is listed in Table \ref{tab:veilIR}}
 \tablefoottext{c}{ We used the convention of positive equivalent width for emission lines and negative values for absorption lines. The equivalent width errors are 0.02\AA\ for the $\mathrm{H\alpha}$ and Pa$\beta$ lines, 0.03\AA\ for the $\mathrm{H\beta}$ line, 0.01\AA\ for the HeI 5876\AA,\ $\mathrm{Ca}$\,II\ 8542$\mathring{\mathrm{A}}$, and NaI doublet lines, 0.08\AA\ for the HeI 10830\AA\ line, and 0.05\AA\ for Pa$\gamma$.}
 \tablefoottext{d}{$\times10^{-9}$}
 }
 \end{table*}

\section{Optical spectroscopy} \label{sec:optical}

\subsection{Optical veiling and radial velocity} \label{sec:optveiling}
Optical veiling of the photospheric lines is one of the accretion effects seen in a spectroscopic analysis. The photospheric lines become shallower than in a non-accreting system when the hot spot is in our line of sight and adds an extra contribution to the stellar continuum \citep[e.g.,][]{2018A&A...610A..40R,1991ApJ...382..617H}. The infrared veiling can have different origins \citep{2006ApJ...646..319E}.

We used the weak line T Tauri star V819 Tau \citep[SpT= K4, M= $1.00\pm0.05\,\mathrm{M_\sun}$, and $vsini=9.5\,\mathrm{km\ s^{-1}}$;][]{2015MNRAS.453.3706D} as a template to measure the veiling of V2129 Oph. We have observations of V819 Tau with both the ESPaDOnS and HARPS spectrographs. We used six photospheric line regions between 4590 and 6400$\,\mathring{\mathrm{A}}$ to compare the spectra of the template to V2129 Oph. We chose spectral regions that do not have many blends between different lines. To compare the spectra of V819 Tau with our system, we broadened the template spectra with the projected rotation velocity of V2129 Oph \citep[$vsini =14.5 \pm 0.3 \,\mathrm{km\ s^{-1}}$;][]{2007MNRAS.380.1297D,2012A&A...541A.116A}. We measured the velocity shift between the spectra, cross-correlating the target spectra and the template. We corrected the template and the target spectra from barycentric velocities before computing the velocity shift between both spectra. After that, we computed the veiling in each spectrum, using the ratio of the continuum excess flux to the stellar photospheric flux as the definition of veiling.

We independently measured the veiling and radial velocity of V2129 Oph using data from ESPaDOnS and HARPS. We found a systematic error between the veiling obtained by the different data sets, which is due to the difference of resolution between the spectrographs or an instrumental effect, which, in turn, is due a long time delay  between the HARPS template spectrum (2004) and the observations (2018). To correct this systematic error, we measured the veiling between the V819 Tau spectra from HARPS and ESPaDOnS (it is not a true veiling as the star is a non-accreting system). We found an offset value of $0.11\pm0.03$ between the HARPS and ESPaDOnS data, which corresponds to a mean and standard deviation of all spectral regions used to measure the veiling. We applied this value to correct the veiling from the HARPS data. We show the individual value of the veiling and radial velocity of each observation in Table \ref{tab:parameters}.

\begin{figure} [!htb]
 \centering
\subfigure[]{\includegraphics[scale=0.40]{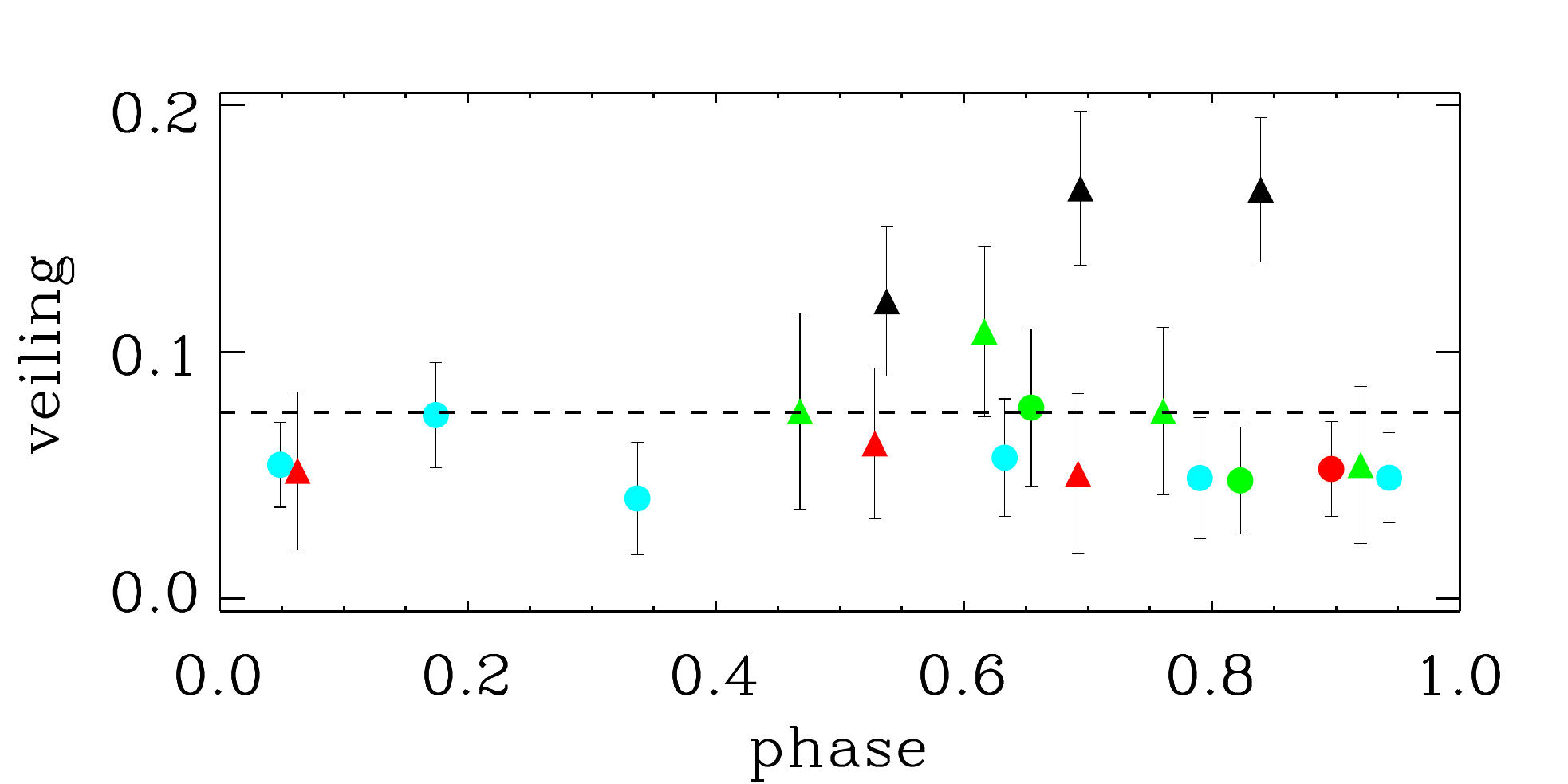}}
\subfigure[]{\includegraphics[scale=0.40]{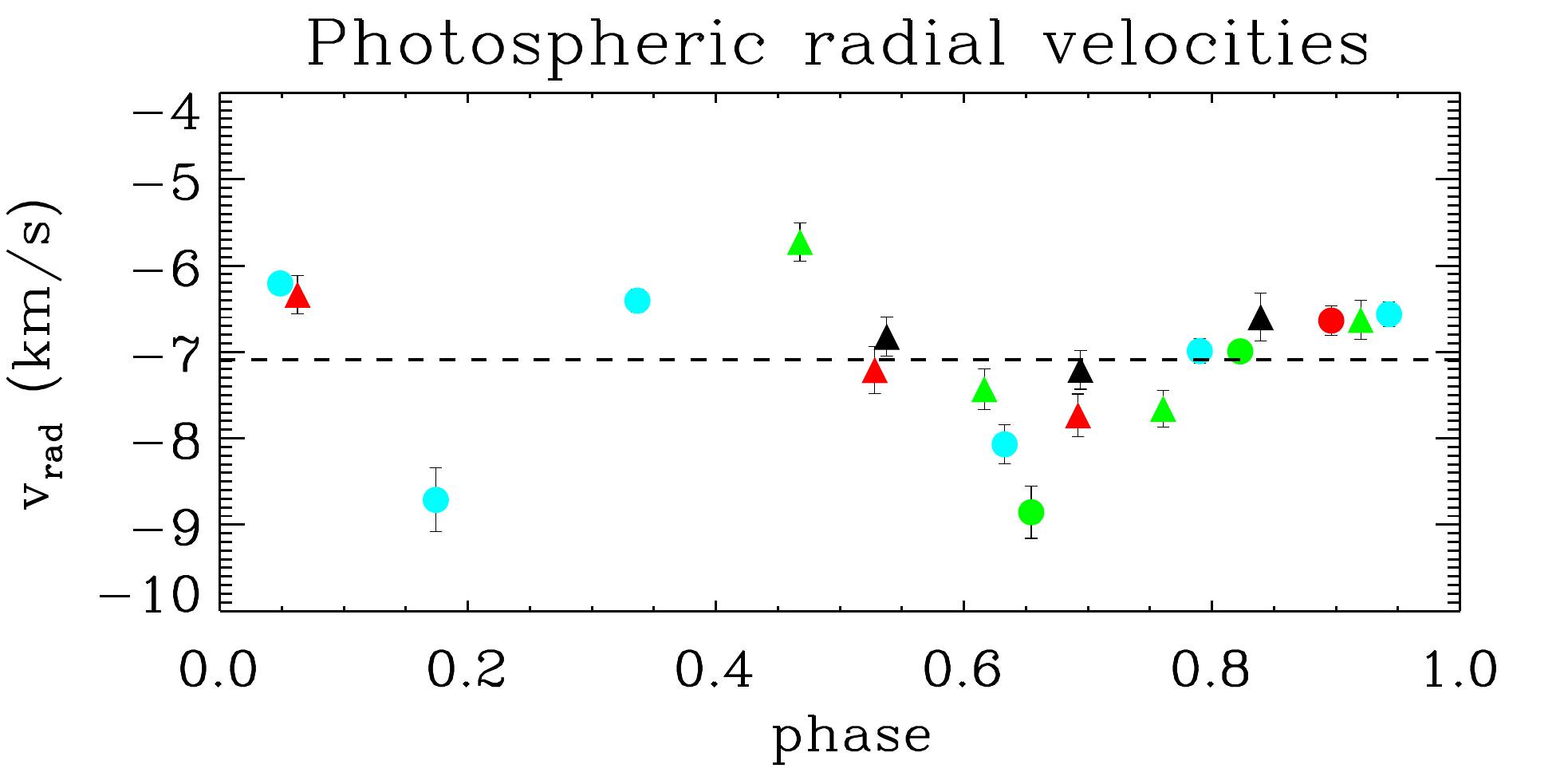}}
\subfigure[]{\includegraphics[scale=0.40]{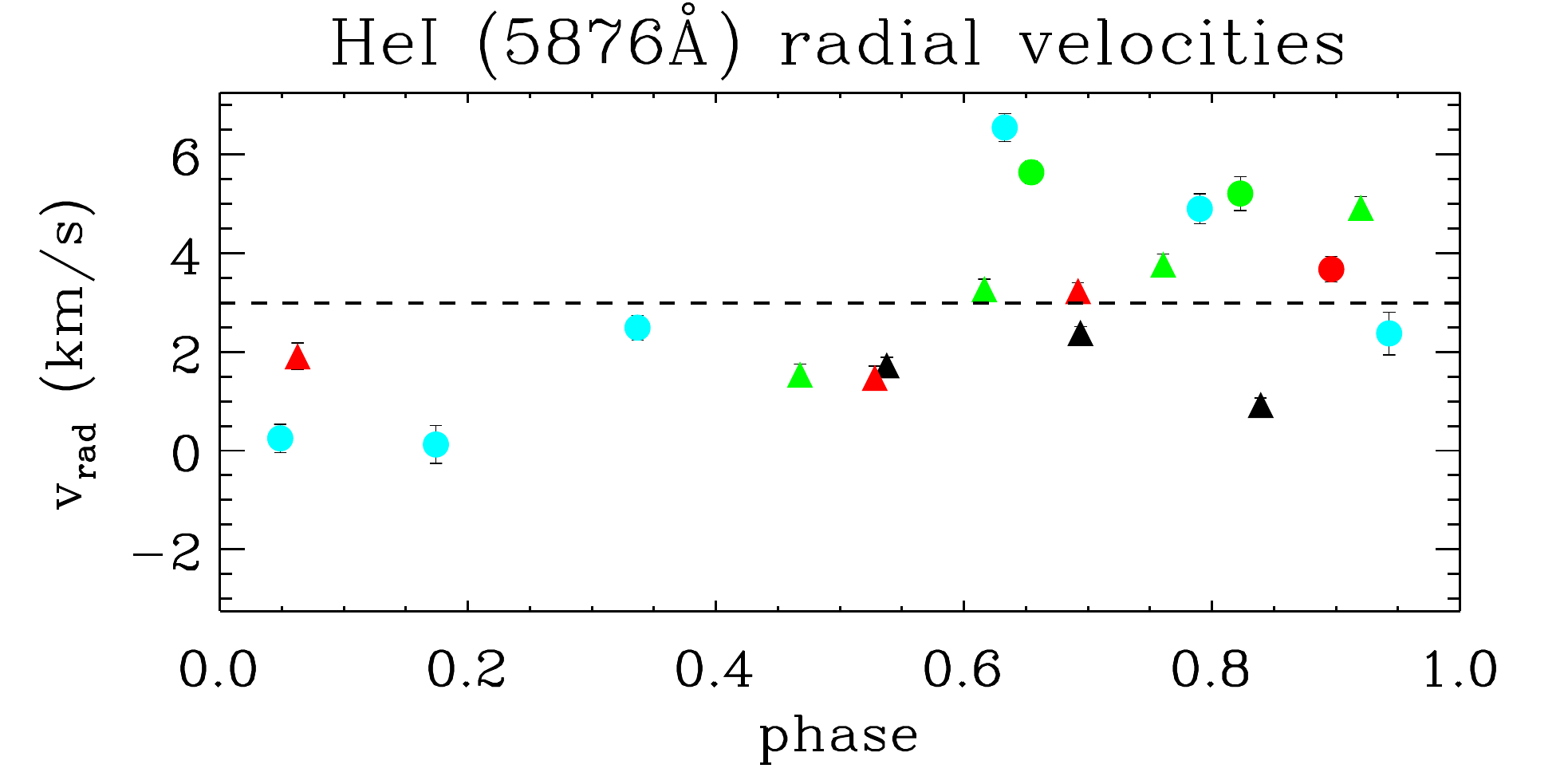}}
\subfigure[]{\includegraphics[scale=0.40]{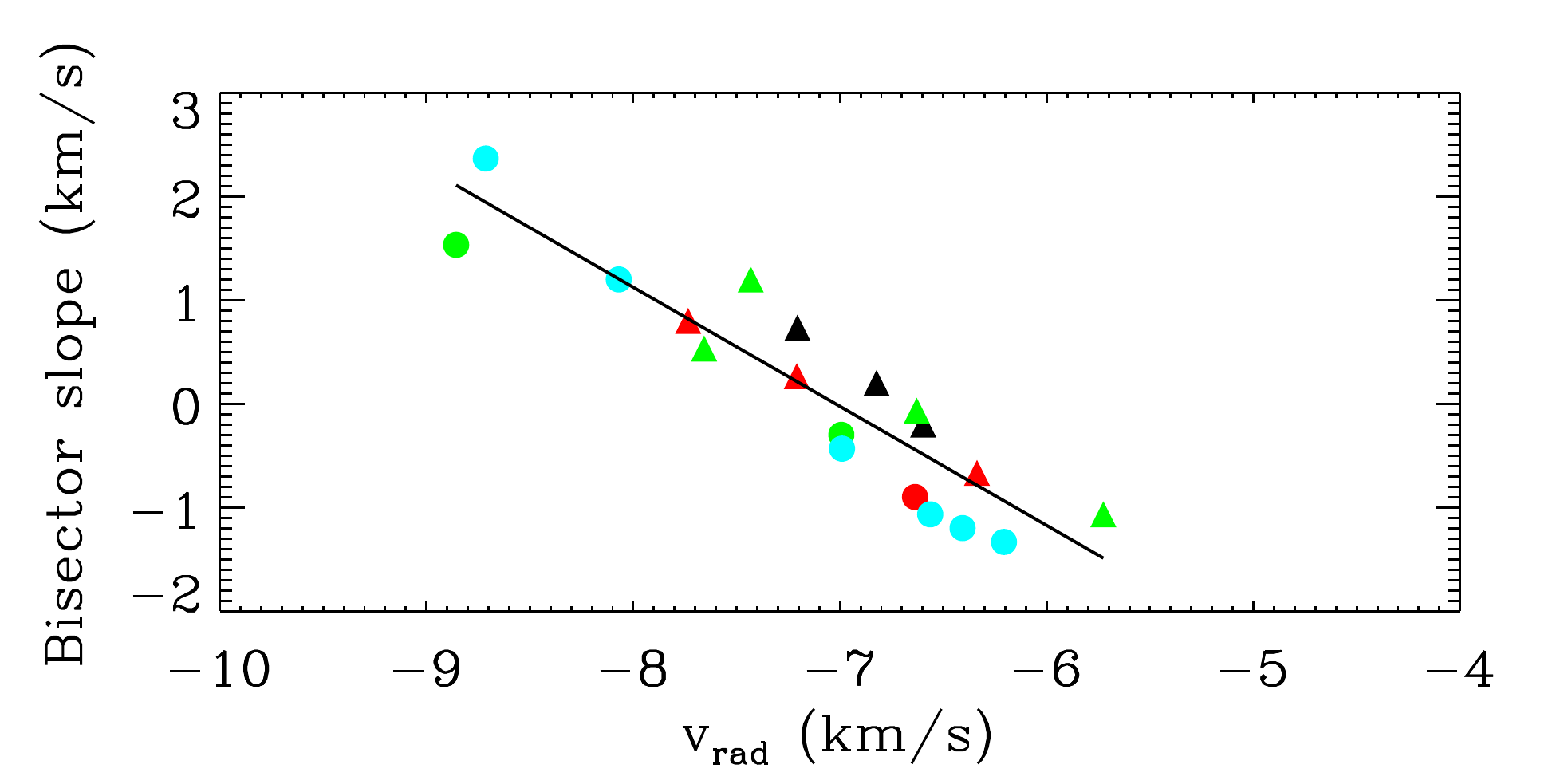}}
\caption{\label{fig:Parm_phase_all} Veiling {\it (a)} and the radial velocity {\it (b)} of photospheric lines in phase with the  ephemeris from \cite{2012A&A...541A.116A} and \cite{2007MNRAS.380.1297D}. The veiling and radial velocity correspond to mean values obtained from all spectroscopic regions used in the calculations. The error bars come from the standard deviations. {\it (c)} $\mathrm{He}$\,I\ 5876$\mathring{\mathrm{A}}$ line radial velocities. {\it (d)} Bisector slope as a function of the radial velocities. The colors represent different cycles: black - cycle 0; green - cycle 1; red - cycle 2; and light blue - cycle 3. The dashed line is the mean value of all observations. Triangles are ESPaDOnS data, and circles are HARPS data.} 
\end{figure}

In Fig. \ref{fig:Parm_phase_all} we present the veiling and radial velocity obtained for V2129 Oph as a function of the rotational phase. Since we did not see a significant veiling variation with wavelength in the analyzed spectral regions, we only show the mean veiling of all the spectral regions.  
V2129 Oph presented a low veiling with modest variations from night to night, except on the first three nights observed by ESPaDOnS (black triangles in Fig \ref{fig:Parm_phase_all}a), which show the largest veiling values.
The radial velocity seems to vary in phase with the rotation of the star, and it shows two minima at different phases, which can be associated with two sets of spots at different longitudes that appear in our line of sight as the star rotates. We computed the periodogram of the radial velocities and found a period of $3.26\,\mathrm{days}$ (false alarm probability [$\mathrm{FAP]} =0.05$), that is, half of the stellar rotation period.

\begin{figure}
 \centering
\includegraphics[scale=0.30]{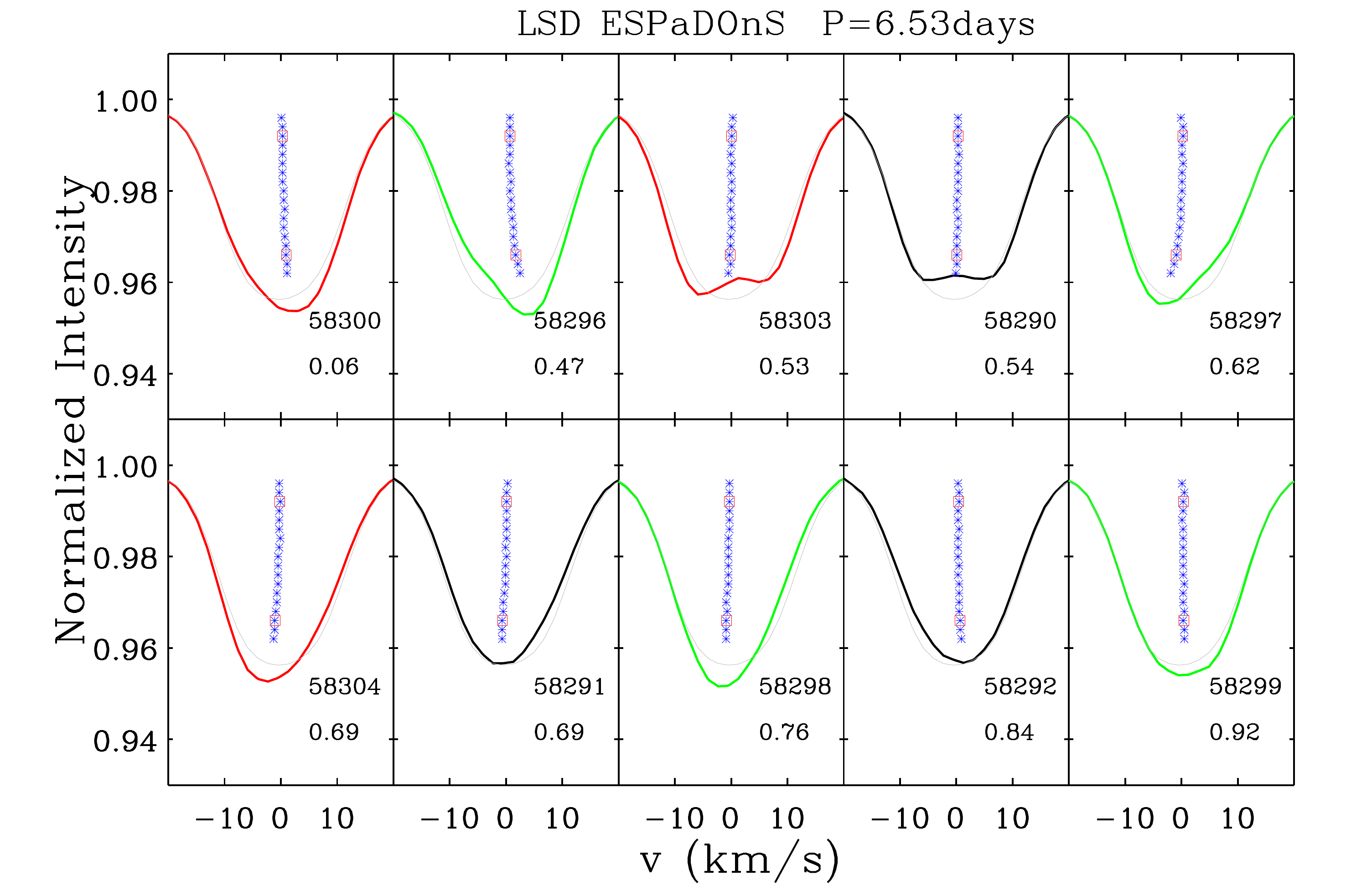}
\includegraphics[scale=0.30]{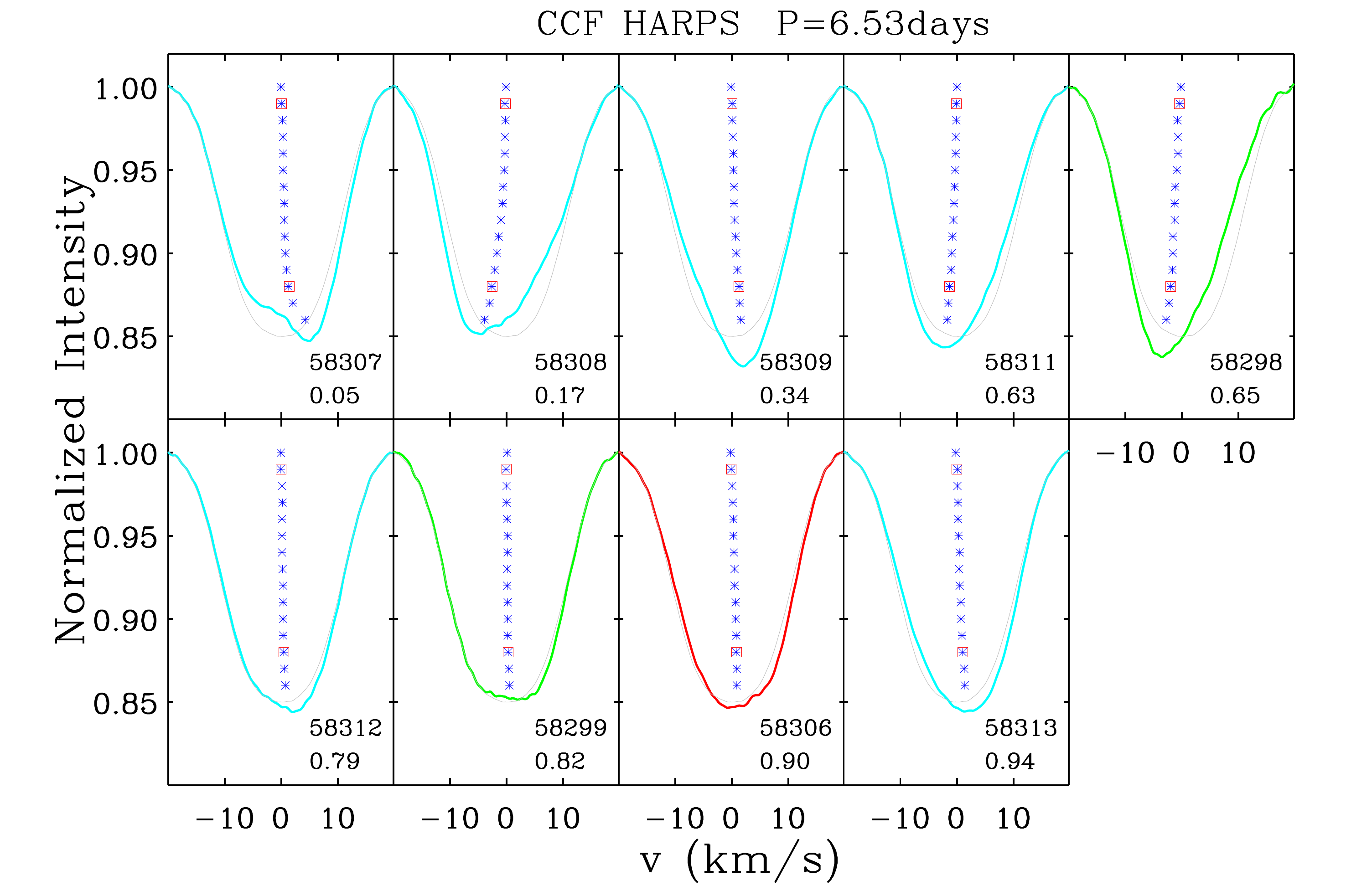}
\caption{\label{fig:lsd} LSD and CCF mean photospheric lines. The blue points are the bisectors, and the red squares represent the points of reference used to measure the bisector slope. The profiles are organized by phase, and the gray profile is the average line over all the observations.} 
\end{figure}

We analyzed the asymmetry of the photospheric lines  using the bisector of the line to determine the origin of the radial velocity variations \citep[e.g.,][]{2001A&A...379..279Q}. We used the photospheric lines obtained through least-squares deconvolution (LSD) profiles for the ESPaDOnS data and cross-correlation function (CCF) profiles for the HARPS data; these photospheric lines are shown in Fig. \ref{fig:lsd}. 
We measured the bisector in each observation and selected two regions at the top and bottom of the bisector line to measure the bisector slope, which is the difference in the velocity between the upper and lower parts of the bisector.
In Fig \ref{fig:Parm_phase_all} we show the bisector slope as a function of radial velocity. The bisector slope is found to vary inversely with the photospheric line radial velocity, which confirms that the radial velocity modulation is due to stellar spots. 

The bisector slope varies with a period of $3.26\,\mathrm{days}$ ($\mathrm{FAP} =0.08$), similar to the period of the radial velocity modulation. 
We conclude that the stellar surface is covered by two main sets of cold spots or polar spot extensions to lower latitudes lying at nearly opposite longitudes, which is in agreement with the photometric minima seen in the optical light curve. We therefore confirm the rotation period of the star ($\mathrm{P} = 6.53\,\mathrm{days}$) from both the photospheric radial velocity variations and the optical light curve of the system at the time of our observations.

\subsection{Optical emission lines} \label{sec:opticallines}
We applied the veiling and the photospheric radial velocity of V2129 Oph described in the last section to the spectra of V819 Tau, which we used as a template to subtract the photospheric component of the spectra of V2129 Oph and obtain residual profiles. As pointed out in the last section, we measured a systematic veiling difference between the HARPS and ESPaDOnS data. For analyze these data sets together, we corrected the HARPS data from this systematic veiling offset.  We show the mean line profiles before and after subtracting the photospheric contribution, together with the mean template profiles, in Fig \ref{fig:resid}.

\begin{figure} 
 \centering
\includegraphics[scale=0.30]{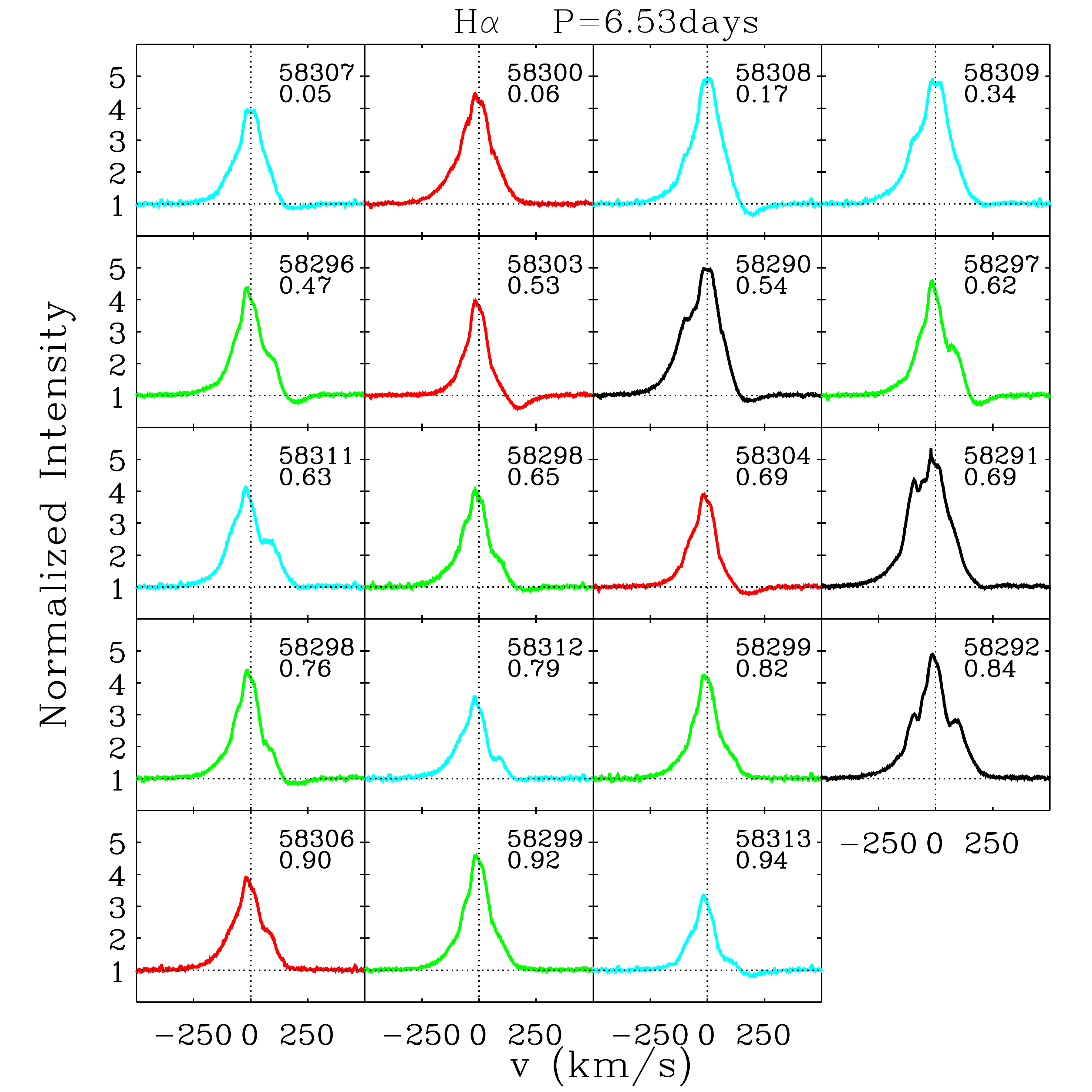}
\caption{\label{fig:Halphaline_all} $\mathrm{H}\alpha$ residual line profiles obtained with the ESPaDOnS and HARPS spectrographs. We organized the profiles by phase. The phase and the JD of the observations are written in each panel. The colors represents different rotational cycles: black - cycle 0; green - cycle 1; red - cycle 2; and light blue - cycle 3.} 
\end{figure}

\begin{figure} 
 \centering
\includegraphics[scale=0.30]{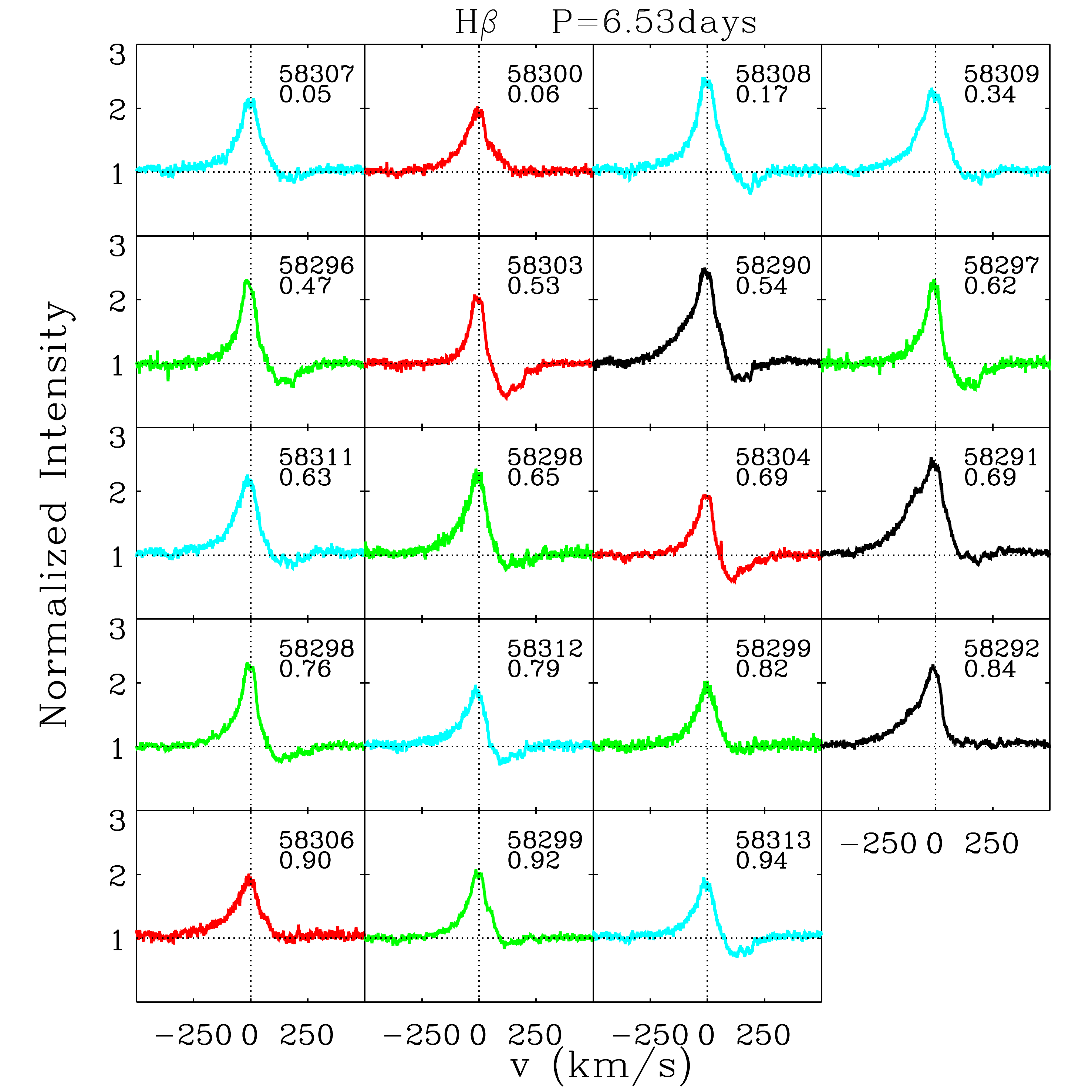}
\caption{\label{fig:Hbetaline_all}$\mathrm{H}\beta$ residual line profiles obtained with the ESPaDOnS and HARPS spectrographs. We organized the profiles by phase. The phase and the JD of the observations are written in each panel. The colors represent different rotational cycles: black - cycle 0; green - cycle 1; red - cycle 2; and light blue - cycle 3.} 
\end{figure}

\begin{figure} 
 \centering
\includegraphics[scale=0.30]{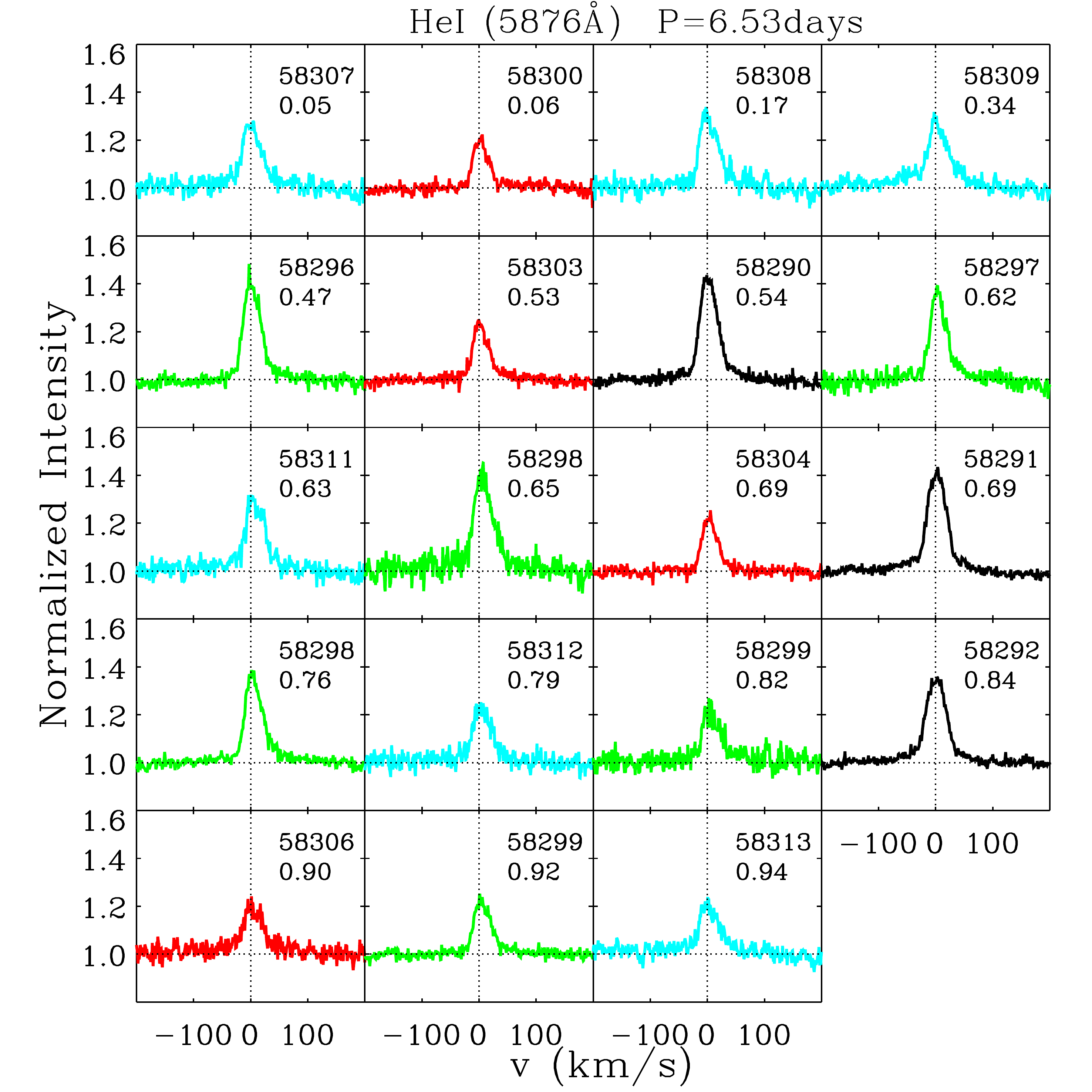}
\caption{\label{fig:HelIline_all}$\mathrm{He}$\,I\ 5876$\mathring{\mathrm{A}}$ residual line profiles obtained with the ESPaDOnS and HARPS spectrographs. We organized the profiles by phase. The phase and the JD of the observations are written in each panel. The colors represent different rotational cycles: black - cycle 0; green - cycle 1; red - cycle 2; and light blue - cycle 3.} 
\end{figure}

We present the residual profiles of the H$\alpha$, H$\beta$, HeI 5876\AA, and NaI D lines, obtained with HARPS and ESPaDOnS, in Figs. \ref{fig:Halphaline_all} to \ref{fig:perline_all}. We corrected all the profiles from the mean radial velocity, $v_{rad}=(-7.1\pm 0.8)\,\mathrm{km\ s^{-1}}$, and organized the profiles by rotational phase defined by Eq. \ref{eq:eph}. V2129 Oph presents a chromospheric emission contribution that affects some emission lines of interest, such as $\mathrm{H}\alpha$, $\mathrm{H}\beta$, and $\mathrm{Ca}$\,II\ (8542$\mathring{\mathrm{A}}$). For the Balmer lines, the chromospheric component, if present, is fainter than the emission due to accretion. We did not attempt to disentangle the chromospheric and emission components in these lines. When subtracting the V819 Tau spectrum from the Balmer line profiles, we set the narrow chromospheric emission present in the Balmer lines of V819 Tau to the continuum level since we do not know whether or not it is the same as in our target (see Appendix~\ref{sec:residual}). Therefore, the residual profiles of the V2129 Oph Balmer lines are a composition of chromospheric and circumstellar emission. On the other hand, the narrow chromospheric emission is significant, compared to the broad emission component, in the observed $\mathrm{Ca}$\,II\ (8542$\mathring{\mathrm{A}}$) line profiles. In this case, we fitted  the $\mathrm{Ca}$\,II\ (8542$\mathring{\mathrm{A}}$) emission line with two Gaussians and subtracted the narrow chromospheric component from the observed profiles. 

The $\mathrm{H}\alpha$ and $\mathrm{H}\beta$ emission lines of V2129 Oph (Figs. \ref{fig:Halphaline_all} and \ref{fig:Hbetaline_all}, respectively) are broad, intense, and variable from night to night, which are characteristics of accreting systems \citep[e.g.,][]{2001ApJ...550..944M}. The lines present redshifted absorptions below the continuum at some rotational phases, commonly associated with the passage of the accretion funnel across our line of sight. It is easier to see the redshifted absorption in the $\mathrm{H}\beta$ line as this line profile presents less intense emission than that of $\mathrm{H}\alpha$  \citep[e.g.,][]{2013MNRAS.431.2673K}.  
The redshifted absorptions appear continuously from phases 0.47 to 0.79 and then sporadically at phases 0.17 and 0.94, which suggests that several independent funnel flows cross our line of sight.

In Fig. \ref{fig:HelIline_all} we show the $\mathrm{He}$\,I emission line. This line can present both a narrow and a broad component \citep[e.g.,][]{beristain01}.  It is believed that the narrow component forms in the post-shock region, and its detection is an accretion diagnostic; the broad component, when present, is associated with the accretion column or the base of a hot wind \citep[e.g.,][]{beristain01}. The $\mathrm{He}$\,I line of V2129 Oph shows mostly a narrow component (full width at half maximum [$FWHM]=36\pm7\,\mathrm{km\ s^{-1}}$). At the base of the profile, there is some sign of a broad component, especially in the profiles shown in black in Fig. \ref{fig:HelIline_all}; these profiles correspond to the first observed rotational cycle, which also presents the highest veiling values. We probably witnessed an accretion burst episode in the first cycle that might also have affected the inner disk wind and the magnetosphere. However, as the intensity of the broad emission components in the  $\mathrm{He}$\,I line is very low, we treated this line as if it only had a narrow component. In \cite{2012A&A...541A.116A}, this line also presents mostly a narrow component.

We fitted a Gaussian function to the narrow component of the $\mathrm{He}$\,I line to get the parameters of the line. We used the position of the maximum intensity of the Gaussian profile as the radial velocity of this line in the stellar rest frame, and we show its night-to-night variability in Fig. \ref{fig:Parm_phase_all}c. The mean radial velocity of the $\mathrm{He}$\,I line is $v_{rad_{HeI}}=(3\pm2)\,\mathrm{km\ s^{-1}}$; this is a similar result to that obtained by \cite{2007MNRAS.380.1297D,2011MNRAS.412.2454D}, which shows that the profile is redshifted. In the literature, this line often appears redshifted, which is in agreement with a formation in the post-shock region \citep[e.g.,][]{beristain01}.

The NaI doublet lines ($5889.950\,\mathrm{\mathring{A}}$ and $5895.924\,\mathrm{\mathring{A}}$) present a deep and broad photospheric contribution, which masks most of the circumstellar emission. As for the previous lines, we removed the photospheric absorptions, using the V819 Tau spectra as a template. We also corrected the lines of telluric absorptions using the TAPAS web-based tool \citep{2014A&A...564A..46B}. 

The NaI D line can form in distinct locations in the system, such as the accretion funnels, outflows, or stellar photosphere, and it can also be present in interstellar components \citep[e.g.,][]{2015ApJ...814...14P}. Figure \ref{fig:perline_all} shows that the residual NaI $5889.950\,\mathrm{\mathring{A}}$ profile appears in emission and is variable in intensity. The NaI $5895.924\,\mathrm{\mathring{A}}$ line exhibits a similar behavior.  
The center of the NaI line displays a narrow, deep, and stable absorption component, most likely due to absorption by the interstellar medium \citep[][]{2015ApJ...814...14P}. 
The blue wing of the line exhibits several absorption components that are clearly variable. These blueshifted absorptions are also present in several active T Tauri stars \citep[e.g.,][]{1984ApJ...280..749M,2013AJ....145..108C,2020A&A...642A..99P}. \cite{1984ApJ...280..749M} analyzed a sample of T Tauri stars that showed small blueshifted absorptions at high velocities, which indicates the presence of an accelerated wind with a velocity of up to a few hundred $\mathrm{km\ s^{-1}}$. In the NaI line of V2129 Oph, we see at least two blueshifted absorption components, located around $-40$ and $-80\,\mathrm{km\ s^{-1}}$, among other higher-velocity components ($<-100\,\mathrm{km\ s^{-1}}$) that only appear on some nights,  as well as may represent transient outflows. 

We also analyzed the residual profiles of the $\mathrm{Ca}$\,II\ 8542$\mathring{\mathrm{A}}$ line. This line presents two components: a broad component commonly associated with the magnetosphere and a narrow stable component of chromospheric origin. The intensity of the broad component varies from night to night without any detectable periodicity, and, at some phases, the profile presents a redshifted absorption component.

The equivalent width of the emission lines carries accretion diagnostics and is frequently used to distinguish between accreting and non accreting systems \citep[e.g.,][]{2003ApJ...582.1109W,2009A&A...504..461F}. We measured the equivalent widths of the emission lines using both ESPaDOnS and HARPS data. We estimated the equivalent width uncertainties caused by the quality of the data and the uncertainty on the continuum level of the spectra \citep{1995chst.conf..207E,1988IAUS..132..345C}. 
The values are listed in Table \ref{tab:parameters}, where we adopted the convention that positive and negative equivalent widths represent emission and absorption lines, respectively. The line equivalent widths do not vary in phase with the stellar rotation period (see Fig. \ref{fig:EWlines_all2}). They also do not present a significant correlation with the veiling, probably due to the small veiling variation along the observational campaign, with the exception of an increase in equivalent width in most lines during the first three observations, which also corresponds to the highest veiling measurements in our data.

\subsection{Circumstellar line periodicity} \label{sec:period}
We degraded the HARPS data to the ESPaDOnS resolution to analyze the variability of the 19 observations all together.  With this data set, we computed the emission line bidimensional periodograms to check if the variability of the circumstellar lines were periodic or not  across the line profiles \citep[e.g.,][]{1995ApJ...449..341J,2000A&A...362..615O,2002ApJ...571..378A,2005MNRAS.358..671K,2007A&A...463.1017B,2016A&A...586A..47S}. We divided each emission line into small velocity intervals ($0.5\,\mathrm{km\ s^{-1}}$ each), and for each interval we applied the Lomb-Scargle periodogram, as modified by \cite{1986ApJ...302..757H}. We searched for periodicity from $2$ to $10\,\mathrm{days}$ to compare with the rotational period of V2129 Oph, which is $6.53\,\mathrm{days}$ \citep[][]{2012A&A...541A.116A}.  

\begin{figure*}
 \begin{center}
{\includegraphics[width=4.5cm]{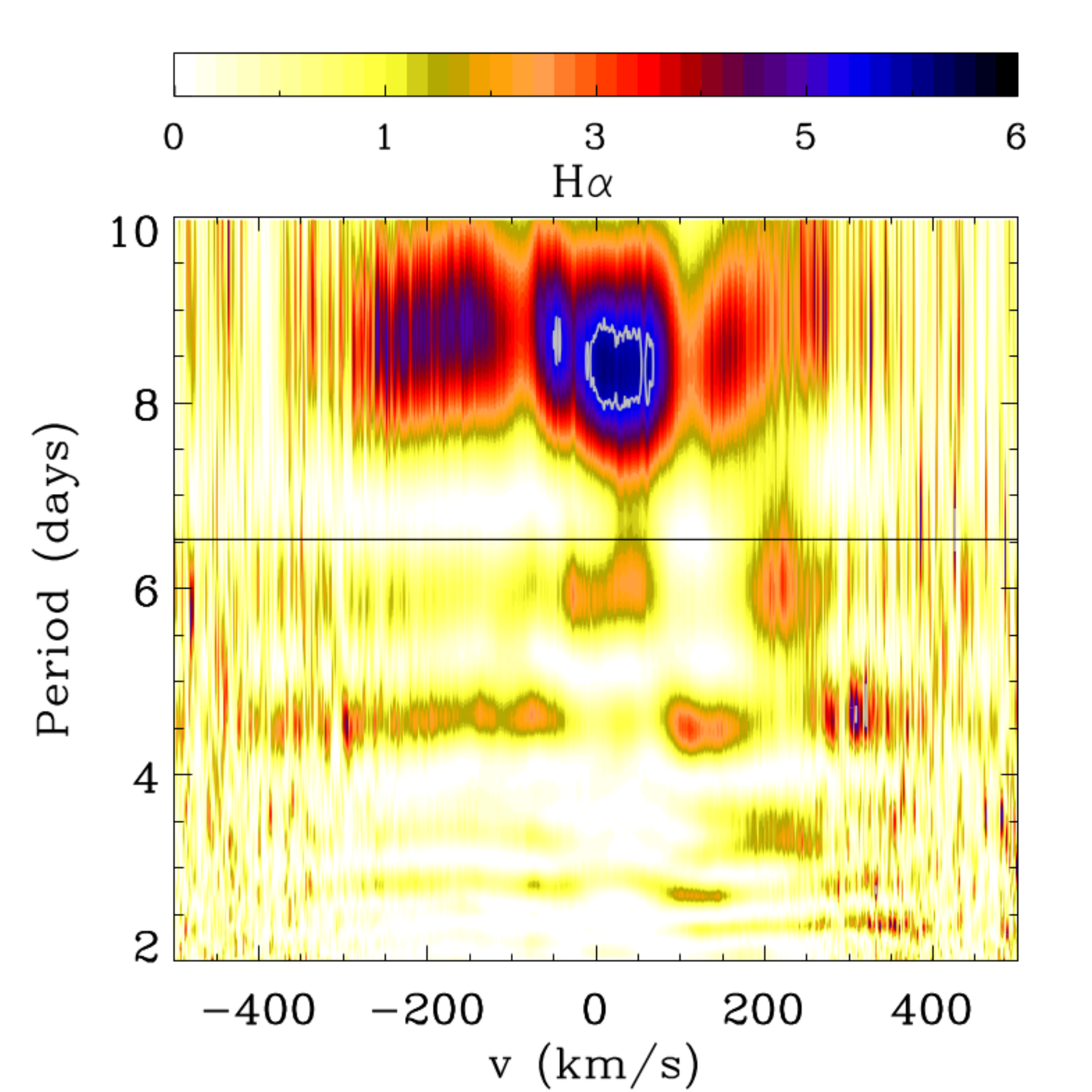}}
{\includegraphics[width=4.5cm]{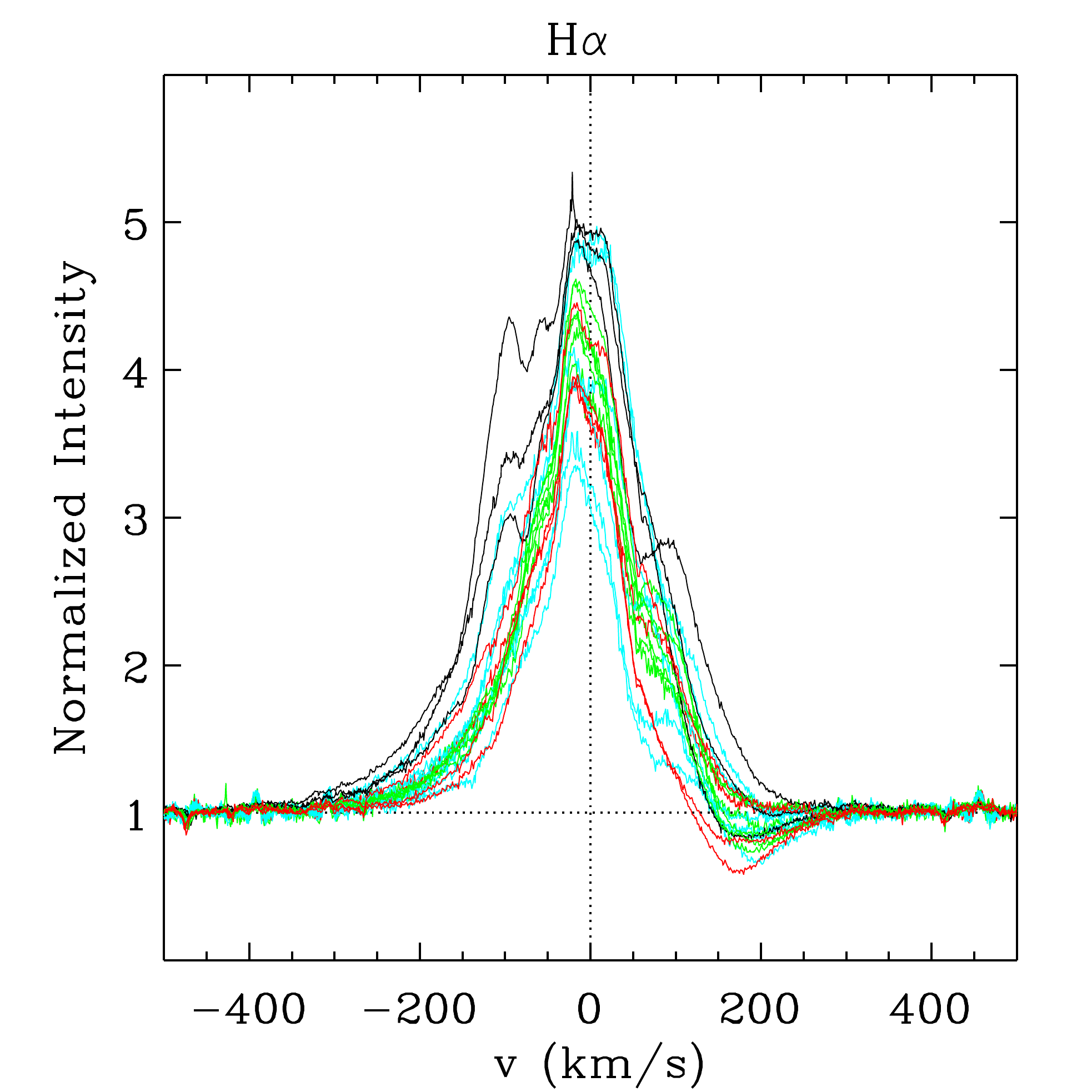}}
{\includegraphics[width=4.5cm]{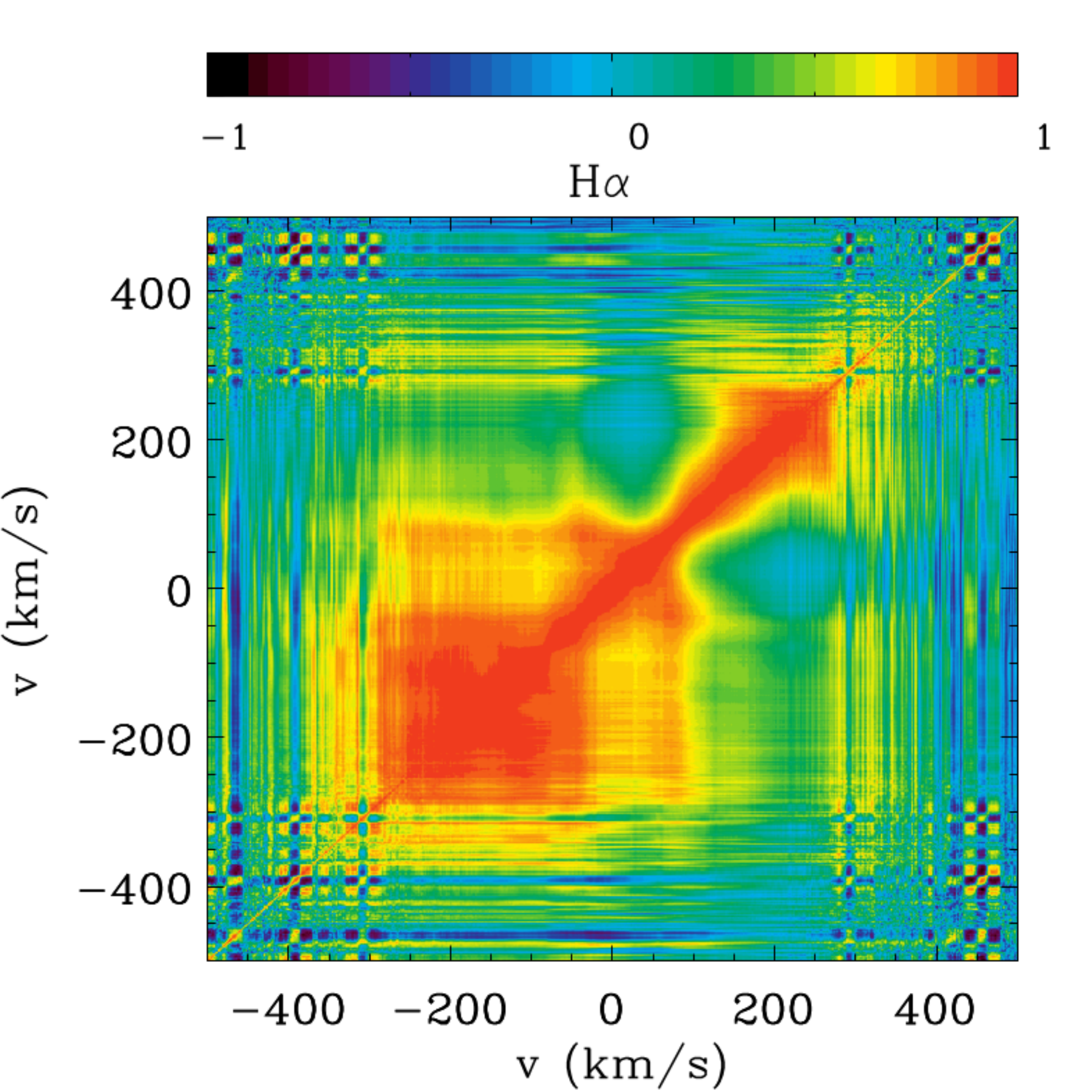}}\\
{\includegraphics[width=4.5cm]{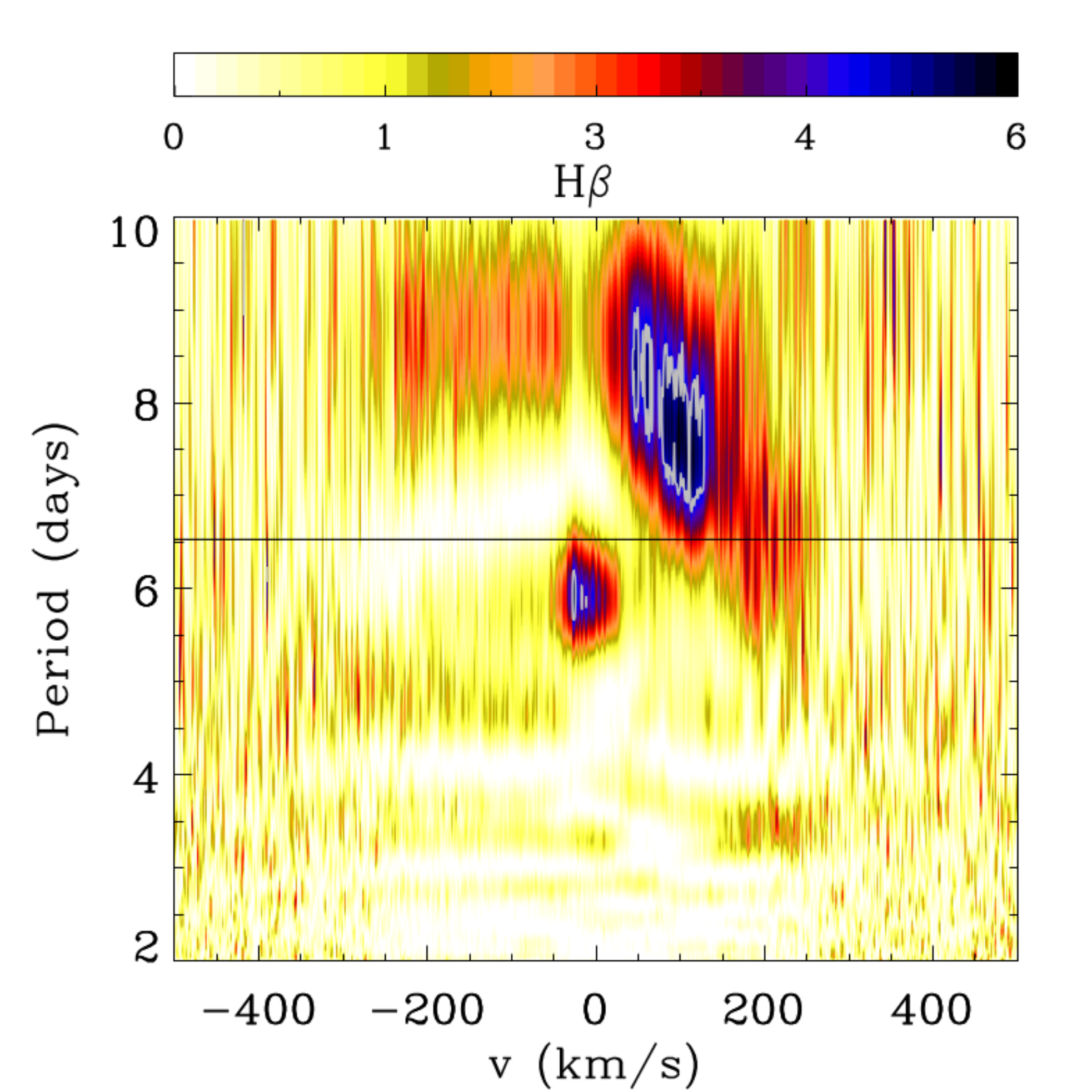}}
{\includegraphics[width=4.5cm]{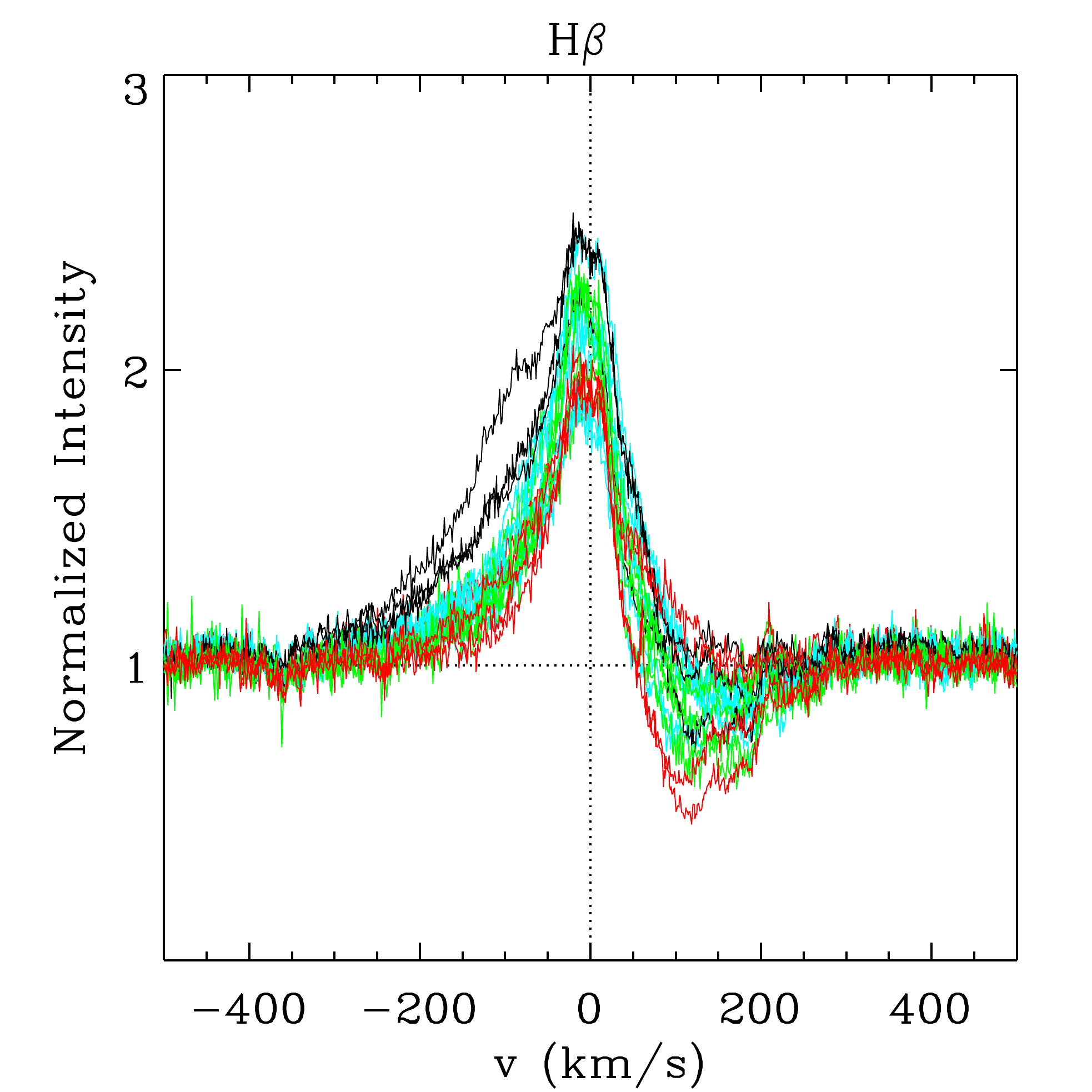}}
{\includegraphics[width=4.5cm]{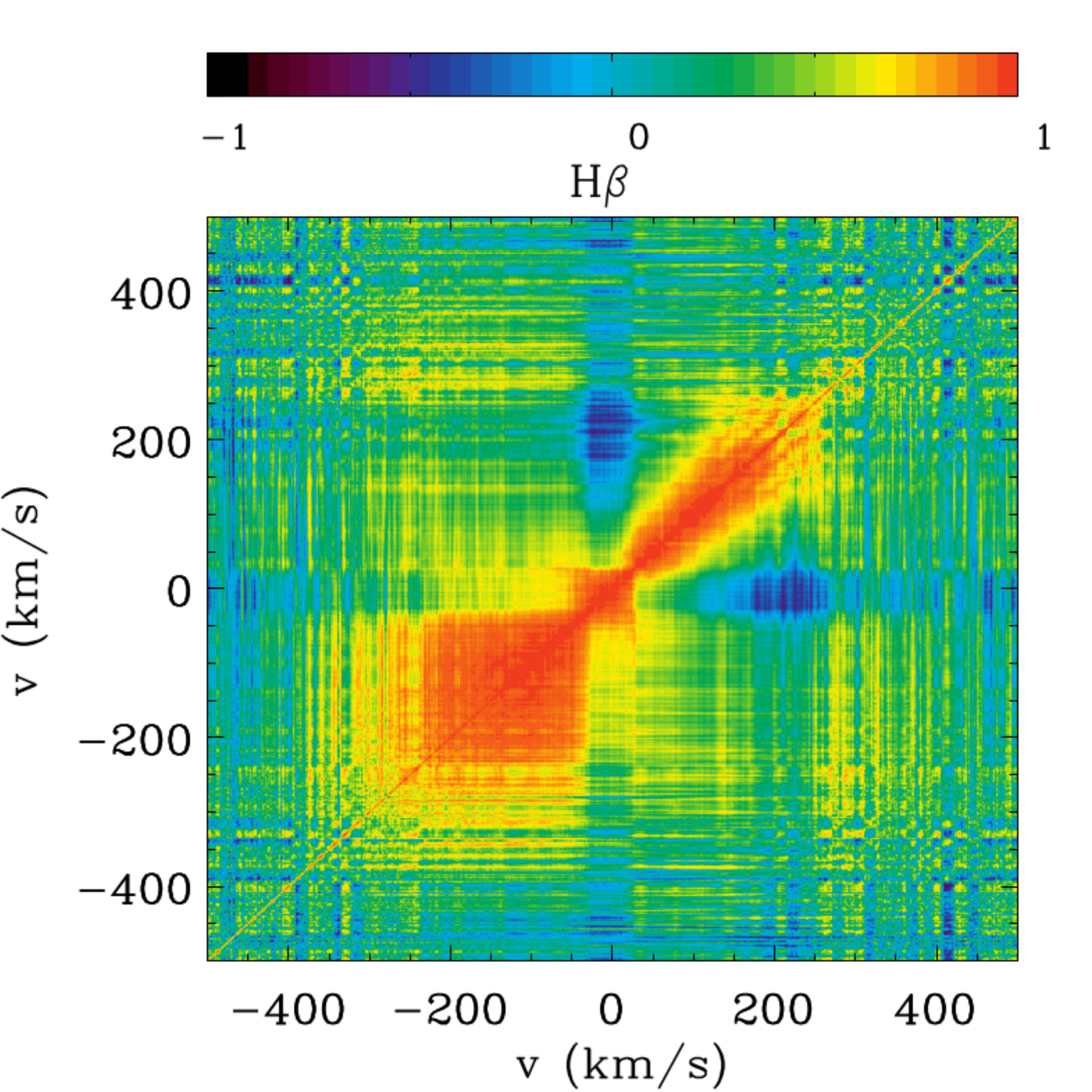}}\\
{\includegraphics[width=4.5cm]{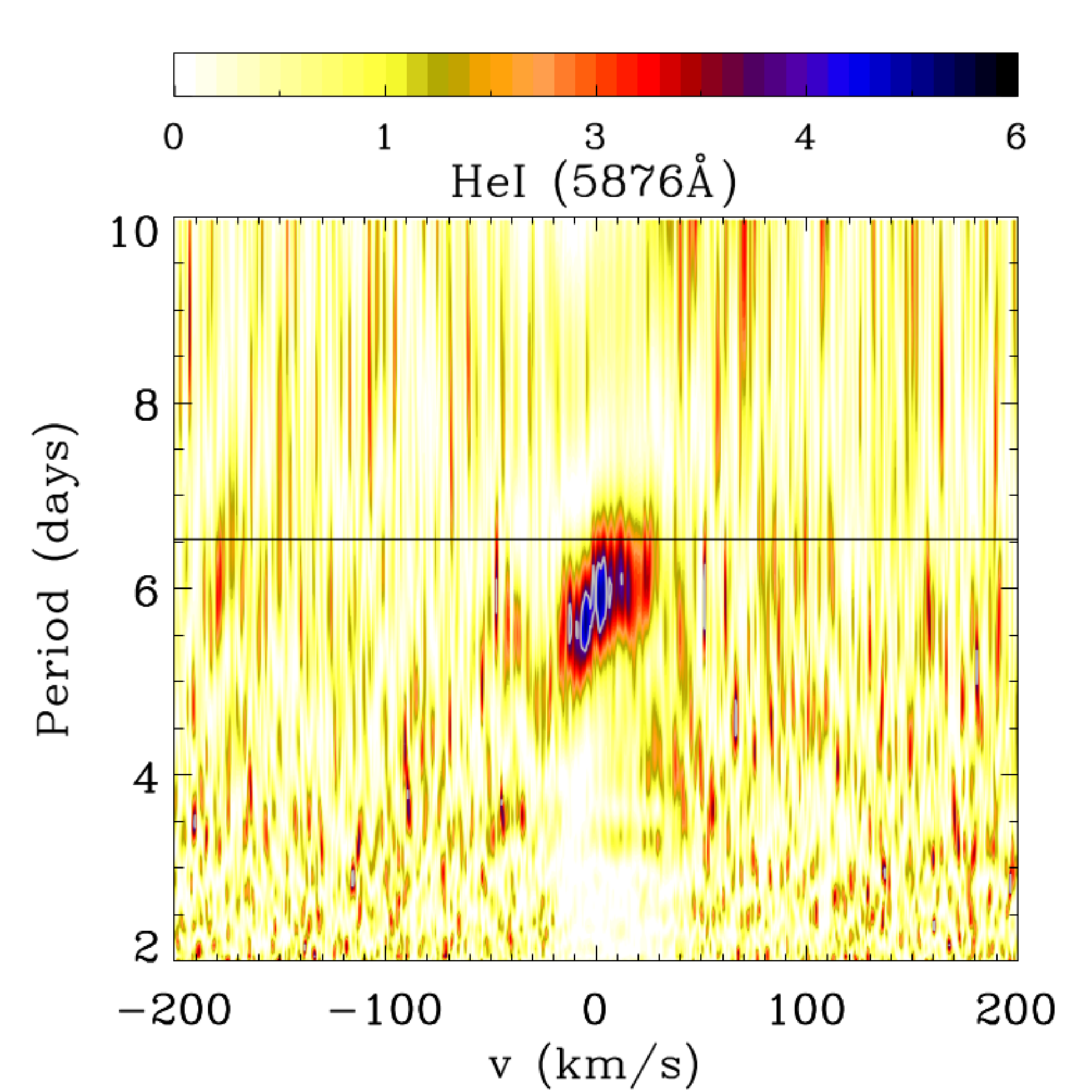}}
{\includegraphics[width=4.5cm]{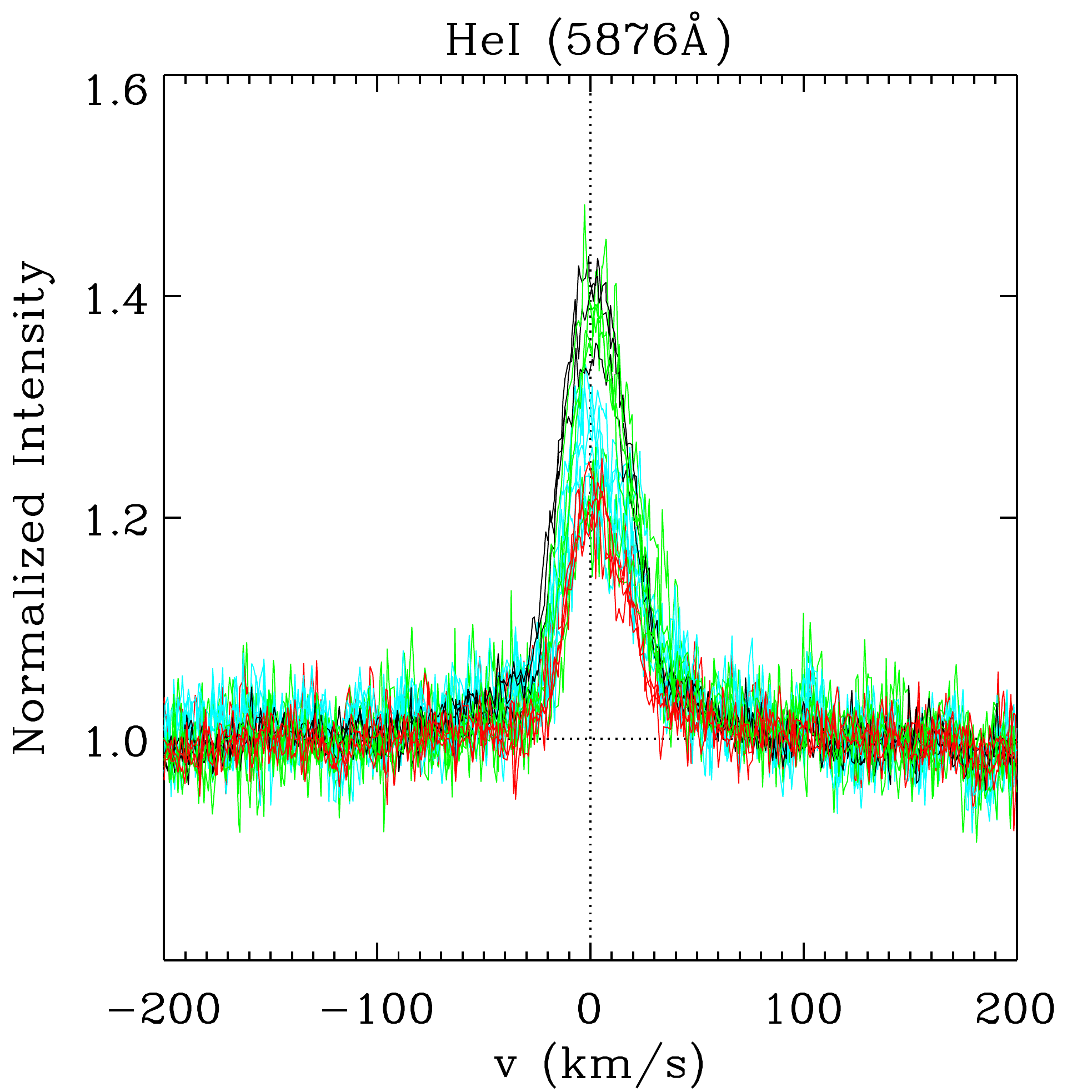}}
{\includegraphics[width=4.5cm]{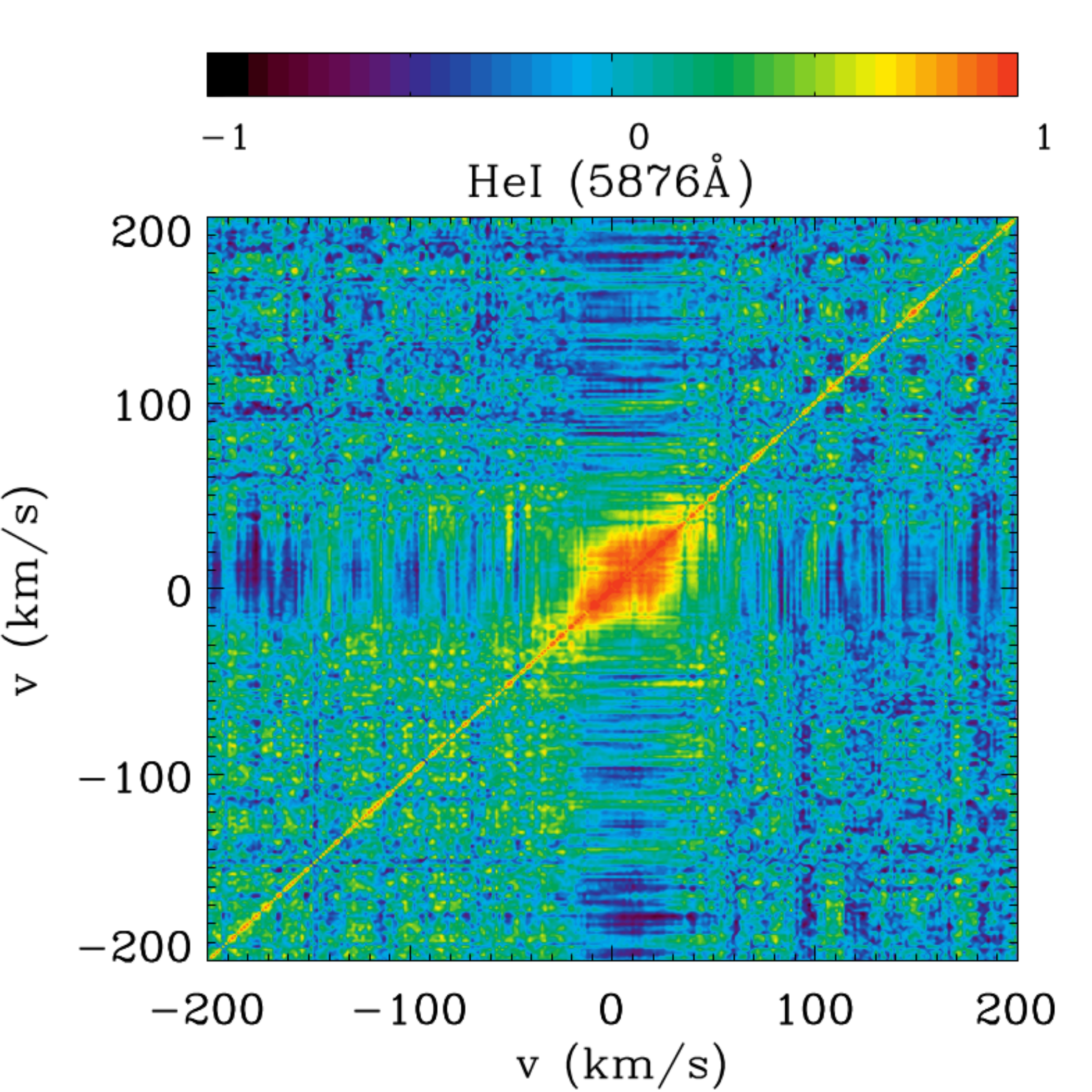}}\\
{\includegraphics[width=4.5cm]{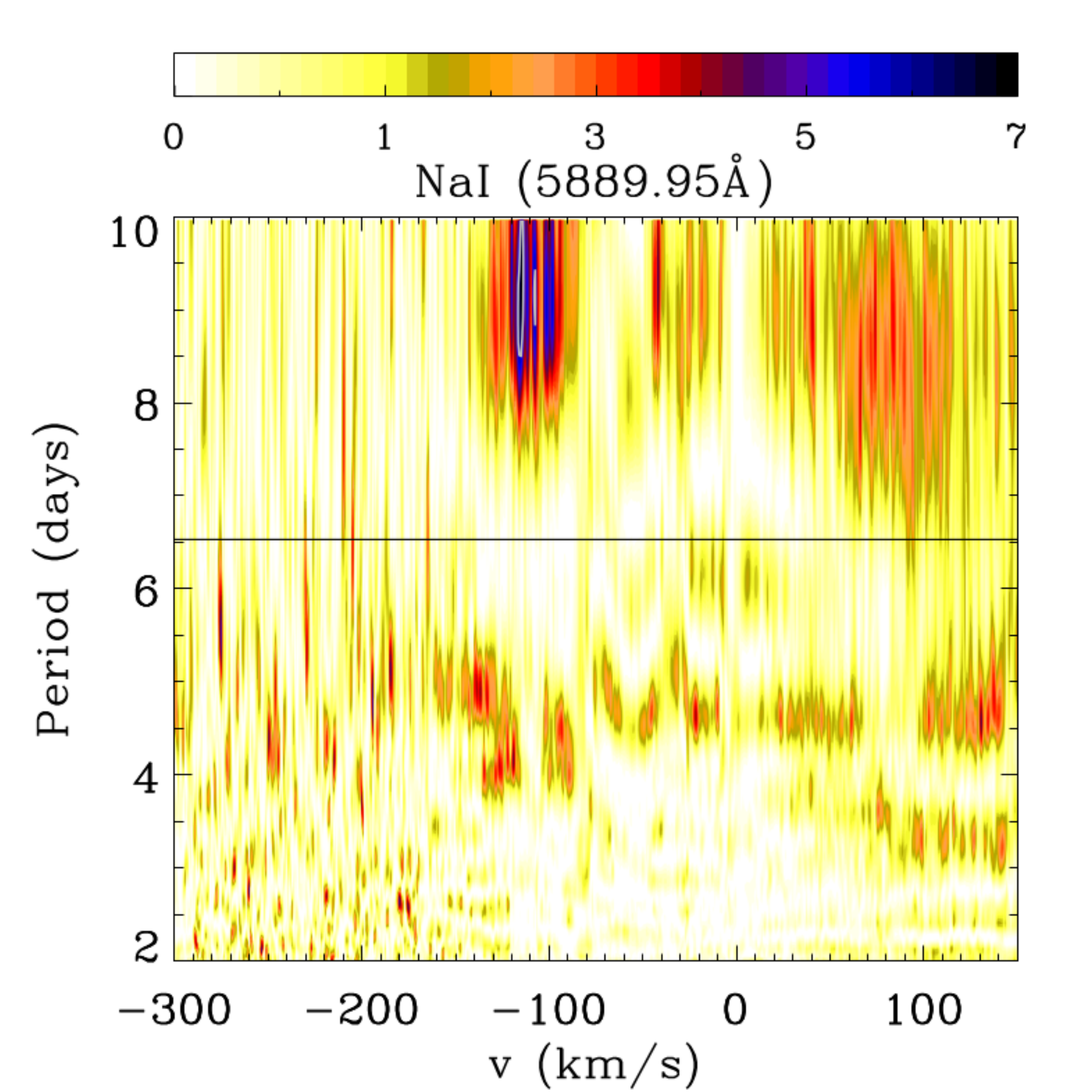}}
{\includegraphics[width=4.5cm]{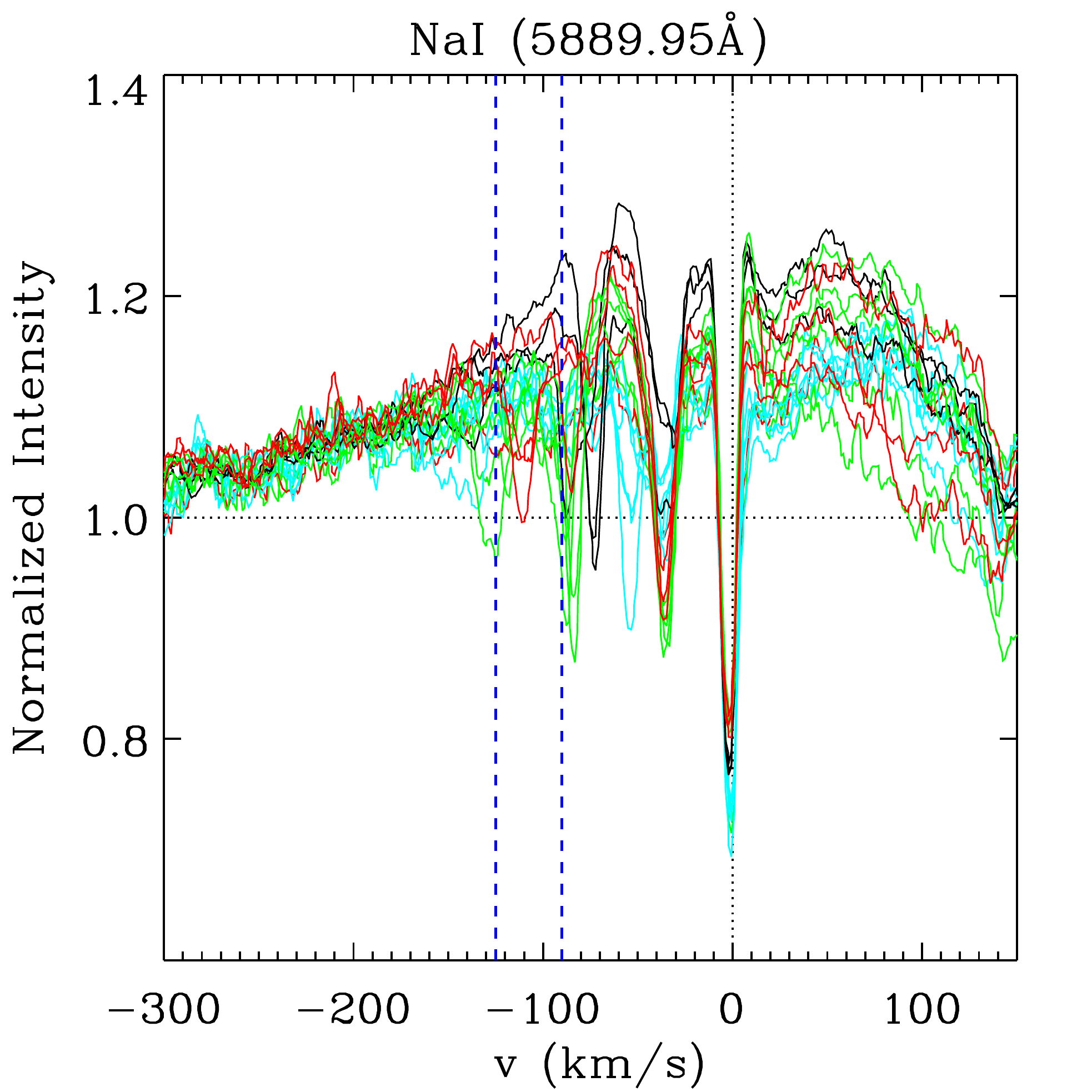}}
{\includegraphics[width=4.4cm]{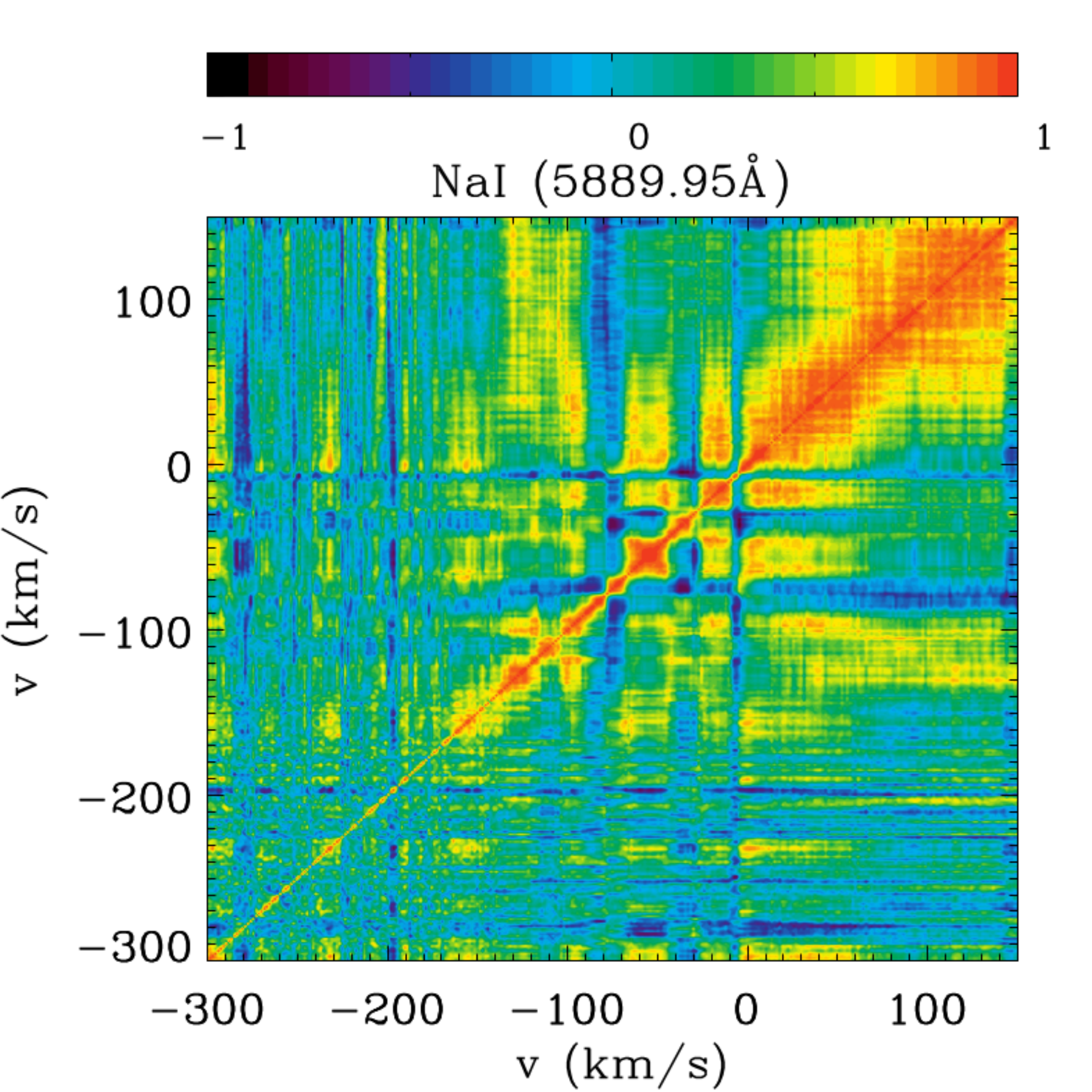}}\\
 \end{center}
\caption{\label{fig:perline_all}Time series analysis of circumstellar emission lines. {\it Left}: Bidimensional periodograms.\ The color range corresponds to the power of the periodogram, and the horizontal line represents the star's rotation period, $P=6.53\,\mathrm{days}$. The white contour represents the FAP level of about $95\%$ confidence. {\it Middle}: Line profiles obtained with the ESPaDOnS and HARPS spectrographs. The colors correspond to different rotation cycles. The two blue vertical lines delimit the periodic region of the Na I line.  {\it Right}: Correlation matrices. The color range corresponds to the value of the linear correlation coefficient. Perfect anticorrelation corresponds to -1 (black), no correlation to 0 (light blue), and a perfect correlation to 1 (orange).}
\end{figure*}

We display the bidimensional periodograms in Fig. \ref{fig:perline_all}, where the periodogram's  power is shown in colors. We also show a contour plot that represents the FAP level of about $95\%$  confidence. To check if the periodicities obtained for these lines are true or not, we performed two tests: We randomly removed two observing nights, and we randomly removed seven spectra over the 19 and redid the periodogram analysis. We repeated both processes 50 times for each line. Although these tests increased the periodogram noise, in almost all of these cases the periodograms still have the maximum power on the period seen on the plots in Fig \ref{fig:perline_all}.

The HeI line variability shows a periodicity ($\sim 6\,\mathrm{days}$; see Fig \ref{fig:perline_all}e) close to the stellar rotation period, although smaller by~0.5 days. The narrow component of the HeI line is supposed to form in the post-shock region; therefore, we would expect the helium line to present a periodicity at the stellar rotation period if the system were in a stable accretion process, where accretion occurs through two main accretion funnels, one in each hemisphere \citep[e.g.,][]{2008MNRAS.385.1931K,2013MNRAS.431.2673K}. %Due to differential rotation?  

The $\mathrm{H}\alpha$ and $\mathrm{H}\beta$ lines can form in different places as the accretion column and  disk wind. The periodicity of these lines therefore reflects the variability of these regions. The $\mathrm{H}\alpha$ line shows a period of $\sim 8.5\,\mathrm{days}$ along the profile, with the strongest power near the line center from $\sim -100$ to $100\, \mathrm{km\ s^{-1}}$. This period is longer than the star's rotation period, indicating that the region where the line forms rotates more slowly than the star and should be outside the corotation radius  \citep[$\mathrm{R_{co}}=7.7\,\mathrm{R}_\ast=0.07\,\mathrm{au}$;][]{2012A&A...541A.116A} of the star-disk system. 

The periodogram of the $\mathrm{H}\beta$ line shows several periods. The central emission of the line, from $-50$ to $25\,\mathrm{km\ s^{-1}}$, presents a periodicity of $\sim 6\,\mathrm{days}$, which is the same period seen in the $\mathrm{He}$\,I line. We can associate this period with the hot spot, as in the $\mathrm{He}$\,I line. 
Another period is detected in the redshifted wing of the line. The maximum power of the periodogram goes from $\sim8.5\,\mathrm{days}$ (the same period seen in the $\mathrm{H}\alpha$ line) close to the center of the line to $\sim6.5\,\mathrm{days}$ (the stellar rotation period) in the redshifted absorption. The chromospheric component of V2129 Oph probably contributes to the central emission of the \ha\ and \hb. Although this component is typically faint and narrow, compared to the broad and intense emission due to accretion in CTTSs, it could interfere with the measured periodicity in the center of these lines (see Appendix~\ref{sec:residual}).

In Fig. \ref{fig:perline_all} we present the periodogram of the Na I $5889.950\,\mathring{\mathrm{A}}$ line. This component only shows a period, with broad power, of around $9\,\mathrm{days}$ in a small part of the blue wing. This region corresponds to a narrow variable absorption of unknown origin that disappears in some observations. This higher-velocity absorption commonly appears in the blue wing of the NaI of T Tauri stars \citep[][]{1984ApJ...280..749M} as a result of stellar and disk winds. We highlight this region in Fig. \ref{fig:perline_all}. The periodogram of the other component of the NaI line ($5895.924\mathring{\mathrm{A}}$) shows the same periodicity at the same velocity range but with lower power.

\cite{2012A&A...541A.116A} also analyzed the periodogram of the $\mathrm{H}\alpha$ line of V2129 Oph. They found a periodicity at the stellar rotation period in the redshifted absorptions. On the blue side of the line, they found a more extended period of $8.3\,$ days, similar to our results. In Appendix \ref{sec:data2012}, we show the periodogram and the circumstellar lines of V2129 Oph from the data analyzed by \cite{2012A&A...541A.116A}, including the periodograms of $\mathrm{He}$\,I and $\mathrm{H}\beta$ lines, which they did not present in their paper. The $\mathrm{He}$\,I and $\mathrm{H}\beta$ lines from the observations of \cite{2012A&A...541A.116A} present variability at the star's rotational period. 

The comparison between the line periodograms obtained from the \cite{2012A&A...541A.116A} data and the ones presented here show that something changed in the circumstellar environment of V2129 Oph between the two epochs of observation. The lines present the same shape as before, but the profiles are less intense in our observations. The periods measured in the $\mathrm{He}$\,I line in both works are different, which is surprising since they should correspond to the hot spot period and the star’s rotation period is not expected to vary on a timescale of 9 years \citep[e.g.,][]{1989AJ.....97..873H}. 
We used photometric data obtained at the Maidanak observatory  between 1986 and 2003 \citep{2008A&A...479..827G} and ASAS data secured between 2001 and 2009 to check the long-term photometric variability of the system. By examining phase-folded light curves, we find the photometric modulation to be consistent with a period of 6.53 days at all epochs, and we do not find evidence for a period as short as 6.0 days at any epoch. 
The extended period around 8.5 days seen in $\mathrm{H}\alpha$ was already present in the previous work, although only close to the line's center. It indicates that the periodic structure beyond the corotation radius was already contributing to the line profile. We discuss possible explanations for the origin of the 8.5-day period in Sect. \ref{sec:discussion}.

\begin{figure} 
 \centering
\subfigure[]{\includegraphics[scale=0.14]{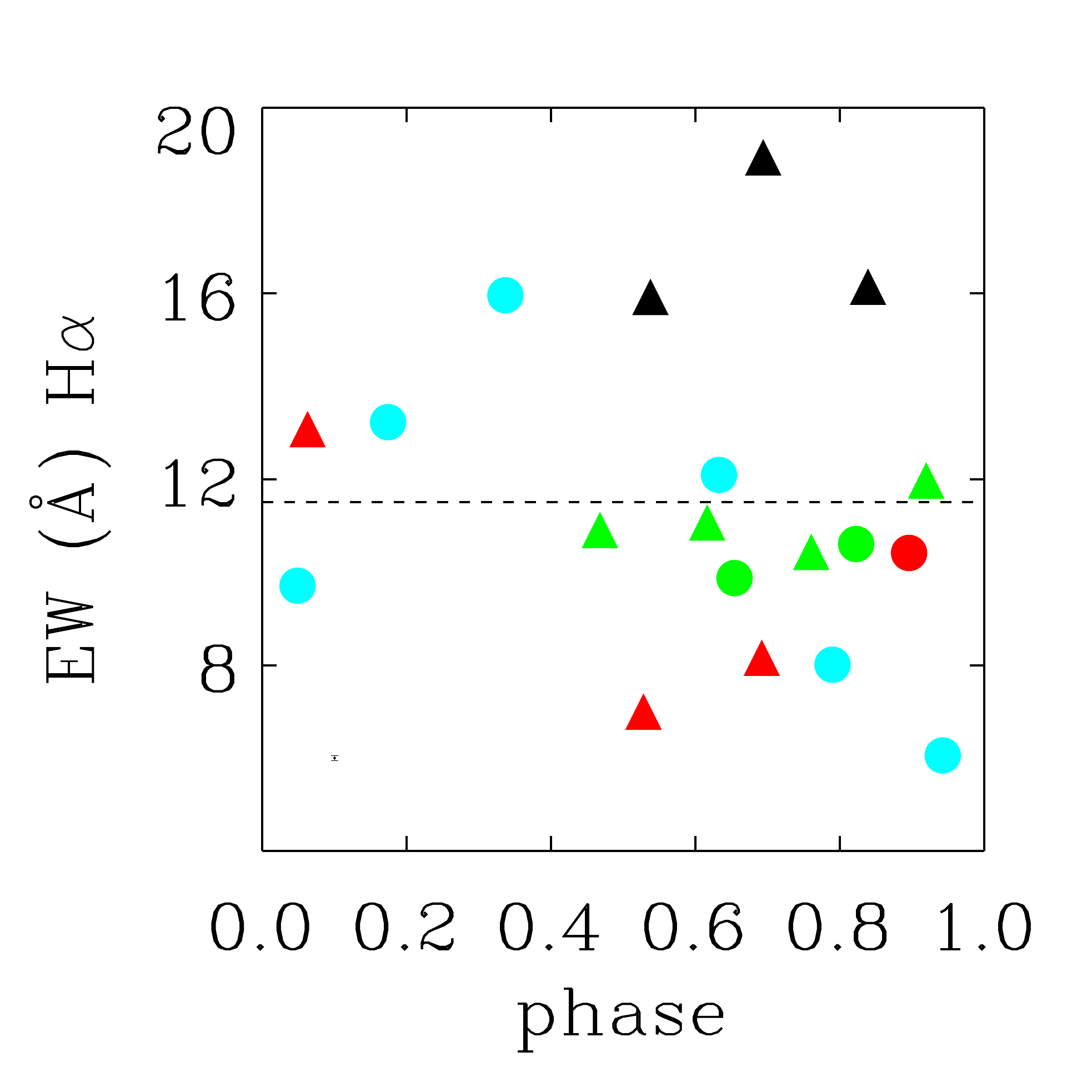}}
\subfigure[]{\includegraphics[scale=0.14]{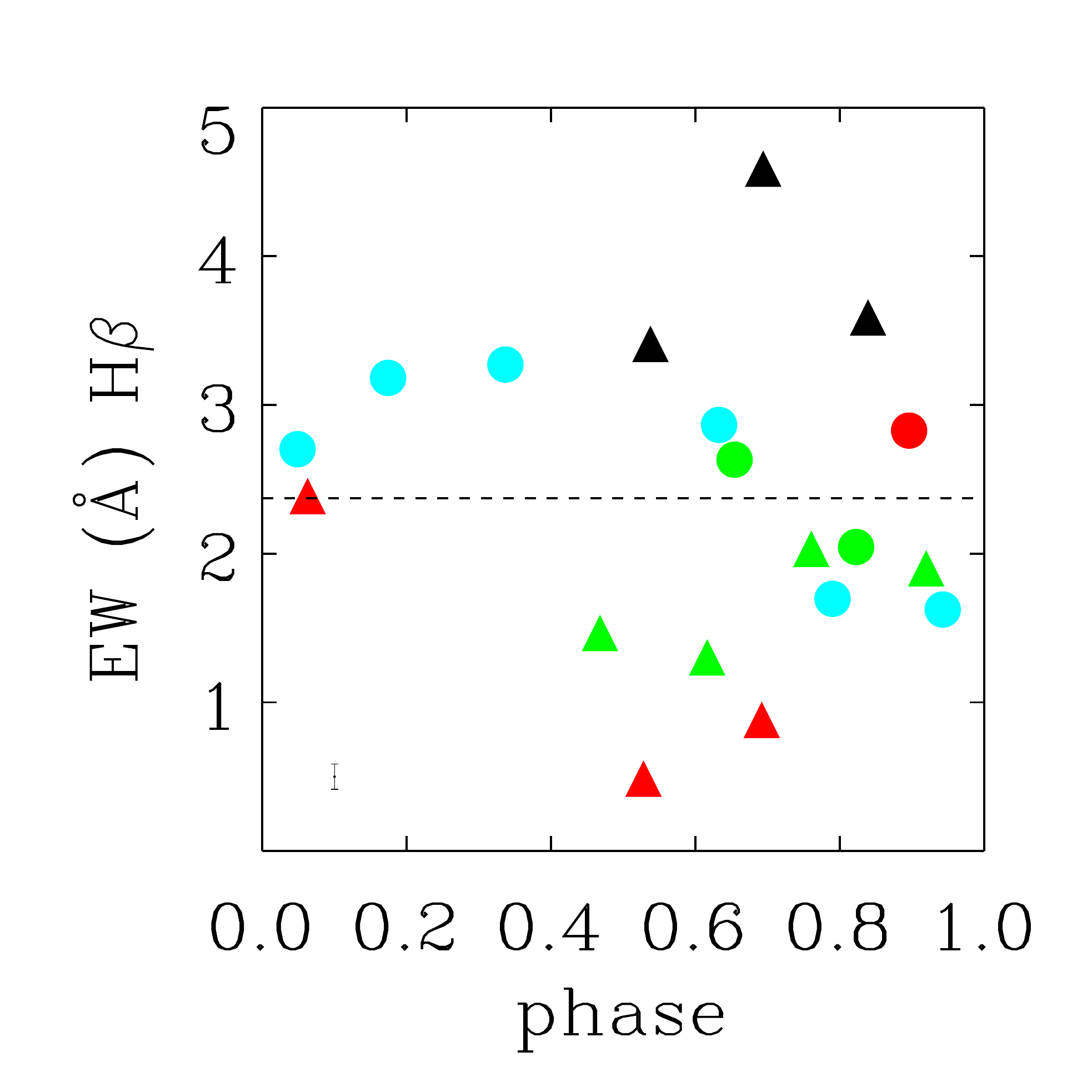}}
\subfigure[]{\includegraphics[scale=0.14]{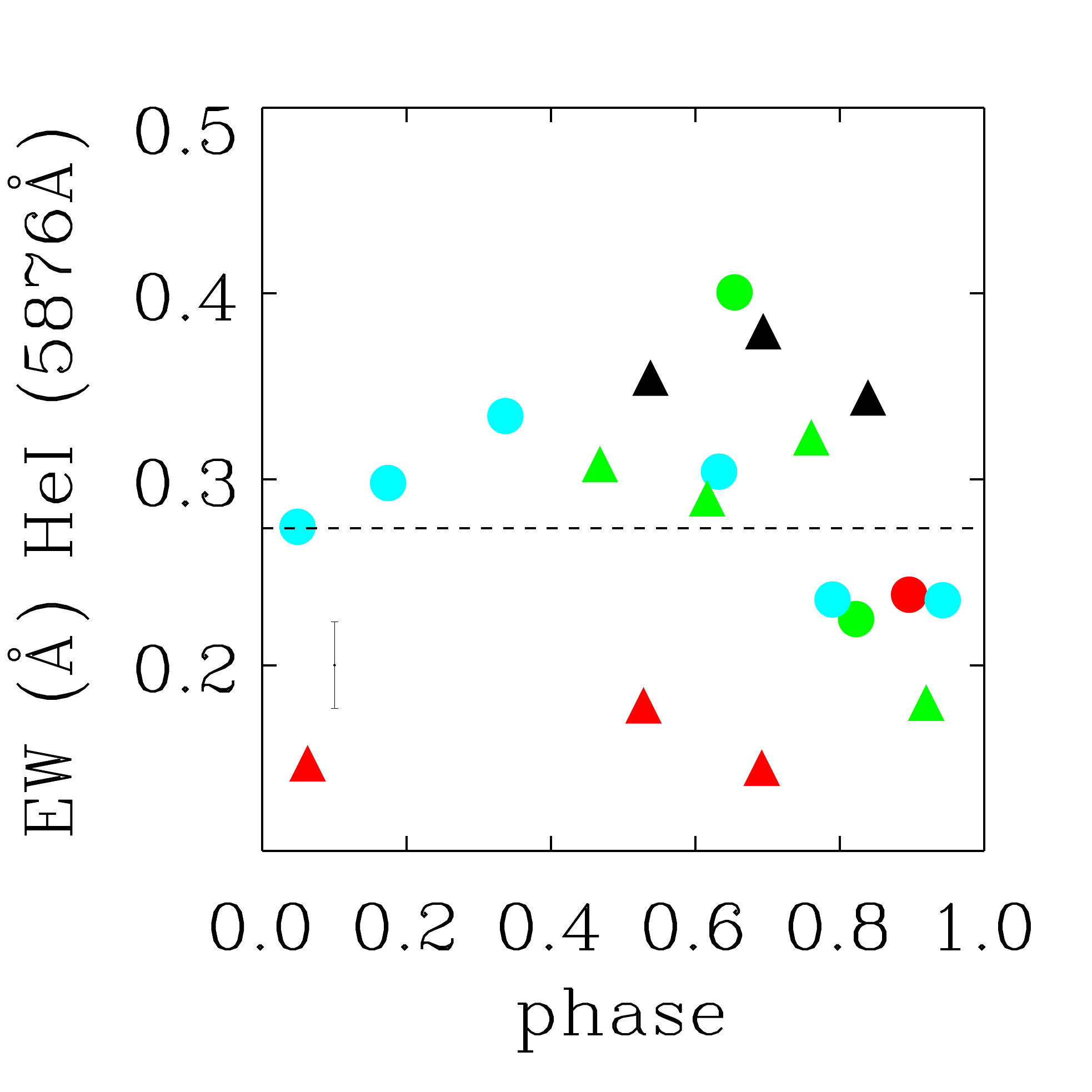}}\\
\subfigure[]{\includegraphics[scale=0.14]{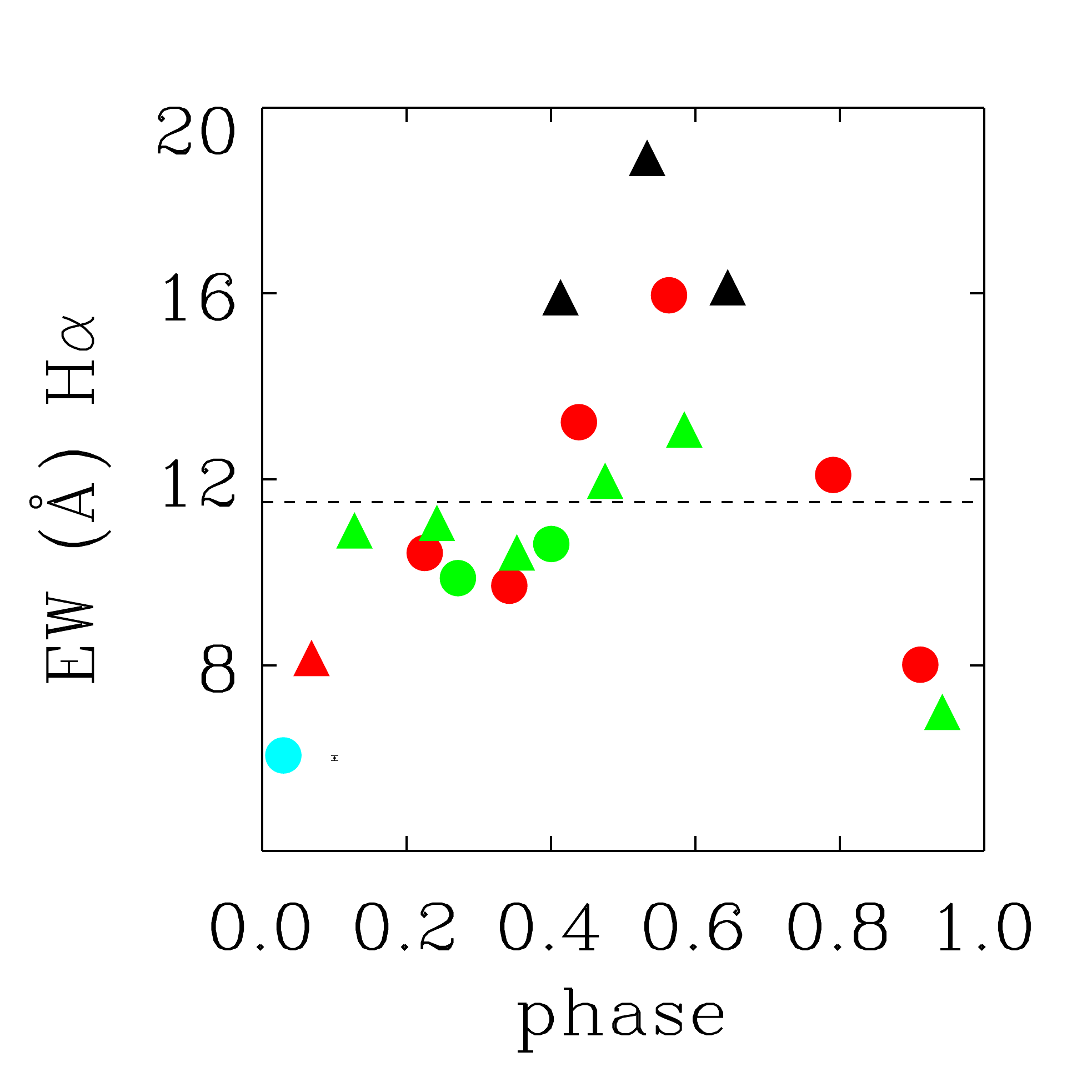}}
\subfigure[]{\includegraphics[scale=0.14]{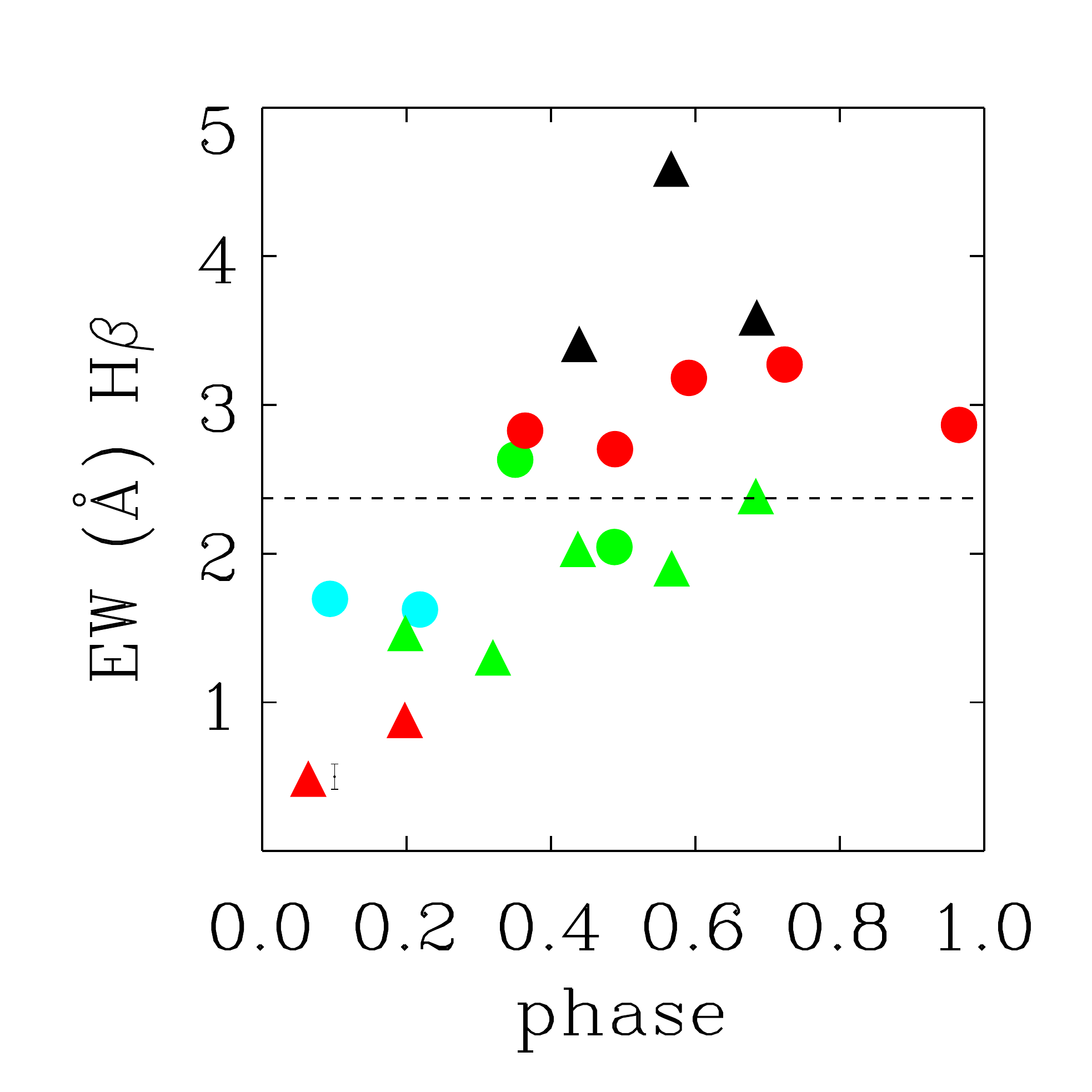}}
\subfigure[]{\includegraphics[scale=0.14]{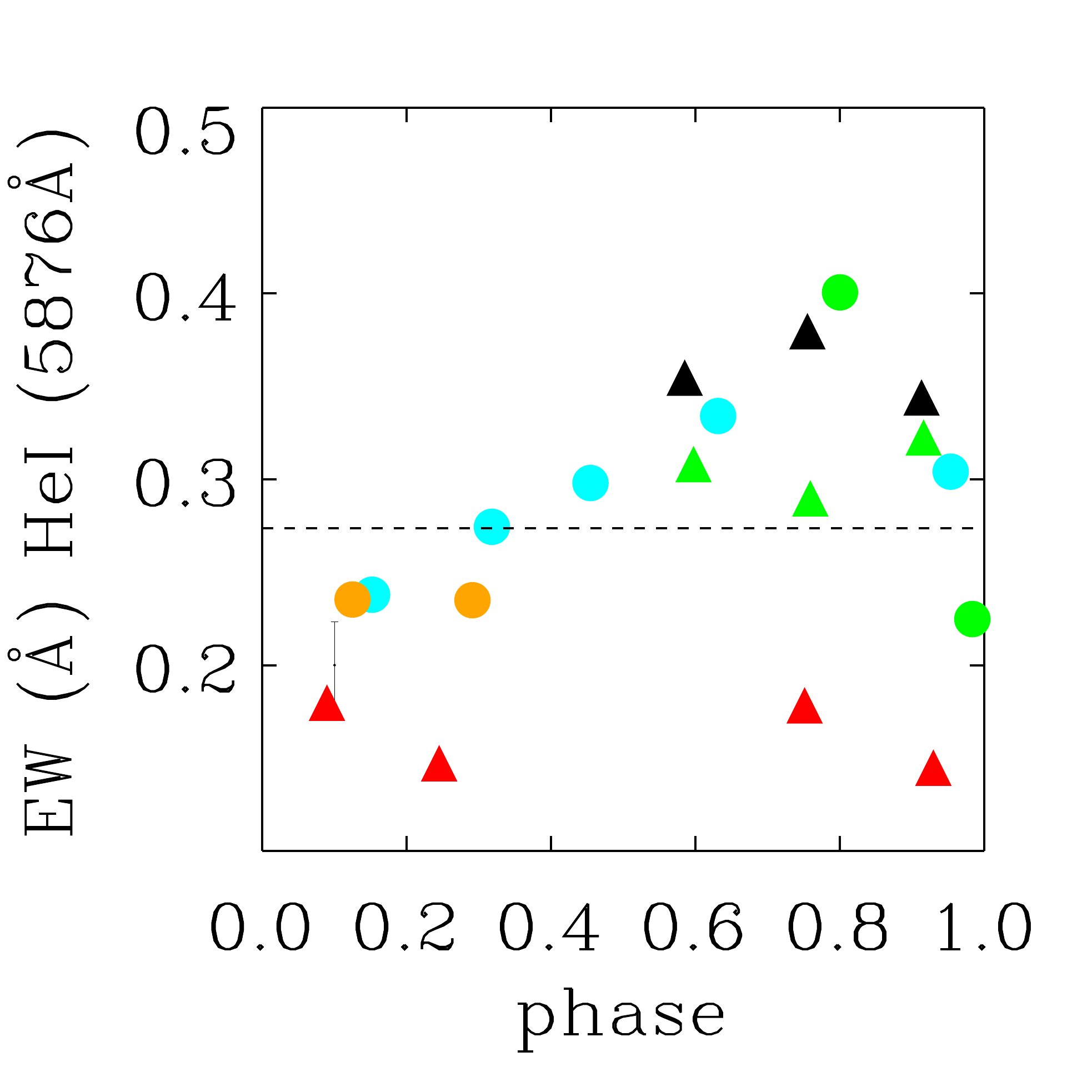}}
\caption{\label{fig:EWlines_all2} Equivalent widths of emission lines as a function of the rotational phase. {\it Top:} Phases computed with the stellar rotational period ($6.53\,\mathrm{days}$). {\it Bottom:} Periods of 8.5, 8.0, and 6.0 days used to reconstruct the cycles for, respectively, $\mathrm{H}\alpha$, $\mathrm{H}\beta$, and $\mathrm{He}$\,I\ 5876$\mathring{\mathrm{A}}$. Colors represent different rotational cycles. We note that the phase and the duration of a cycle are different in the top and bottom plots. The error bar on the plots are the mean of  the $3\sigma$ equivalent width uncertainties (see text). } 
\end{figure}

As mentioned above, the equivalent widths of the emission lines do not show a clear modulation at the rotational period of the star. We recalculated the rotational phase (Eq. \ref{eq:eph}) using the strongest periodicity of each line (8.5, 8.0, and 6.0 days, respectively, for $\mathrm{H}\alpha$, $\mathrm{H}\beta$, and $\mathrm{He}$\,I), instead of the stellar rotational period. The results are shown in Fig. \ref{fig:EWlines_all2}, where the equivalent widths seem to vary approximately in phase.

\subsection{Circumstellar line correlation matrices} \label{sec:optical_cor}
Analyzing the circumstellar line variability can help us understand the physical processes occurring in the CTTSs, and we can find out where the line have formed. One procedure to investigate the line variability is through the analysis of correlation matrices \citep{1995AJ....109.2800J,2005A&A...440..595A,2016A&A...586A..47S}. 

We divided each emission line into small velocity intervals ($1.5\,\mathrm{km\ s^{-1}}$ each) and computed the linear correlation coefficient, $r(i,j)$, between each  i and j  velocity interval. The results go from a perfect correlation between $i$ and $j$ velocity intervals if  $r(i,j)=1$ to no correlation for  $r(i,j)=0$ and $r(i,j)=-1$ for a perfect anticorrelation If $i=j$, $r(i,j)=1$ and correlation matrices present correlation along the main diagonal. The autocorrelation matrices are also symmetric with respect to the diagonal since $r(i,j)=r(j,i)$. We present the results in a bidimensional diagram where the linear correlation coefficients are shown in color. We show the autocorrelation matrices in Fig. \ref{fig:perline_all} for the $\mathrm{H}\alpha$, $\mathrm{H}\beta$, $\mathrm{He}$\,I, and Na I $5889.950\,\mathrm{\mathring{A}}$ lines.

The $\mathrm{He}$\,I line variability presents hints of a correlation along the line profile, indicating that the same physical process should dominate the profile. However, the red and blue parts of the profile are not well correlated, indicating that at least two independent processes control the emission and dynamics of the $\mathrm{He}$\,I line. 
The narrow component of this line is expected to form in accretion shocks only; however, the broad component, despite being faint, should affect the blue wing of the line. The correlation matrix of the $\mathrm{H}\alpha$ line is more complex than that of the $\mathrm{He}$\,I line. The $\mathrm{H}\alpha$ line does not present a correlation between red and blue wings, which is indicative that the red and blue wings are dominated by different physical processes. The absorption in the red wing of the line self-correlates well and could be associated with the passage of the funnel flow across our line of sight. 

Similar to the $\mathrm{H}\alpha$ line, the $\mathrm{H}\beta$ profile self-correlates well in each wing but does not present a significant correlation between the red and blue wings, which is an indication that the red and blue wings are predominantly influenced by different circumstellar features. Closer to the center of the line, it is possible to see a small region that is anticorrelated with the redshifted absorption, which indicates that when the central emission is more intense, the high-velocity redshifted absorption is deeper. However, this anticorrelation is too weak to be significant. \cite{2007MNRAS.380.1297D} found this anticorrelation for V2129 Oph using eight observations obtained with ESPaDOnS.

In Fig. \ref{fig:perline_all}, we only present the correlation matrix of the NaI $5889.950\,\mathrm{\mathring{A}}$ component since the NaI $5895.924\,\mathrm{\mathring{A}}$ component shows similar features. The redshifted wing of the line shows a hint of self-correlation but is polluted by an additional unrelated process. The blueshifted wing of the line does not show a clear correlation; this is a reflection of the complexity of this line, which has many components, most of which are sporadic absorptions with unknown origin.  

\begin{figure}
 \centering
{\includegraphics[width=4.4cm]{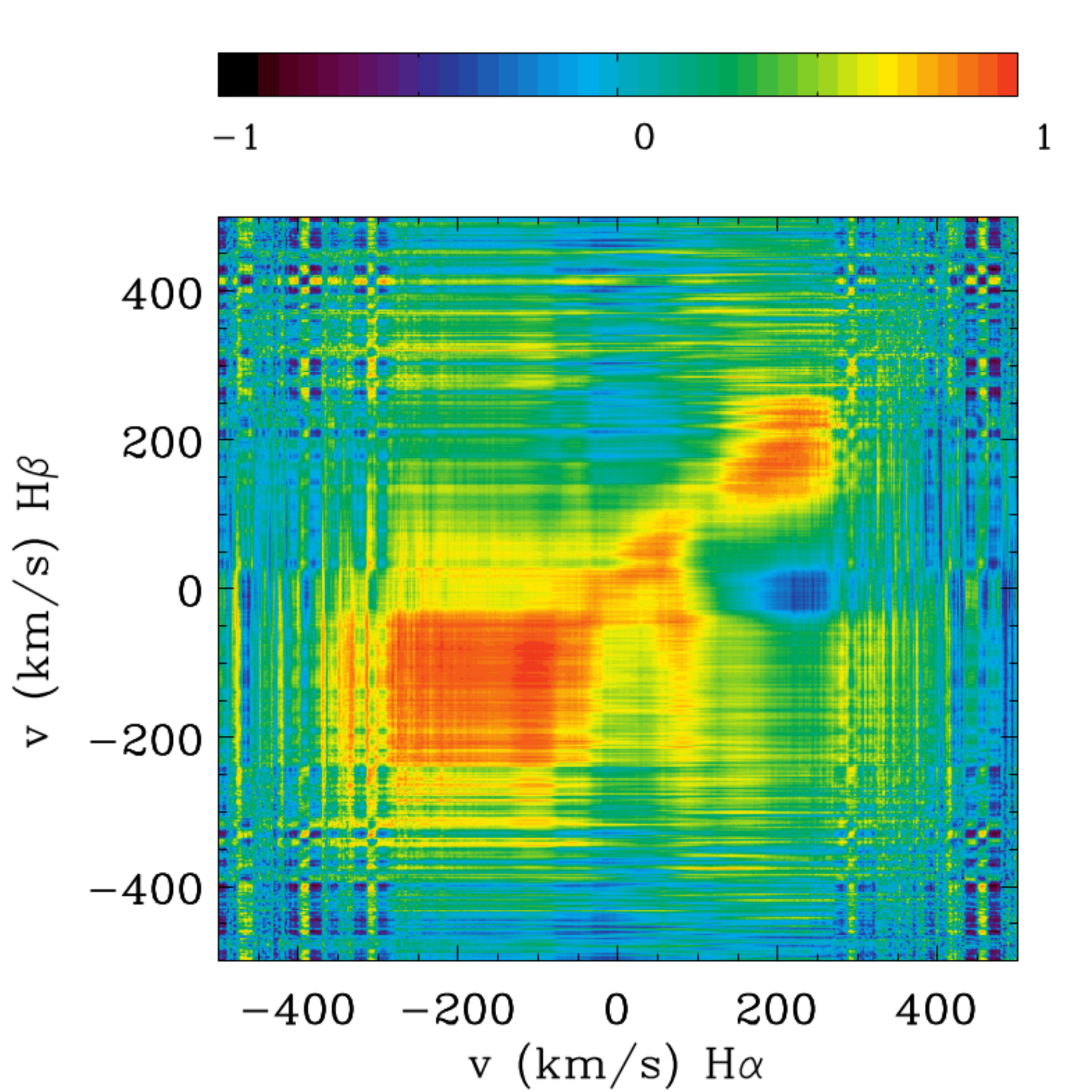}}
{\includegraphics[width=4.4cm]{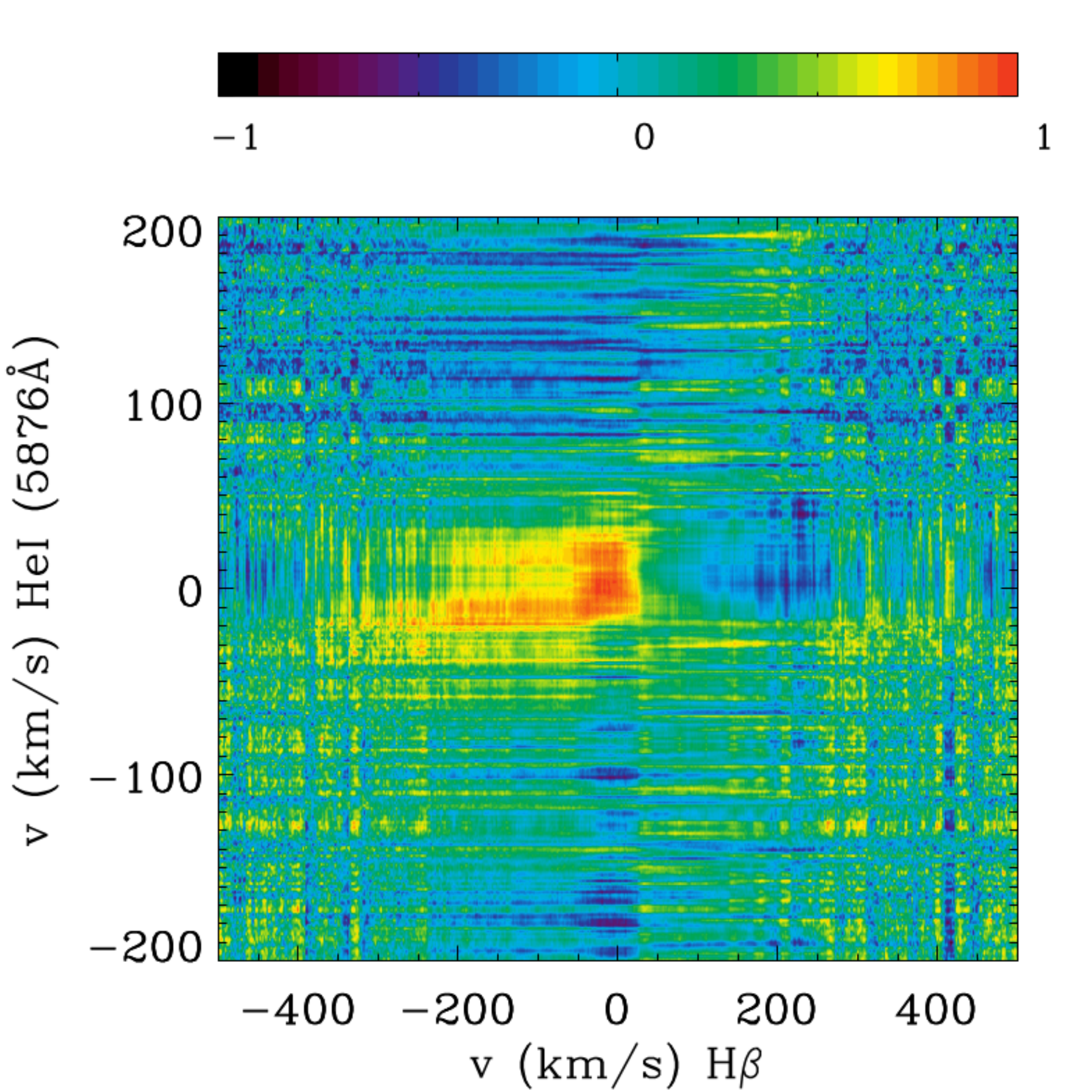}}
\caption{\label{fig:MatcHa_all} Correlation matrices between different lines.  The color range corresponds to the value of the linear correlation coefficient between the different velocity bins of the line profiles. Perfect anticorrelation corresponds to -1 (black), no correlation to 0 (light blue), and a perfect correlation to 1 (orange).}
\end{figure}

The correlation of the variability between different lines can help us understand if these lines originate from the same physical process and form at the same place. We calculated the cross-correlation matrices between different lines, observed by both ESPaDOnS and HARPS. In Fig. \ref{fig:MatcHa_all} we can see the correlation analysis between the $\mathrm{H}\alpha$ and $\mathrm{H}\beta$ lines. The blue wings of the $\mathrm{H}\alpha$ and $\mathrm{H}\beta$ profiles are well correlated. The high-velocity redshifted absorptions of both lines also correlate well. We also show the correlation between the $\mathrm{He}$\,I and $\mathrm{H}\beta$ lines in Fig. \ref{fig:MatcHa_all}. The central emission of $\mathrm{H}\beta$ and the $\mathrm{He}$\,I line are correlated. The $\mathrm{H}\beta$ central emission also presents the same periodicity as the $\mathrm{He}$\,I line. Since the $\mathrm{He}$\,I line is thought to form in the accretion shock, the observed correlation indicates that the central emission of $\mathrm{H}\beta$ also comes from this region. We did not find significant results in the correlation matrices between the other optical lines.

\section{Infrared spectroscopy} \label{sec:infrared}
  
\subsection{Infrared veiling and radial velocity} \label{sec:irveling}

Measuring the veiling and the radial velocity in the infrared is slightly more complicated than in the optical for two reasons: The photospheric lines are weak, and the veiling value varies significantly with wavelength. Therefore, we computed the radial velocity using the CCF profiles generated by the SPIRou pipeline, employing a numerical mask corresponding to a K2 spectral type. This target is sufficiently bright, and the telluric OH emission lines do not affect our results. The radial velocity obtained for each night is the average of four measurements taken per night, and the uncertainty is its standard deviation. 
We followed the method described in Sect. \ref{sec:optveiling} to calculate the veiling in the infrared. We used an observation obtained with SPIRou of the weak-line T Tauri star (WTTS) V819 Tau  \citep[spectral type K4 and $vsini=9.5\,\mathrm{km\ s^{-1}}$;][]{2015MNRAS.453.3706D} as a template to compare with the infrared spectra of V2129 Oph. We applied the velocity shift between the template and the stellar spectra to the standard before estimating the veiling. We measured veiling values in adjacent regions to each emission line of interest. We used the spectral ranges $10620-10820\,\mathring{\mathrm{A}}$, $10864-10920\,\mathring{\mathrm{A}}$, $12830-12890\,\mathring{\mathrm{A}}$, and $22600-22690\,\mathring{\mathrm{A}}$,  which are close to the $\mathrm{He}$\,I, Pa$\gamma$, Pa$\beta$, and Br$\gamma$ lines, respectively.

\begin{figure} 
 \centering
\includegraphics[scale=0.40]{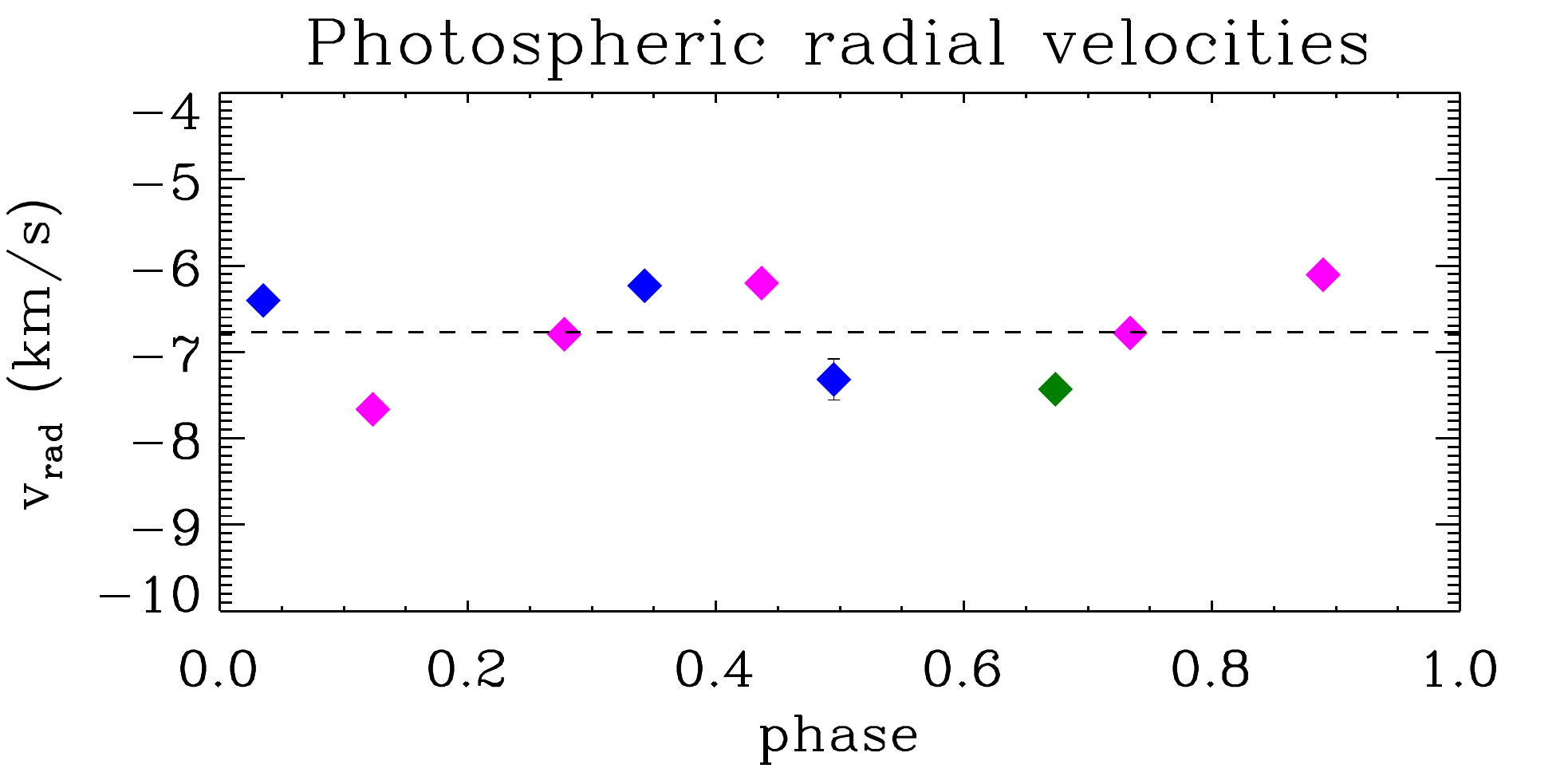}
\caption{\label{fig:Parm_phase_spirou}Radial velocity of photospheric lines obtained with SPIRou data and in phase with the same ephemeris from \cite{2012A&A...541A.116A} and \cite{2007MNRAS.380.1297D}. The radial velocity is the mean of all the radial velocities measured on the same night, and the error bar is its respective standard deviation. The colors represent different cycles: dark green - cycle 5; purple - cycle 6; and blue - cycle 7. The dashed line is the mean value of all observed nights.} 
\end{figure}

\begin{figure} 
 \centering
\includegraphics[scale=0.40]{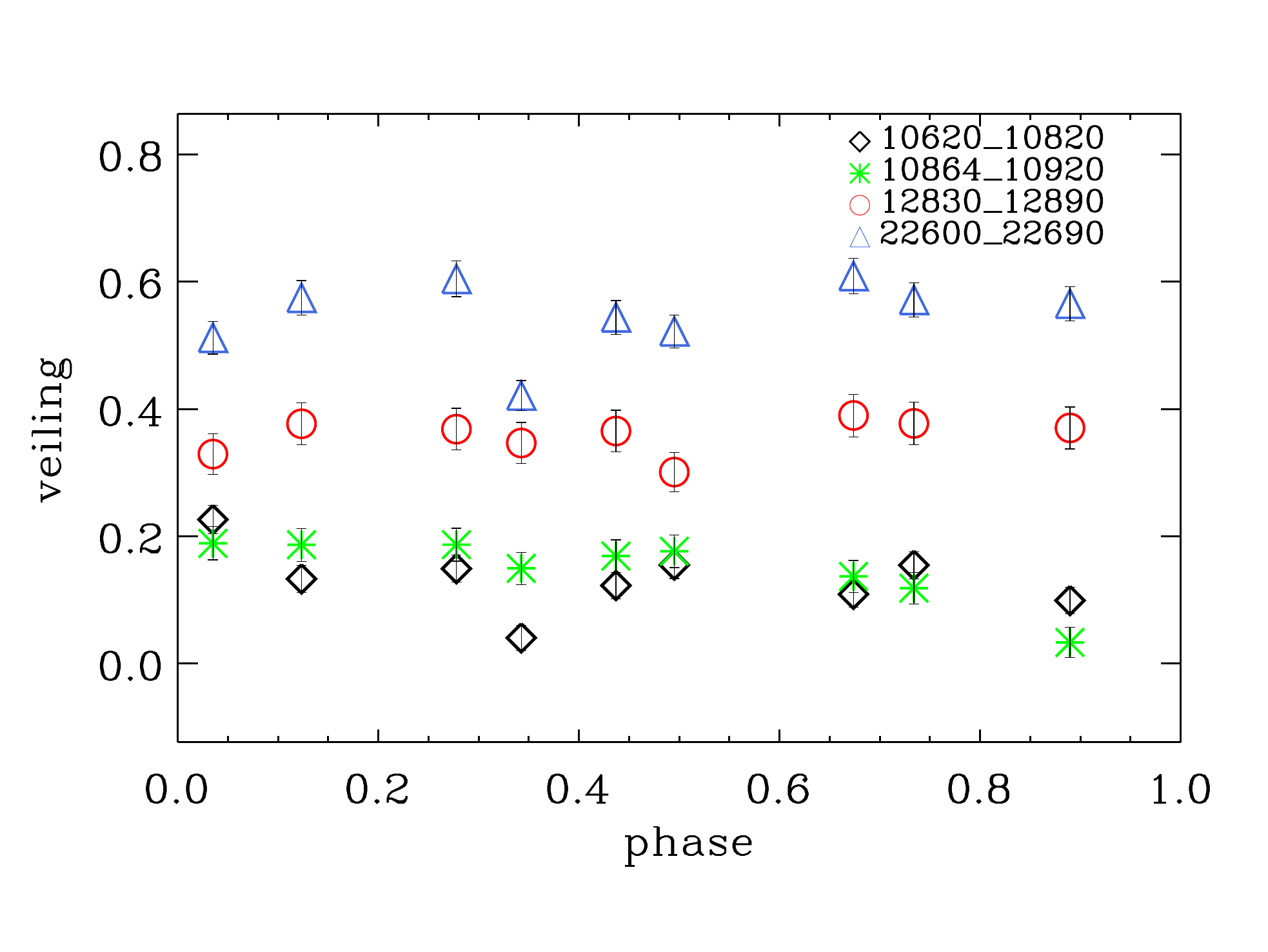}
\caption{\label{fig:veiling_spirou} Veiling of photospheric lines obtained with SPIRou data as a function of the stellar rotation phase. Each symbol shows the veiling measured in different spectral regions.} 
\end{figure}

In Table \ref{tab:veilIR} and Figs. \ref{fig:Parm_phase_spirou} and \ref{fig:veiling_spirou}, we show the results of the mean radial velocity and veiling we obtained for each region. We derived $v_{rad}=(-6.8\pm0.6)\,\mathrm{km\ s^{-1}}$, which corresponds to the mean radial velocity over all the observed nights and its respective standard deviation. The shape of the infrared radial velocity variations is similar to the radial velocity variations we obtained from optical spectra (see Fig. \ref{fig:opirvrad}). As expected from spot modulation, the amplitude of the radial velocity variations is much smaller in the near infrared than in the optical, about two times smaller for V2129 Oph. The sampling of the infrared radial velocities is relatively scarce, especially toward the extrema. The variability of the infrared radial velocity shows a period of $3.13\,\mathrm{days}$ (FAP=0.2), slightly smaller than half the stellar rotation period. We also looked for periodicities using the infrared and optical radial velocities together, and we found consistent results. 

\begin{table}[htb!]
\tiny
\addtolength{\tabcolsep}{-4.5pt}  
\caption{\label{tab:veilIR}  Infrared veiling obtained with the SPIRou data.}
\begin{center}
\begin{tabular}{llllll}
  \hline\hline 
 JD\tablefootmark{a} &$\phi$ & $\mathrm{veil}_{10620\_10820}$ &$\mathrm{veil}_{10864\_10920}$ &$\mathrm{veil}_{12830\_12890}$ & $\mathrm{veil}_{22600\_22690}$  \\
\hline
8324.36  &  5.67  &  0.11  $\pm$  0.01  &  0.14  $\pm$  0.01  &  0.39  $\pm$  0.02  &  0.61  $\pm$  0.01  \\
8327.30  &  6.12  &  0.13  $\pm$  0.01  &  0.19  $\pm$  0.01  &  0.38  $\pm$  0.02  &  0.57  $\pm$  0.01  \\
8328.30  &  6.28  &  0.15  $\pm$  0.01  &  0.19  $\pm$  0.01  &  0.37  $\pm$  0.02  &  0.60  $\pm$  0.01  \\
8329.34  &  6.44  &  0.12  $\pm$  0.01  &  0.17  $\pm$  0.01  &  0.37  $\pm$  0.02  &  0.54  $\pm$  0.01  \\
8331.28  &  6.73  &  0.15  $\pm$  0.01  &  0.12  $\pm$  0.01  &  0.38  $\pm$  0.02  &  0.57  $\pm$  0.01  \\
8332.30  &  6.89  &  0.10  $\pm$  0.01  &  0.03  $\pm$  0.01  &  0.37  $\pm$  0.02  &  0.57  $\pm$  0.01  \\
8333.25  &  7.04  &  0.23  $\pm$  0.01  &  0.19  $\pm$  0.01  &  0.33  $\pm$  0.02  &  0.51  $\pm$  0.01  \\
8335.26  &  7.34  &  0.04  $\pm$  0.01  &  0.15  $\pm$  0.01  &  0.35  $\pm$  0.02  &  0.42  $\pm$  0.01  \\
8336.25  &  7.50  &  0.15  $\pm$  0.01  &  0.18  $\pm$  0.01  &  0.30  $\pm$  0.02  &  0.52  $\pm$  0.01  \\
\hline
\end{tabular}
\end{center}
 \tablefoot{This table is ordered according to the observation date.}\\
 \tablefoottext{a}{JD-$2\,450\,000$}
 \end{table}
 
The infrared veiling seems to increases with wavelength (see Fig. \ref{fig:veiling_spirou}), as found in previous works for different T Tauri systems \citep[e.g.,][]{1999A&A...352..517F}.  The infrared veiling in the $J$-band is similar to the values obtained in the optical region. However, the veiling in the optical and infrared can have different origins. The optical veiling is directly associated with the accretion process and is due to the continuum emitted by the hot spot on the stellar surface \citep[e.g.,][]{1998ApJ...509..802C,1991ApJ...382..617H}, while the infrared veiling %is not well understood. The IR veiling 
is too high to be explained by the accretion shock alone, and the dust in the innermost disk emission may contribute to the veiling in this spectral region \citep{2006ApJ...646..319E}.

\subsection{Infrared emission lines}\label{sec:infraredlines}
\begin{figure} 
 \centering
\includegraphics[scale=0.30]{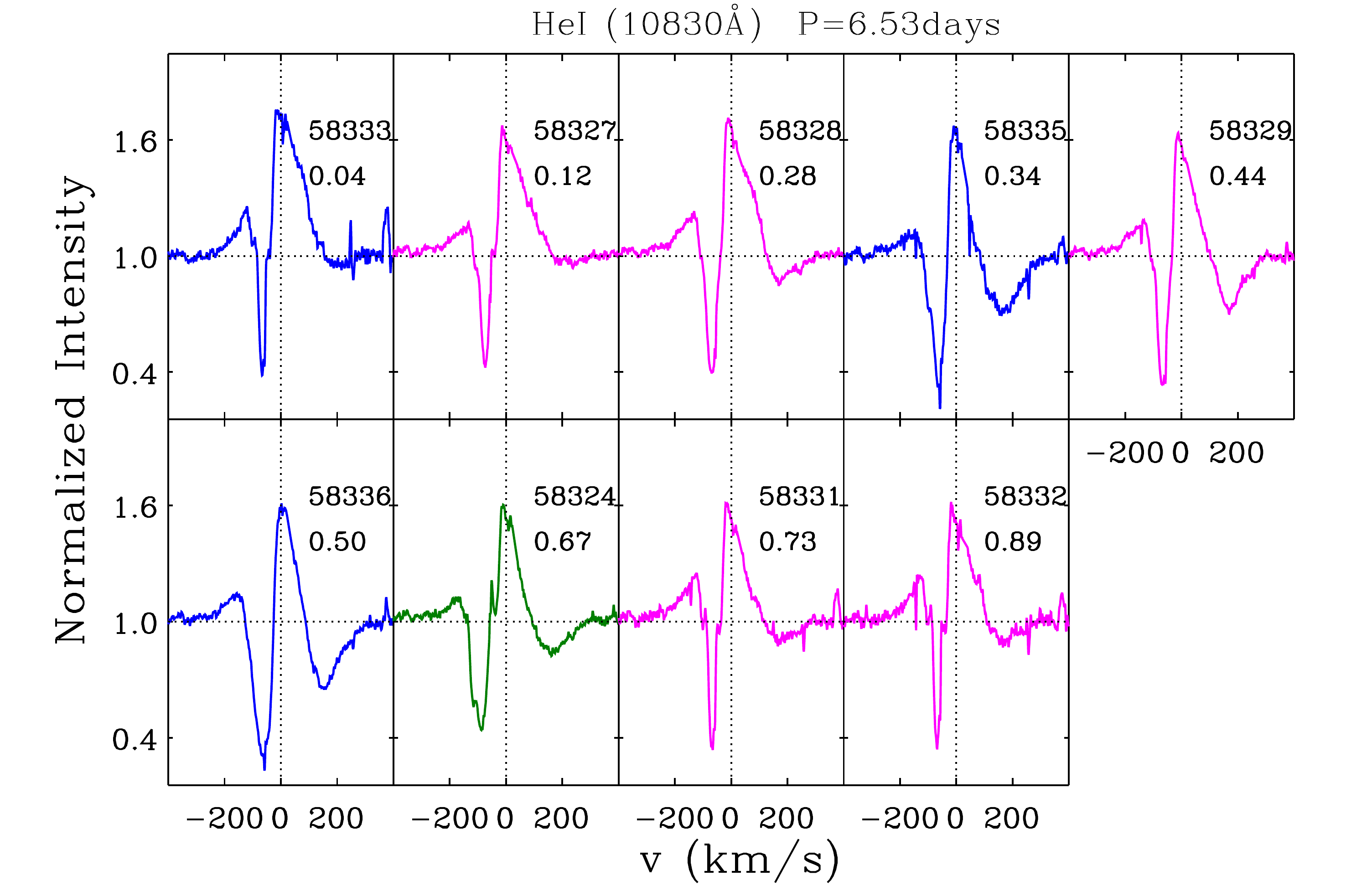}
\caption{\label{fig:HelI3line} $\mathrm{He}$\,I\ 10830$\mathring{\mathrm{A}}$  emission line profiles obtained with the SPIRou spectrograph, ordered by phase. The phase and the JD of the observations are written in each panel. The colors represent different rotational cycles. } 
\end{figure}

\begin{figure}
\centering
\includegraphics[scale=0.30]{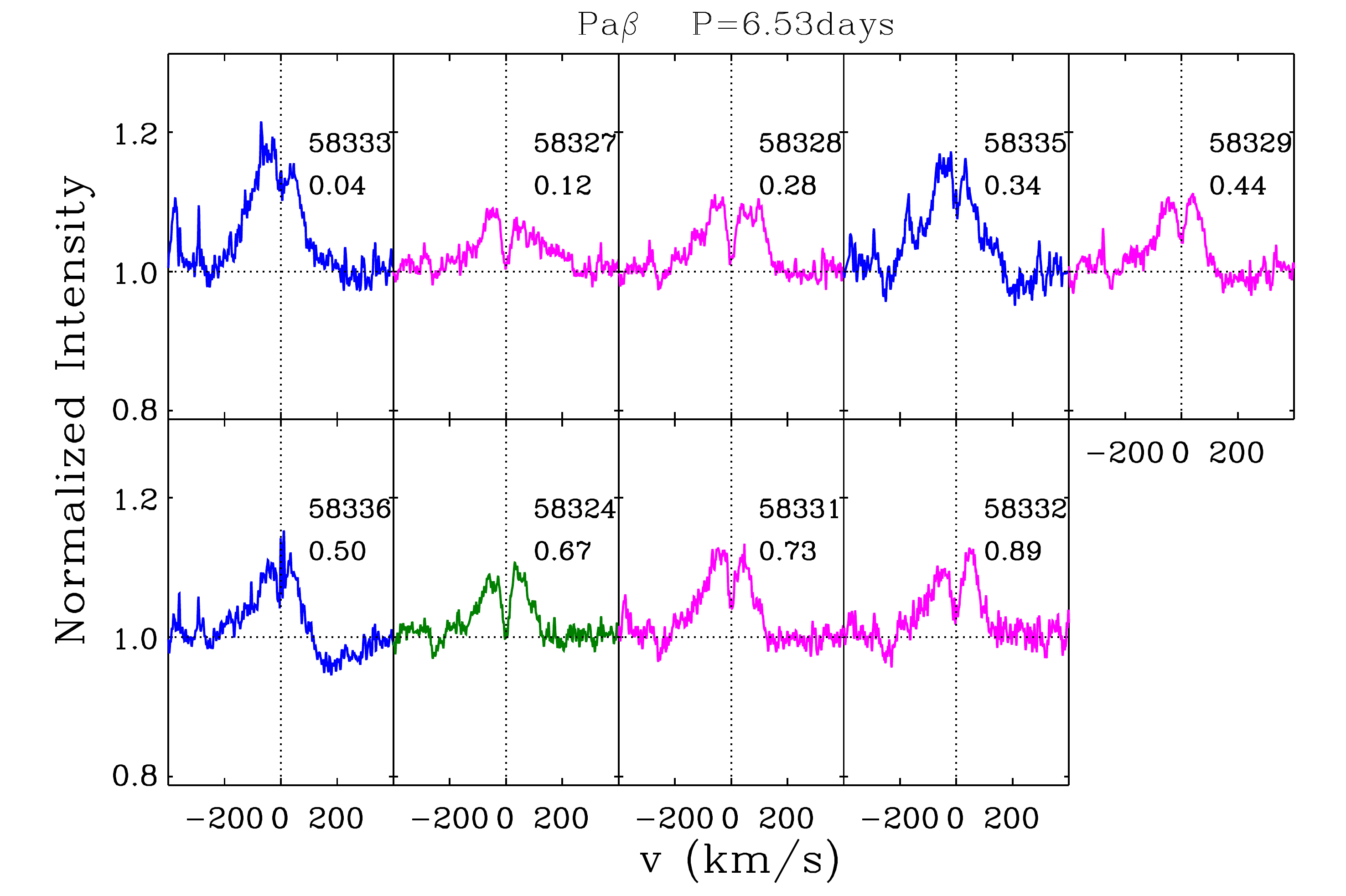}
\caption{\label{fig:Pabetaline} Pa$\beta$ emission line profiles obtained with the SPIRou spectrograph, organized by phase. The phase and the JD of the observations are written in each panel. The colors represent different rotational cycles.} 
\end{figure}

We subtracted the photospheric lines from the emission line profiles of V2129 Oph, using V819 Tau as a template, which we broadened to the rotation of the target, veiled, and velocity-shifted. We present the residual profiles of the HeI $10830\,$\AA\ and Pa$\beta$ line profiles in Figs. \ref{fig:HelI3line} and \ref{fig:Pabetaline}. In Fig. \ref{fig:resid} we show the emission lines before and after removing the photospheric contributions.  

The $\mathrm{He}$\,I infrared triplet line around 10830$\,\mathring{\mathrm{A}}$\footnote{In the paper we quote the air wavelength of the infrared lines despite the fact that SPIRou data are calibrated on vacuum wavelengths.} that is seen in young stars carries traces of accretion and ejection processes \cite[e.g.,][]{2020ApJ...892...81T,2008ApJ...687.1117F,2006ApJ...646..319E}. In Fig. \ref{fig:HelI3line} we present the $\mathrm{He}$\,I profiles obtained with SPIRou data, organized by phase. This line is variable from night to night and presents a broad redshifted absorption below the continuum level in most of the observations. The line also presents a blueshifted absorption component, which is persistent in all phases and varies in depth and width as the system rotates. Usually, the redshifted absorptions are associated  with the accretion funnel \citep[e.g.,][]{2008ApJ...687.1117F}, while the blueshifted absorptions are characteristic of ejection processes, such as a disk or stellar wind. The stellar wind contribution is generally more significant in high mass accretion rate systems \citep[e.g.,][]{2007ApJ...657..897K}.

The maximum radial velocity of the blueshifted absorption of the $\mathrm{He}$\,I line varies from -65 to -93 $\mathrm{km\ s^{-1}}$, while the maximum radial velocity of the redshifted absorption varies from 90 to 160 $\mathrm{km\ s^{-1}}$. This shows that the redshifted absorption comes from gas accelerated to higher velocities than the blueshifted one, which is compatible with the free-fall velocity expected in the magnetosphere \citep{koenigl1991disk}. The blueshifted absorption may be a mix of stellar and disk wind components. 

As in the optical spectra that we discussed in the previous sections, in the infrared  we also find emission lines from the hydrogen series. In the SPIRou wavelength range, T Tauri stars usually present Pa$\beta$ $12818\mathring{\mathrm{A}}$, Pa$\gamma$ $10938\mathring{\mathrm{A}}$, and Br$\gamma$ $21660\mathring{\mathrm{A}}$ emission lines \cite[e.g.,][]{2020A&A...643A..99B,2006ApJ...646..319E,2001A&A...365...90F}.  The residual spectra of V2129 Oph, however, do not present a Br$\gamma$ signature, not even in absorption. V2129 Oph does show Pa$\beta$ and Pa$\gamma$ in emission, and the Pa$\beta$ profile is shown organized by phase in Fig. \ref{fig:Pabetaline}. The spectral region of the Pa$\gamma$ line is noisier than the Pa$\beta$ region, and we do not show the results of this line. 

The Pa$\beta$ line presents a central absorption on most of the observed nights. However, the line center is affected by chromospheric emission, which is also present in the template that we used to remove the photospheric contribution. For this reason, we preferred not to analyze the central region of Pa$\beta$. The residual Pa$\beta$ profile shows a redshifted absorption that is compatible with an accretion signature between phases 0.34 and 0.50. At these phases, the $\mathrm{He}$\,I line shows the deepest redshifted absorptions. 

\begin{figure} 
 \centering
 {\includegraphics[scale=0.13]{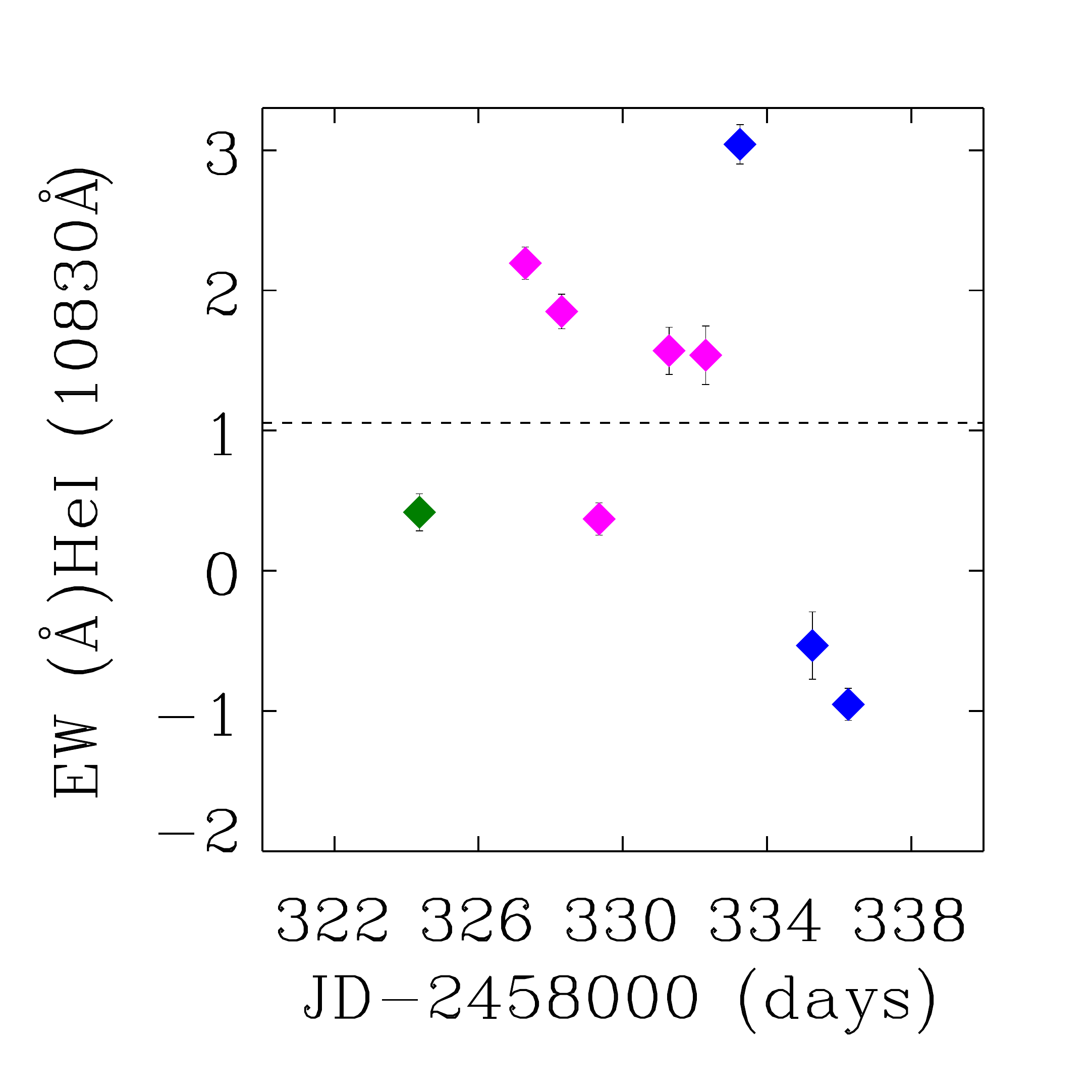}}
{\includegraphics[scale=0.13]{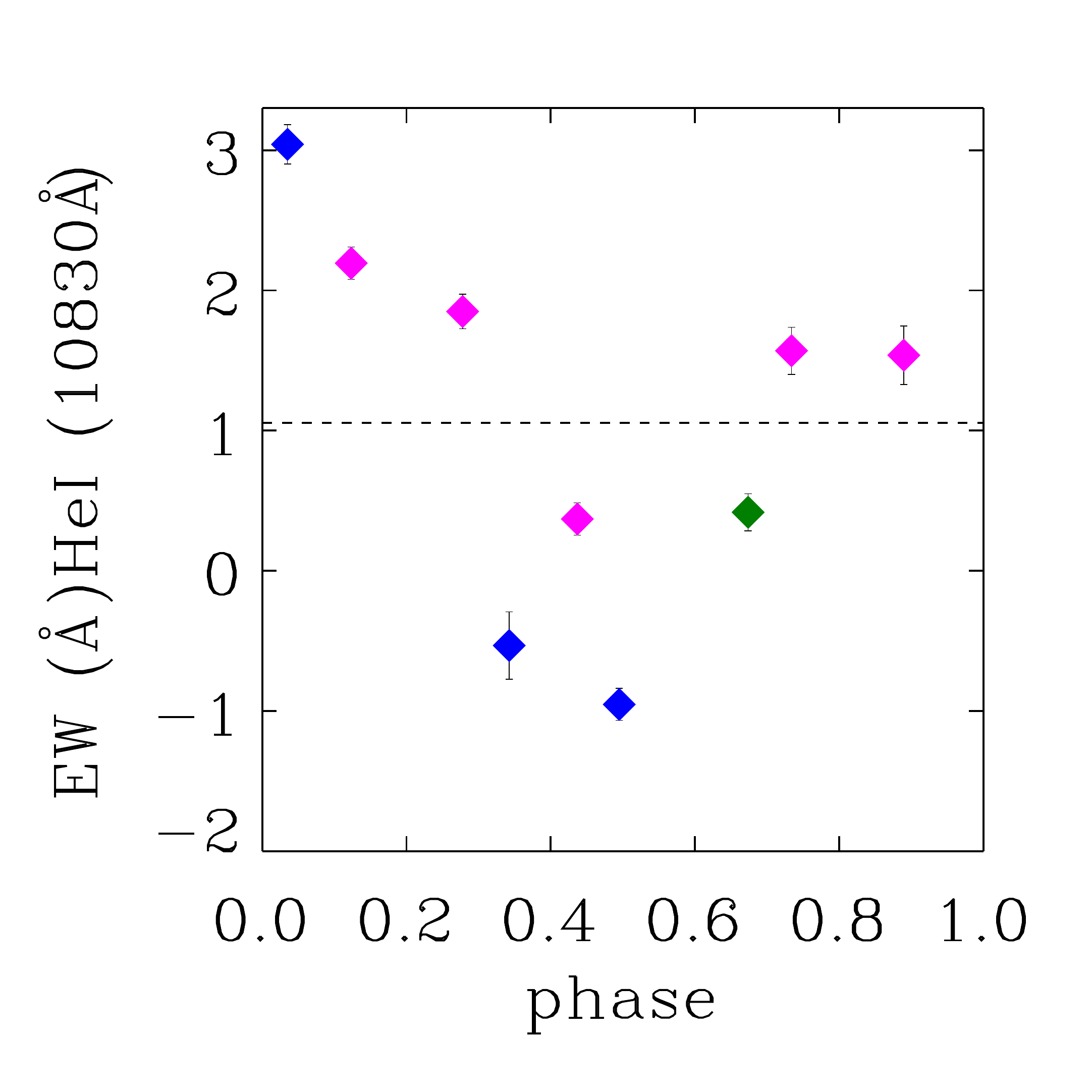}}
{\includegraphics[scale=0.13]{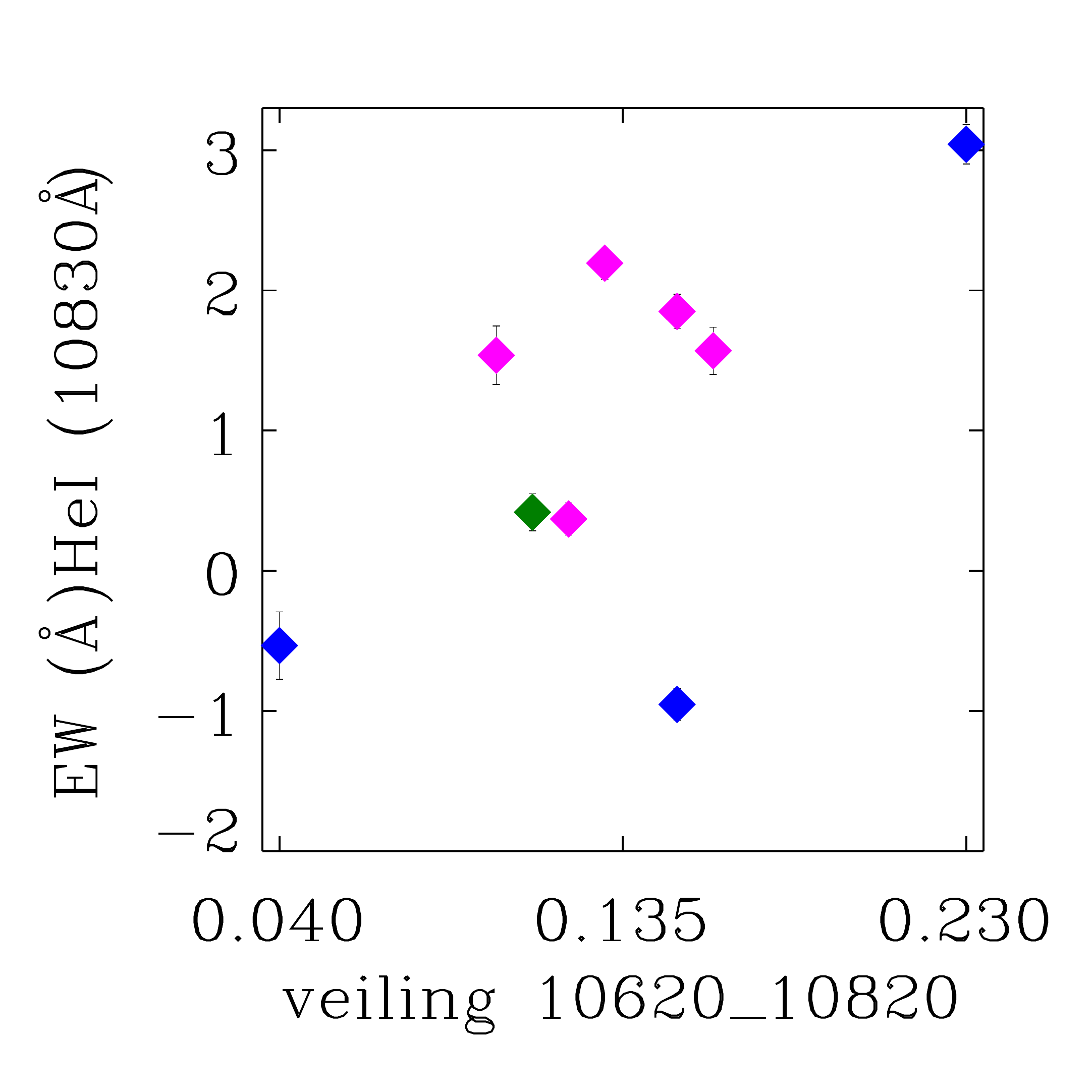}}
{\includegraphics[scale=0.13]{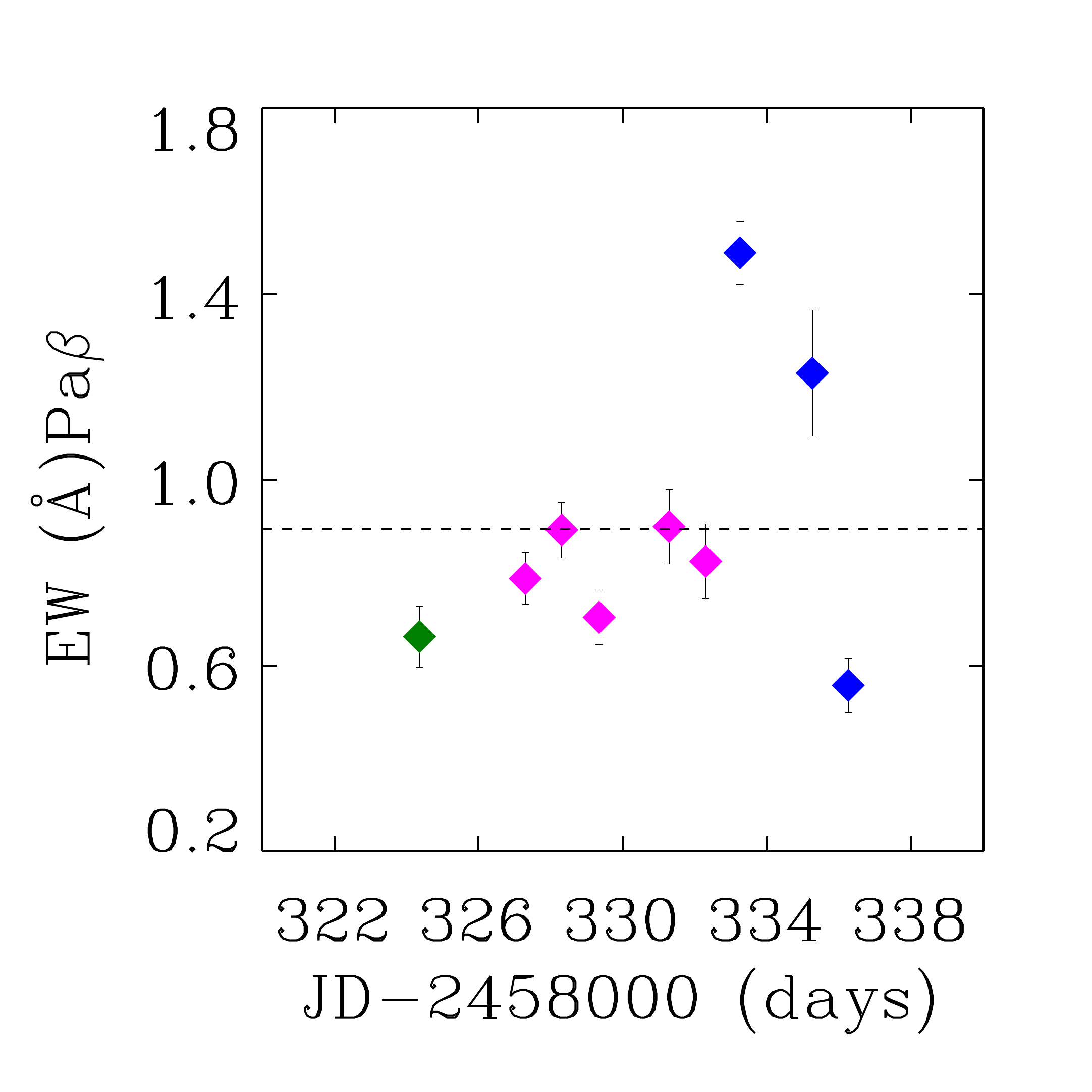}}
{\includegraphics[scale=0.13]{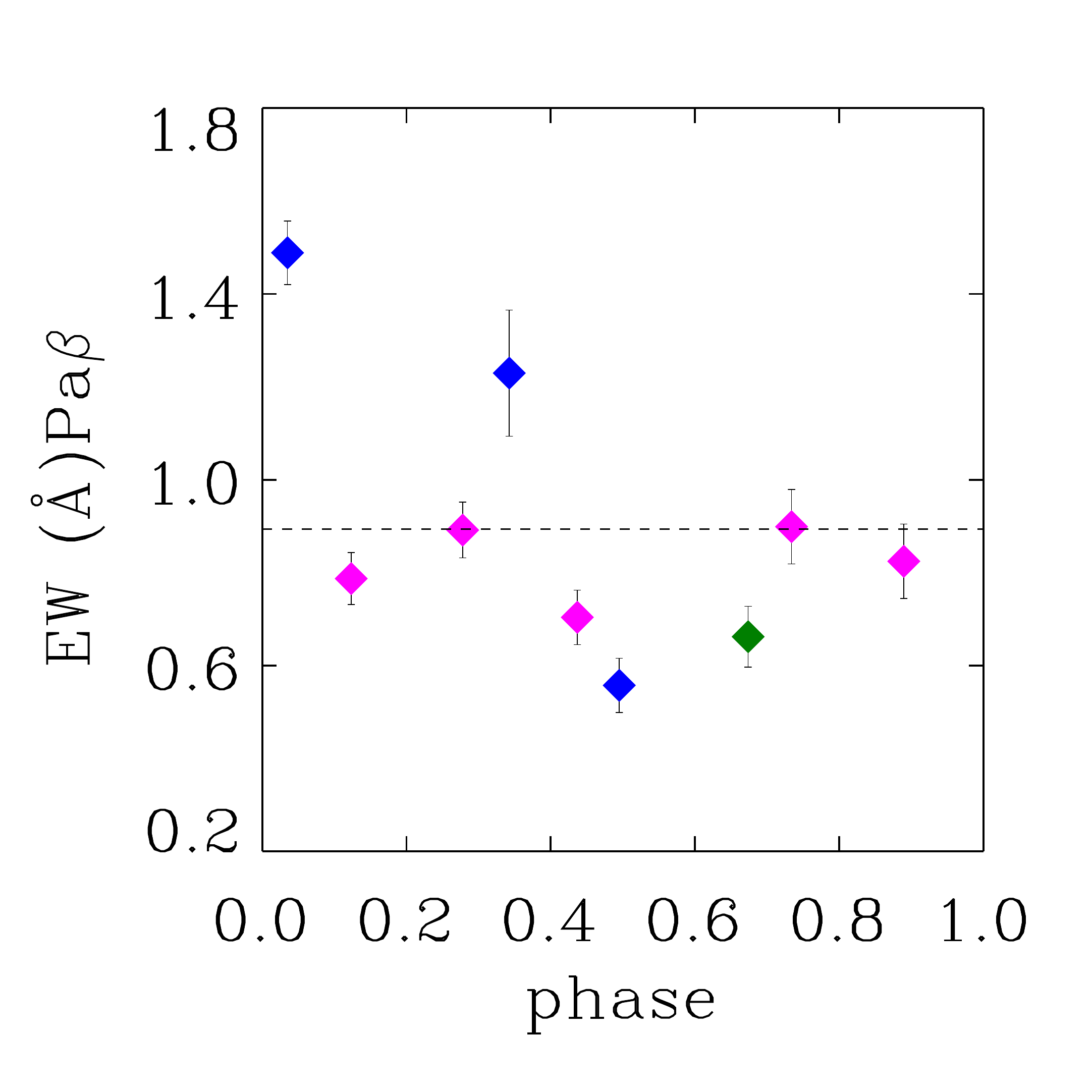}}
{\includegraphics[scale=0.13]{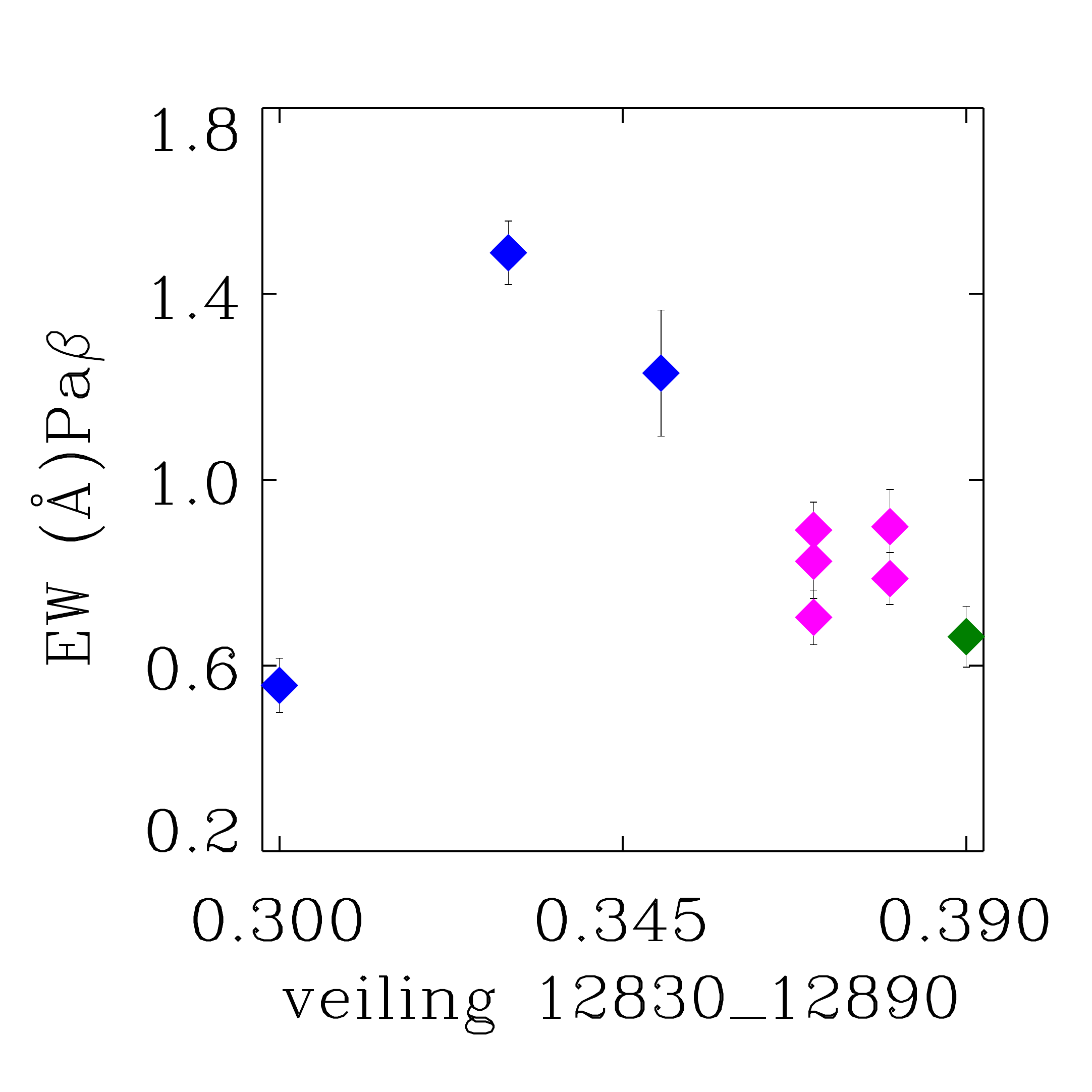}}
\caption{\label{fig:EWlines_spirou} Equivalent widths of infrared emission lines as a function of the observational date ({\it left}), rotational phase ({\it middle}), and veiling ({\it right}). The error bars correspond to $2\sigma$ equivalent width uncertainties. Colors represent different rotational cycles: dark green - cycle 5; purple - cycle 6; and blue - cycle 7. }  
\end{figure}

We measured the equivalent width of the infrared lines and present the results in Table \ref{tab:parameters} and in Fig. \ref{fig:EWlines_spirou} as a function of JD, rotational phase, and the veiling measured close to each line. Only the equivalent width of the $\mathrm{He}$\,I line seems to vary in phase with the rotation of the star. It also increases with the veiling, probably driven by the absorption components. The absorption components of this line are so strong that they dominate the emission profile in some observations and the equivalent width becomes negative. 

\begin{figure} 
 \centering
{\includegraphics[scale=0.13]{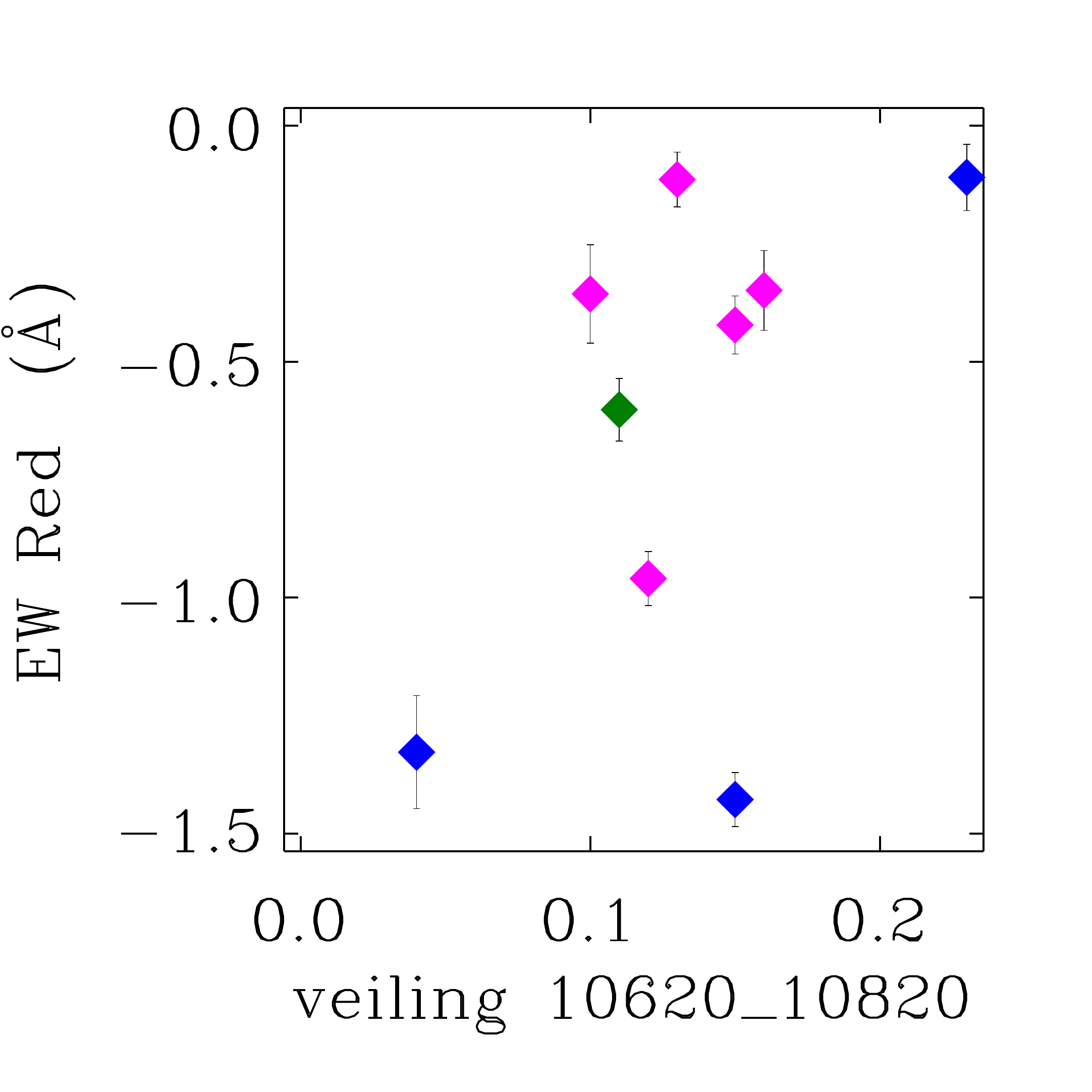}}
{\includegraphics[scale=0.13]{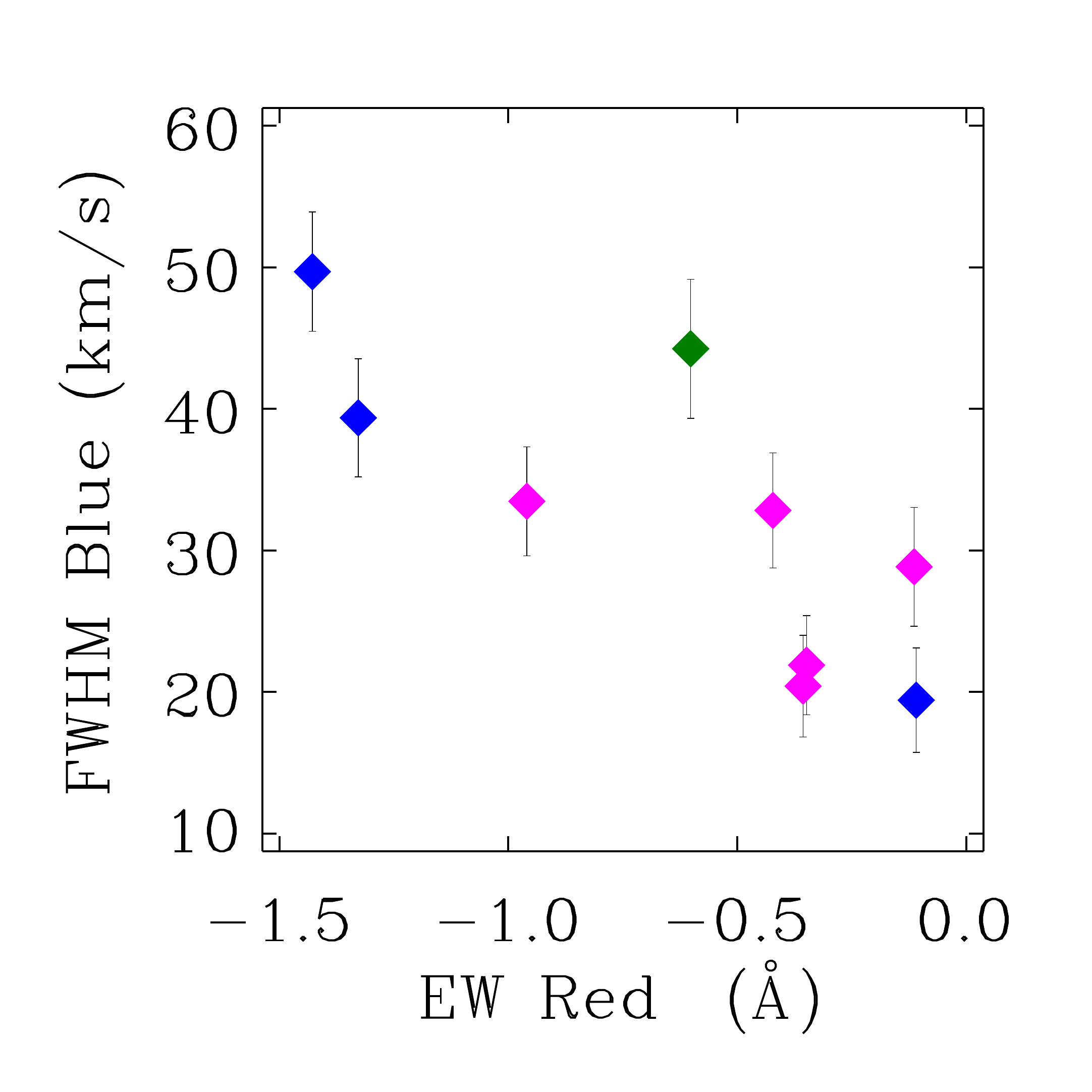}}
\caption{\label{fig:EWHeI_spirou} Parameters of the $\mathrm{He}$\,I\ 10830$\mathring{\mathrm{A}}$ line. {\it Left}: Equivalent width of redshifted absorption as a function of the veiling measured close to the line.   {\it Right}: FWHM of the blueshifted absorption. %, and equivalent width of the blueshifted absorption ({\it right}). 
Colors represent different rotational cycles: dark green - cycle 5; purple - cycle 6; blue - cycle 7. }  
\end{figure}

We analyzed the parameters of the absorption components of the $\mathrm{He}$\,I line and show the results in Fig. \ref{fig:EWHeI_spirou}. We considered only the line below the continuum level to compute the blueshifted and redshifted absorptions equivalent widths and the FWHM.

Despite the limited number of observations, we find a potential relationship between the veiling and the equivalent width of the redshifted absorption of $\mathrm{He}$\,I. The infrared veiling is the lowest when the redshifted absorption is the deepest (i.e., when the hot spot is on our line of sight), while the infrared veiling is at its maximum when the redshifted absorption vanishes (i.e., the hot spot disappears from view). Additional data points would be necessary to confirm this relationship. Nonetheless, this result supports the idea that the infrared veiling comes from the inner disk that is irradiated by the hot spot \citep{2006ApJ...646..319E}. When the shock-illuminated inner disk edge faces us, (i.e., when the accretion hot spot is opposite to our line of sight), we observe maximum infrared veiling. We emphasize that this effect is opposite to what occurs in the optical, where the veiling is at its maximum when the hot spot faces the observer (see Sect. \ref{sec:optveiling}) since the optical veiling is due the extra continuum emitted by the hot spot itself.

Another interesting and even tighter relationship between the FWHM of the blueshifted absorption component and the equivalent width of the redshifted absorption component  is seen in Fig. \ref{fig:EWHeI_spirou}: the stronger the redshifted absorption, the wider the blueshifted absorption. This behavior is also clearly seen in Fig. \ref{fig:HelI3line}. This indicates that, as the funnel flow crosses the line of sight, we simultaneously probe a deeper path along the velocity gradient of the outflow that is responsible for the blueshifted absorption component. In other words, this relationship suggests that the outflow arises from a region that is closely connected to the accretion funnel flow, such as an inner disk or an interface wind \citep{2007ApJ...657..897K}.  

\subsection{Infrared circumstellar line periodicity}

\begin{figure*}
 \begin{center}
\subfigure[]{\includegraphics[width=4.5cm]{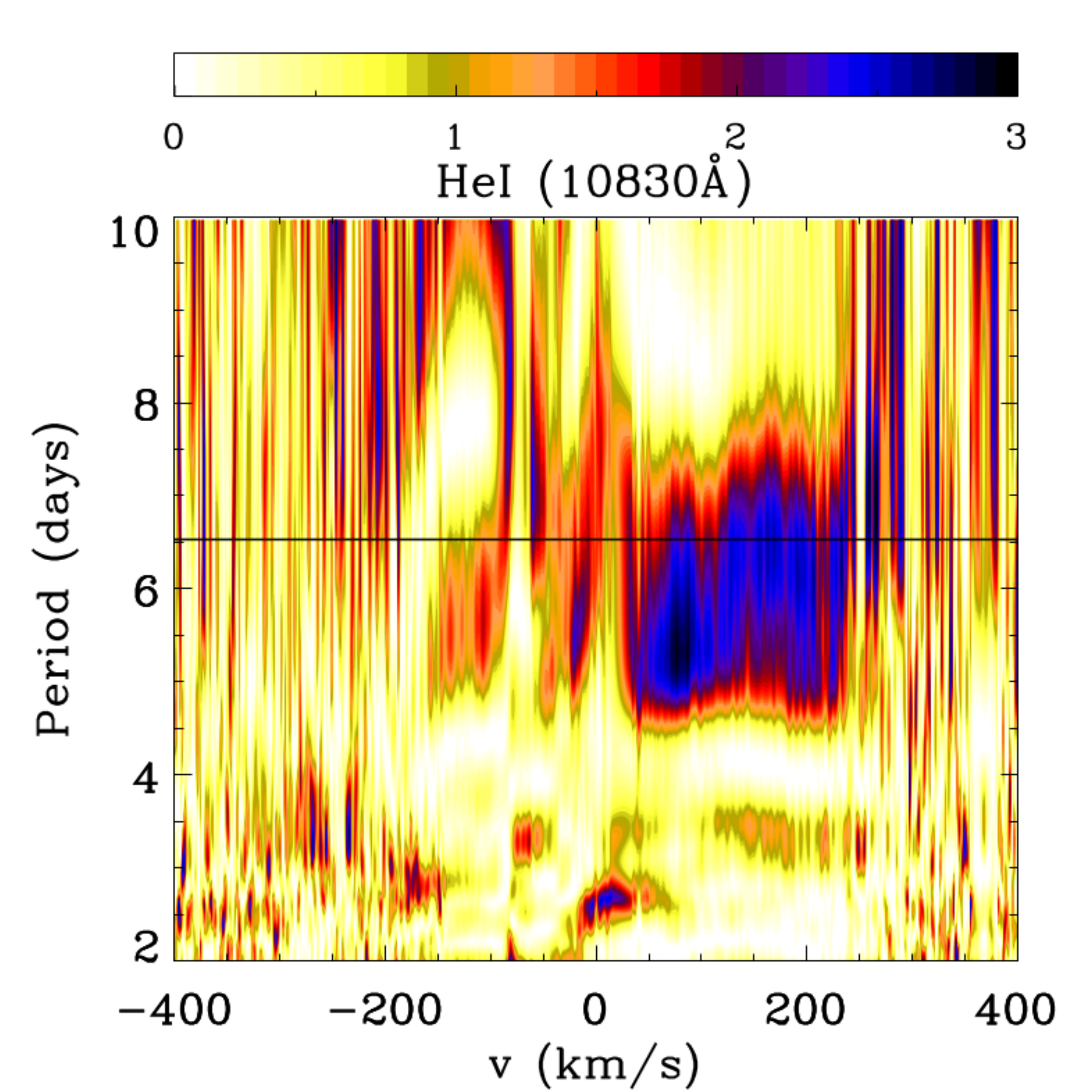}}
\subfigure[]{\includegraphics[width=4.5cm]{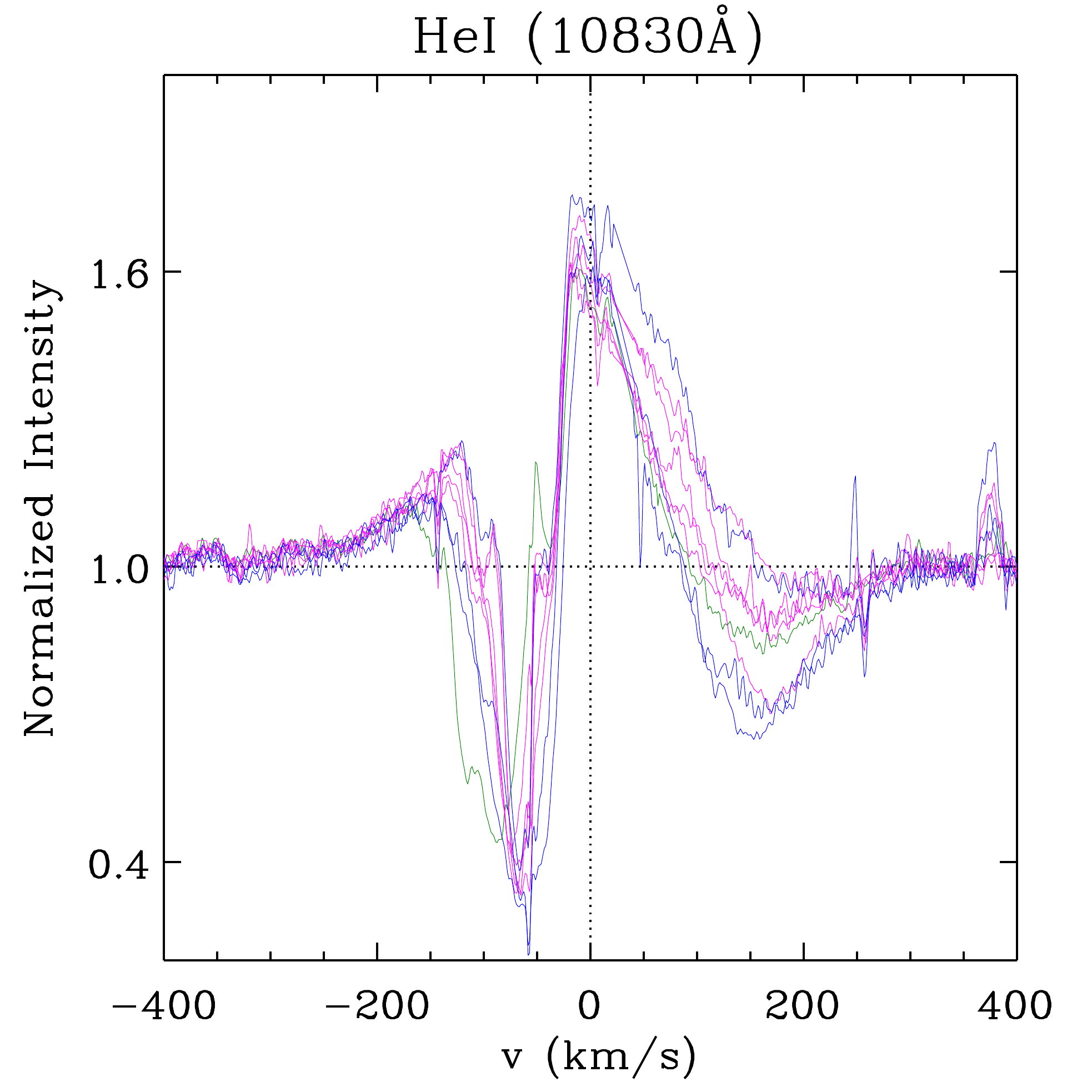}}
\subfigure[]{\includegraphics[width=4.5cm]{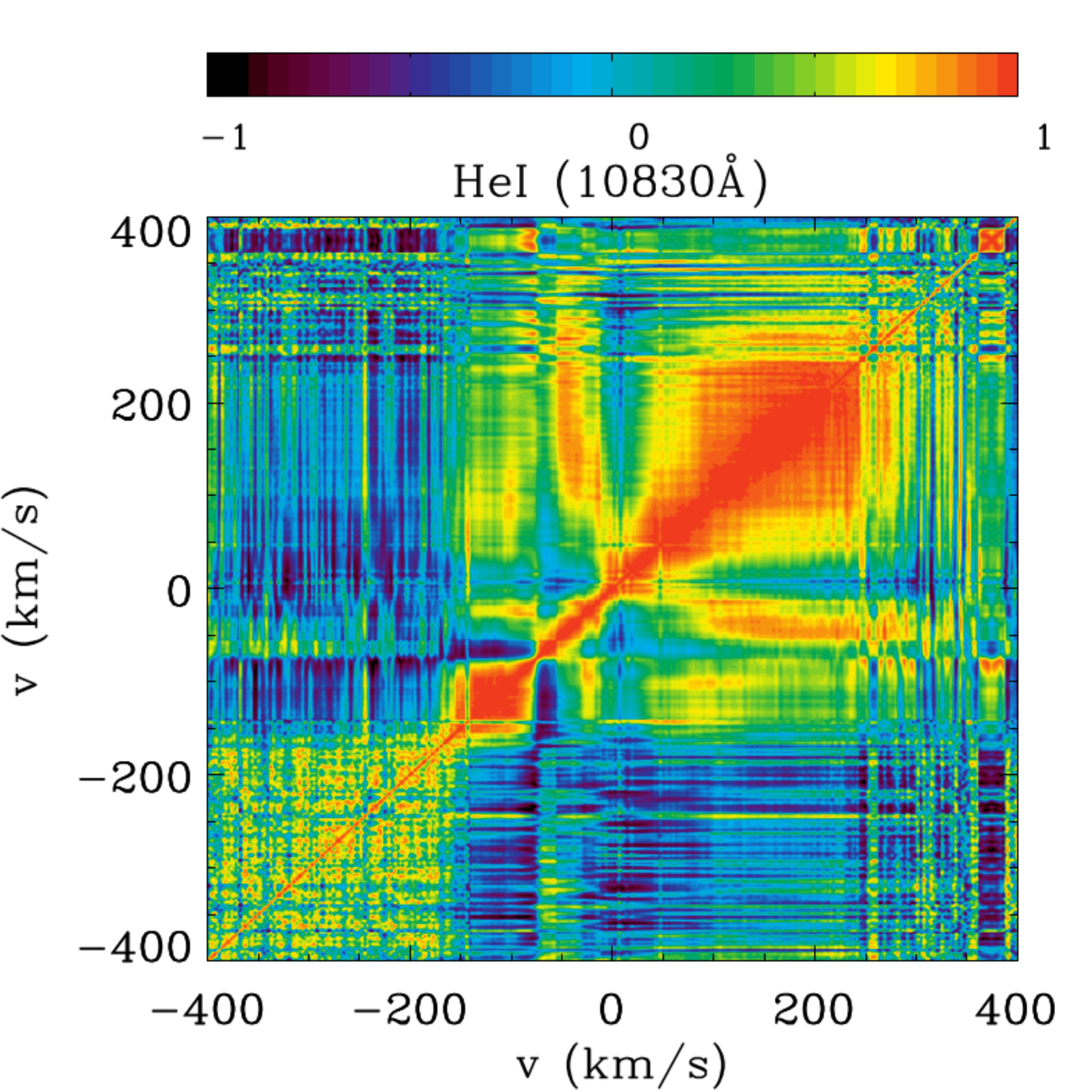}}\\
\subfigure[]{\includegraphics[width=4.5cm]{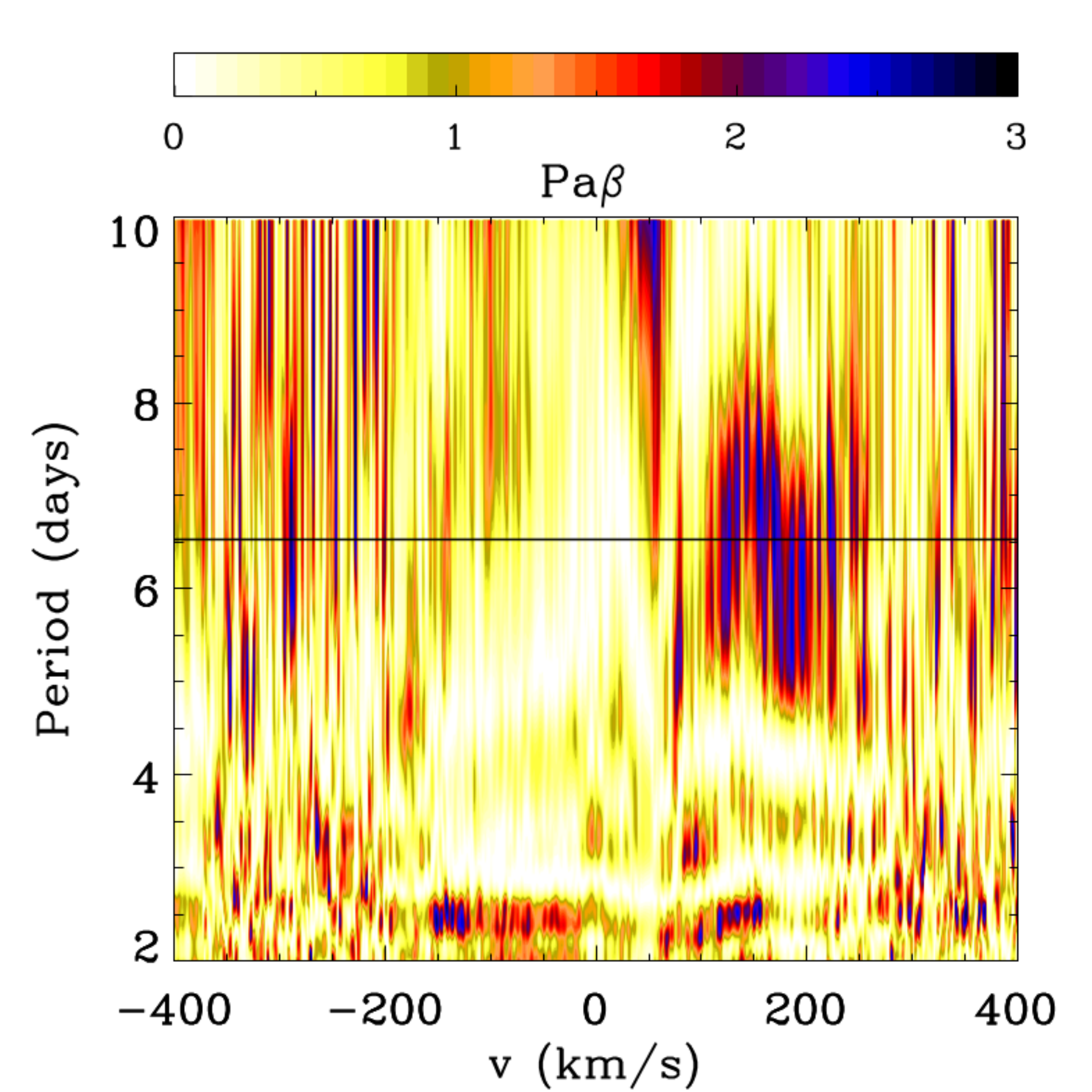}}
\subfigure[]{\includegraphics[width=4.5cm]{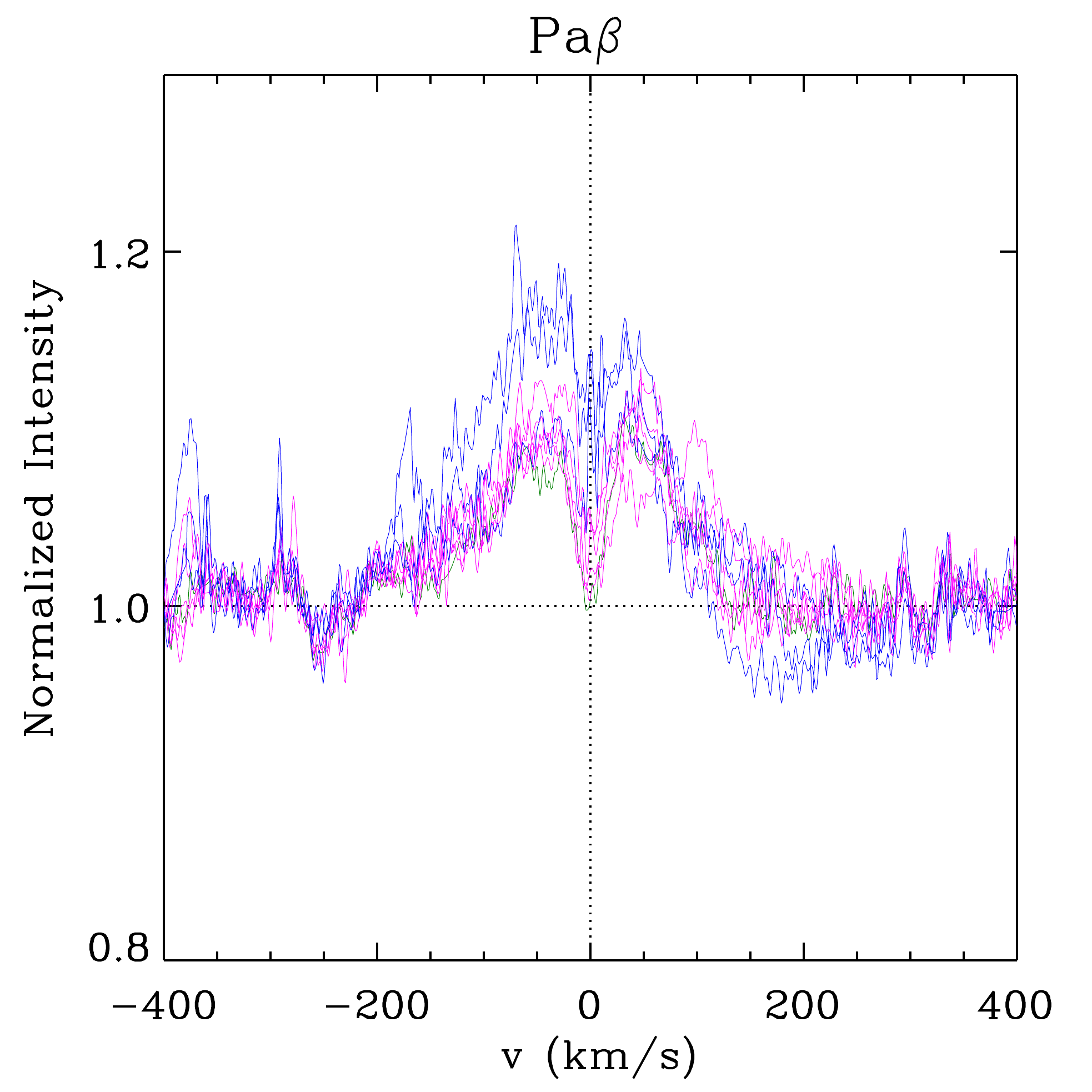}}
\subfigure[]{\includegraphics[width=4.5cm]{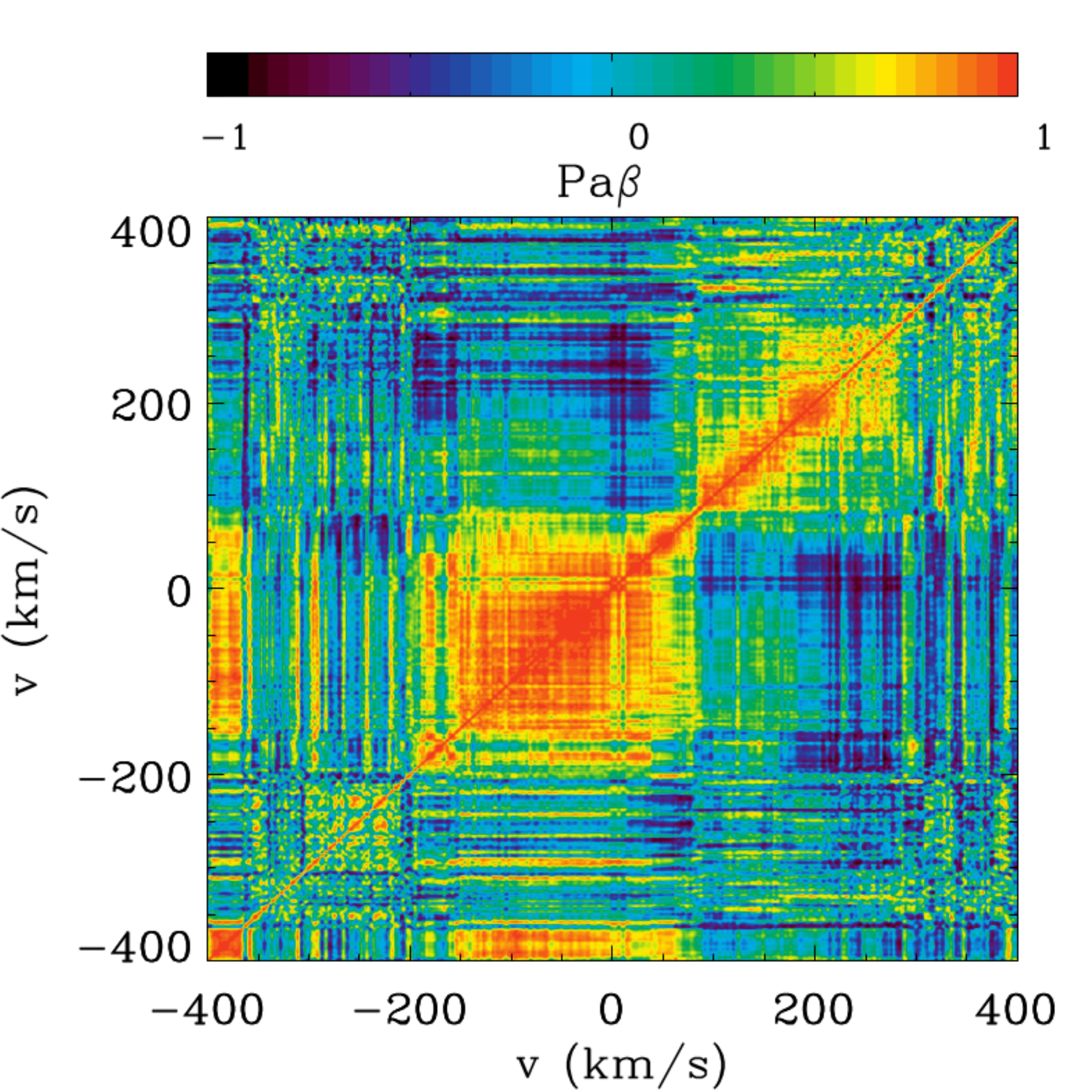}}
 \end{center}
\caption{\label{fig:perline_spirou}Time series analysis of infrared circumstellar emission line variability. Shown are the bidimensional periodogram ({\it left}), the corresponding line profiles ({\it middle}), and the correlation matrices ({\it right}). In the {\it right} panels, the color range corresponds to the normalized power of the period, and the horizontal solid line represents the rotation period of the star, $P=6.53\,\mathrm{days}$. In the {\it middle} panels, different colors correspond to different rotation cycles: black - cycle 0; green - cycle 1; red - cycle 2; and light blue - cycle 3. In the {\it left} panels, the color range represents the linear correlation coefficient between different velocity bins of the line profiles. Perfect anticorrelation corresponds to -1 (black), no correlation to 0 (light blue), and  perfect correlation to 1 (orange).}
\end{figure*}

We only have nine observations obtained with the SPIRou spectrometer, which is not ideal for measuring a periodicity of a few days. Using the same methodology to get the period in optical lines, we computed the bidimensional periodogram of the infrared lines. We present the results in Fig. \ref{fig:perline_spirou} for $\mathrm{He}$\,I and Pa$\beta$.  

The $\mathrm{He}$\,I shows a period of about 6.0 days in the redshifted wing.
However, this period presents a wide spread  in the periodogram, which reaches the stellar rotation period, due to the small number of observations. This periodic region corresponds to the redshifted absorption below the continuum level. The period is similar to that detected in the optical $\mathrm{He}$\,I line and in the central emission of the $\mathrm{H}\beta$ line (see Sect. \ref{sec:period}). The red wing of the Pa$\beta$ lines exhibits a period close to the 6.53-day stellar rotation period, apparently decreasing down to 6.0 days at the reddest velocities. We did not detect periodicity in the blue wing of the $\mathrm{He}$\,I and Pa$\beta$ line profiles. 

\subsection{Infrared circumstellar line correlation matrices} \label{sec:Incorr}

We measured the correlation coefficients of the variability along the infrared lines, applying the same methodology explained in Sect. \ref{sec:optical_cor}, and we present the results in Fig. \ref{fig:perline_spirou}. 

The red wing of $\mathrm{He}$\,I is composed of a broad absorption and presents a self-correlated behavior that is dominated by a single physical process. The high velocity part of the blue wing (from -70 to -150$\,\mathrm{km\ s^{-1}}$) seems to predominantly form in a different region of the system and is not correlated with the main source of variability of the red wing. However, the blue wing close to zero velocity (from 0 to -70$\,\mathrm{km\ s^{-1}}$), which corresponds to part of the blueshifted absorption component below the continuum level, correlates well with the redshifted absorption. 
We have presented evidence for a relationship between the width of the blueshifted absorption component and the equivalent width of the redshifted absorptions (Fig. \ref{fig:EWHeI_spirou}). The redshifted absorption is expected to come from the hot spot emission being absorbed by the accretion column as it crosses our line of sight. As reported above, the blueshifted absorption becomes wider as the redshifted absorption deepens. The wings of the blueshifted absorption thus follow the variation of the redshifted absorption. The correlation between the variability of the redshifted absorption and part of the blueshifted components would then reflect the topology of the magnetic field in the inner disk region, simultaneously driving accretion funnel flows and an inner magnetized wind.

The correlation matrix of the Pa$\beta$ line is reminiscent of that of Balmer lines. The red and blue wings are each self-correlated, but there is no correlation between them. The high-velocity redshifted absorption ($\mathrm{v}>150\,\mathrm{km\ s^{-1}}$) slightly anticorrelates with the blue wing, as could be expected by a transiting funnel flow. It is likely that most of this line profile forms in the accretion column, with the red wing variability being dominated by funnel flow self-absorption, as indicated by the inverse P Cygni profiles observed at specific phases. 

\section{Mass accretion rates}\label{sec:accrate}

Using the equivalent width of the optical and infrared emission lines, we estimated the mass accretion rate of V2129 Oph. We do not have photometric observations obtained simultaneously for all the observation nights of ESPaDOnS and HARPS, and we have none with the SPIRou data.  The V magnitudes of this system do not vary considerably (see Sect. \ref{sec:asassn}); therefore, we estimated the continuum flux of the $\mathrm{He}$\,I\ 5876$\mathring{\mathrm{A}}$ line using the mean V magnitude of the photometry from ASAS-SN. For $\mathrm{H}\alpha$ we computed the continuum flux using R magnitudes obtained from the color $(V-R)$ relation in \cite{2008A&A...479..827G} and converted to the Cousins system \citep{1983PASP...95..782F}. For the infrared, we used the mean J magnitude from REM, discussed in Sect. \ref{sec:photo}. We corrected the magnitudes from reddening using $\mathrm{Av} = 0.6\,\mathrm{mag}$ \citep{2007MNRAS.380.1297D}. We then used the appropriate relations obtained by \cite{2017A&A...600A..20A} to derive the accretion luminosity from $\mathrm{H}\alpha$, $\mathrm{He}$\,I\ 5876$\mathring{\mathrm{A}}$, and Pa$\beta$ lines and obtained the mass accretion rates. 
We found the mean accretion rates of $(1.2\pm0.4)\times10^{-9}\mathrm{M_\odot yr^{-1}}$ using the $\mathrm{He}$\,I\ 5876$\mathring{\mathrm{A}}$ line, $(1.5\pm0.5)\times10^{-9}\,\mathrm{M_\odot yr^{-1}}$ using the $\mathrm{H}\alpha$ line, and $(1.2\pm0.4)\times10^{-9}\,\mathrm{M_\odot yr^{-1}}$ using the Pa$\beta$ line. Taking the errors into account, these accretion rates agree with the mean value of ($1.5\pm0.6\times10^{-9}\,\mathrm{M_\odot yr^{-1}}$) obtained by \cite{2012A&A...541A.116A}. 

\section{Discussion} \label{sec:discussion}

In this section we discuss the circumstellar environment of V2129 Oph and the dynamics of the system over almost one decade, comparing our results with previous works.

\subsection{Previous campaigns on V2129 Oph} % provisory title
The accretion process and the magnetic field of V2129 Oph were previously studied by various authors \citep[e.g.,][]{2012A&A...541A.116A,2011MNRAS.412.2454D,2011MNRAS.411..915R,2007MNRAS.380.1297D}. 
The magnetic field of V2129 Oph is more complex than a simple dipole. Some previous works \citep[e.g.,][]{2008MNRAS.389.1839G,2011MNRAS.411..915R} found that the dipole component truncates the disk, while the octupole dominates on a smaller scale and at the surface of the star. The magnetic field intensity changed between the ESPaDOnS  spectropolarimetric campaigns of 2005 \citep{2007MNRAS.380.1297D} and 2009 \citep{2011MNRAS.412.2454D}: The magnitude of the octupole and dipole  components increased by factors of 1.5 and 3.0, respectively.  A variation in the magnetic field strength can change the accretion properties of the young system considerably. The full derivation of the magnetic properties of the system will be presented in an accompanying paper (Donati et al., in prep.).

The time series analysis performed by \cite{2012A&A...541A.116A} is reproduced and completed in Appendix \ref{sec:data2012}. Using optical spectroscopic data, they recover the $6.53${-day} stellar rotation period in the redshifted absorption components of the Balmer emission lines, as well as in the narrow component of the optical HeI line. These results indicate that, in 2009 when this data set was obtained, the accretion was channeled along one main accretion column. Accordingly, the emission lines vary in phase with the stellar rotation period. They additionally found a period of around $8.3\,\mathrm{days}$ close to the line center in the periodogram of the $\mathrm{H}\alpha$ line, with a smaller power than that of the signal seen at the stellar rotation period in the red wing of the line profile. As this period is longer than the stellar rotation period, they tentatively associated it with a disk wind contribution to line emission.

\cite{2012A&A...541A.116A} modeled the $\mathrm{H}\alpha$ and $\mathrm{H}\beta$ profiles using a magnetohydrodynamic (MHD) model that computes line emission from an inclined magnetosphere. They computed the emission lines for different rotational phases, modeling the magnetic field as only a dipole or as a dipole plus octupole components. For both magnetic field configurations, they obtained $\mathrm{H}\alpha$ and $\mathrm{H}\beta$  profiles with characteristics consistent with the observations, although the redshifted absorptions did not appear at the same phases as in the observations. The theoretical profiles also presented a less extended blue wing than the observed ones, and this difference was initially attributed to the absence of a disk wind component in the code. To test the influence of the wind on the profiles, they modeled the $\mathrm{H}\alpha$ line with an MHD code that combines both magnetospheric and disk wind components. However, the results showed that the disk wind component does not contribute significantly to the $\mathrm{H}\alpha$ profile of V2129 Oph due to the low mass-loss rate of the system. Theoretical models of circumstellar emission lines show that the disk wind component contributes to the intensity of the line in cases of high mass accretion rates \citep[$>10^{-8}$\ms; e.g.,][]{2010A&A...522A.104L} and, consequently, high mass-loss rates. In other words, only a dense disk wind can contribute significantly to the emission line profile.

To summarize,  the variability of  the circumstellar lines  of V2129 Oph in previous works can be explained by stable magnetospheric accretion, even taking the complexity of its surface magnetic field into account. Indeed, except for the central part of the $\mathrm{H}\alpha$ line, most of the line profile variability was modulated on a period consistent with the 6.53-day rotational period of the central star, as expected if the magnetosphere truncates the disk close to the corotation radius.

\subsection{Long-term stability of the magnetosphere of V2129 Oph}
The spectroscopic time series analysis we performed in this work on the photospheric and circumstellar lines of V2129 Oph shows that the system presented multiples periods during our observations. The stellar rotation period of $6.53\,\mathrm{days}$ is recovered in photospheric lines, being clearly modulated by cold surface spots (see Sect. \ref{sec:optical}). We detected a period of 6.0 days in the circumstellar emissions lines, which is shorter than the stellar rotation period by about $\sim0.5\,\mathrm{days}$, as well as another period that was longer than the stellar rotation period, close to $\sim8.5\,\mathrm{days}$, in various components of the emission line profiles.  

This longer periodicity was already present in the peak of the $\mathrm{H}\alpha$ line in the previous analysis of V2129 Oph \citep{2012A&A...541A.116A}. While the origin of this $8.5\,${-day} period is unclear, it appears to be stable on a timescale of 9 years. In contrast, the stellar rotation period of $6.53\,\mathrm{days}$ disappeared almost entirely from the emission line periodograms between the 2009 and 2018 campaigns. We find hints for the stellar rotational period only in part of the red wing of the Pa$\beta$ line profile, and marginally in the reddest velocity channels of the H$\beta$ profile. Yet, the mass accretion rate did not change significantly between these two epochs (see Sect. \ref{sec:accrate}), and the shape of the Balmer and $\mathrm{He}$\,I\ 5876$\mathring{\mathrm{A}}$ line profiles was extremely similar between the two campaigns; the lines were only slightly more intense during the former. It is therefore likely that the contrasting periodicity patterns the emission lines display between the two epochs result from a magnetic structure changing on a timescale of years, as was previously reported for this source \citep{2011MNRAS.412.2454D}.

\subsection{New environment configuration} 
A new configuration of the circumstellar environment of V2129 Oph seems therefore necessary to explain the line variability observed in our data set. We explore below a few possible interpretations.

One possibility to explain the system's observed variability is to assume that the magnetospheric truncation radius is larger than the corotation radius of the disk. In this case, the funnel flow would rotate more slowly than the star at its base, where the low-velocity Balmer line emission originates. As the funnel flow approaches the stellar surface, it is dragged by the faster stellar rotation. A trailing accretion funnel flow thus develops, where the high-velocity redshifted absorption components, which form closer to the star, would rotate faster than the base of the flow. This could account for the continuous decrease in the periodicity, from 8.5 to 6.5 days, seen from the center to the red wing of the H$\beta$ line profile (see Fig. \ref{fig:perline_all}). At the stellar surface, the accretion shock seems to rotate even faster, with a period about 10\% shorter than that of the star itself, as suggested by the 6.0-day periodicity seen in the post-shock HeI emission line. 
As the long-term photometric monitoring of V2129 Oph has shown the stellar period to vary between $6.34$ and $6.59\,\mathrm{days}$ \citep{2008A&A...479..827G}, it would be tempting to assign the shorter $6.0${-day} period to latitudinal differential rotation at the stellar surface, with a high latitude hot spot rotating more rapidly than the stellar equator. However, neither the magnitude nor the direction would be consistent with the differential rotation previously measured for V2129 Oph \citep{2011MNRAS.412.2454D}.

The environment of a CTTS can be more complex than a single accretion column in each hemisphere, with several components contributing to the line profile. \cite{2019ApJ...884...86T} analyzed the accretion process of the CTTS CVSO 1335. The $\mathrm{H}\alpha$ and $\mathrm{H}\beta$ lines of this system present two redshifted absorptions, one close to the line center and another at high velocity. By modeling the circumstellar lines, they constructed an empirical scenario that can explain the system, namely a magnetosphere consisting of two major funnel flows originating at different radii in the inner disk. In such a two-flow magnetosphere, the lower-velocity redshifted absorption component is formed in the outer funnel flow, while the high-velocity redshifted absorption component arises from the more compact magnetospheric component.

Unlike CVSO 1335, V2129 Oph presents only one redshifted absorption component and has a moderate mass accretion rate, while CVSO 1335 is a weak accretor ($\leq 9\times10^{-10}\,\mathrm{M_\odot yr^{-1}}$). Nevertheless, a two-shell accretion scenario could 
 potentially explain some of the results we obtained for V2129 Oph since both accretion funnels could then contribute to the emission lines. The longer period of $\sim 8.5\,\mathrm{days}$ seen in the Balmer emission lines would then be assigned to the external funnel flow, which would have to extend up to $\sim9.2\,\mathrm{R_\ast}$ ($\sim0.09\,\mathrm{au}$) from the star,  beyond the corotation radius located at $7.7\,\mathrm{R_\ast}$ ($0.075\,\mathrm{au}$), assuming Keplerian rotation. The compact accretion funnel flow would be responsible for the high-velocity redshifted absorptions seen in the optical and infrared profiles, which exhibit a period closer to that of the star. The gas in the inner funnel flow would hit the star at lower latitudes than the gas in the outer funnel. The optical $\mathrm{He}$\,I line forms in this region, and this could explain the periodicity of $\sim 6\,\mathrm{days}$ seen in this line, which is smaller than the rotation period of the star. The red wing of the $\mathrm{He}$\,I\ 10830$\mathring{\mathrm{A}}$ and Pa$\beta$ lines as well as the redshifted absorptions in the $\mathrm{H}\alpha$ and $\mathrm{H}\beta$ lines could arise from the inner accretion column. A Keplerian period of $6.0\,\mathrm{days}$ would locate the base of the inner accretion column at $\sim7.3\,\mathrm{R_\ast}$ ($\sim0.071\,\mathrm{au}$) from the star, that is, slightly below the corotation radius. As pointed out by \cite{2019ApJ...884...86T}, a configuration with two accretion flows may still be a simplification of a more complex accretion topology governed by different magnetic field components.

V2129 Oph was observed with the VLTI/GRAVITY interferometer in the K-band at the time of our campaign (GRAVITY Collaboration: Perraut et al., in prep.). Assuming a flux contribution from the disk to the system ranging from 0.30 to  0.38, as follows from the estimate of the veiling in the K-band,
Perraut et al. derived a half-flux radius for the K-band continuum emission of 0.79$^{+0.14}_{-0.13}\,\mathrm{mas}$, that is, $0.103\pm0.017\,\mathrm{au}$ ($\sim10\pm2\,\mathrm{R}_\ast$), at a distance of $131.9\,\mathrm{pc}$. 
The half-flux radius is an estimate of the location of the disk's dusty inner edge and is consistent with a magnetospheric radius of $0.09\,\mathrm{au}$.
 
In these two proposed scenarios, the system is dominated by the magnetosphere (two accretion columns or only a single extended one), and we expect the disk truncation ($\mathrm{r_m}$) to occur beyond the corotation radius of the star-disk system \citep[$R_\mathrm{{co}}$$=7.7\,\mathrm{R}_\ast;$][]{2012A&A...541A.116A}. Accretion is facilitated if $r_\mathrm{m}<R_\mathrm{co}$ \citep[e.g.,][]{2015SSRv..191..339R}, and a system with $r_\mathrm{m}>R_\mathrm{co}$ is expected to be in the propeller regime \citep[e.g.,][]{1999ApJ...514..368L}, where the star rotates faster than the inner disk. In the propeller regime, accretion is still present \citep[e.g.,][]{2018NewA...62...94R}, but the accretion column created outside the corotation radius is unstable and should disappear after a few stellar rotations. A concern for this configuration is the dipole magnetic field intensity required to disrupt the disk at such a distance from the star. If we assume that the truncation radius lies at $9.3\,\mathrm{R}_\ast$, the relationship between the truncation radius and the magnetic field intensity from \cite{2008A&A...478..155B} yields a dipole magnetic field strength of $\sim2.8\,\mathrm{kG}$, which is about three times more intense than the value obtained during the previous campaign \citep{2011MNRAS.412.2454D}.

Another proposal for the observed line variability takes into account a possible evolution in the accretion regime between the two campaigns, where the system would switch from stable accretion -- driven by one major accretion funnel on each stellar hemisphere -- to an unstable accretion regime -- where accretion can occur through multiple random accretion tongues \citep{2013MNRAS.431.2673K,2016MNRAS.459.2354B}. In the 2018 observations, the emission lines show some characteristics that support this scenario. For example, the redshifted absorptions, best seen in the \hb\ line, appear continuously from phase 0.47 to 0.79, as expected from a primary accretion funnel, but they also appear sporadically at different phases, indicative of unstable accretion funnels that can appear and disappear in a single rotation cycle. In the near-infrared \he line, a redshifted absorption component appears continuously below the continuum from phase 0.28 to 0.89, and it is the deepest between phases 0.34 and 0.67.  The intensity of the \hem line in the 2018 data is not modulated at the stellar rotation period (see Fig. \ref{fig:HelIline_all}), as was observed in the 2009 data set \citep{2012A&A...541A.116A}. Unstable accretion is actually facilitated for a small magnetospheric radius \citep{2016MNRAS.459.2354B}.\ Additionally, the mass accretion rate of V2129 Oph did not change between the two data sets; as such, if the magnetic field dipole component decreased in 2018, the inner disk could reach closer to the star, resulting in a smaller magnetosphere. This configuration, however, does not explain the period of 8.5 days seen in the \ha\ and \hb\ lines.

Finally, we cannot exclude that an external disturbance to the magnetospheric accretion-ejection process could be responsible for the occurrence of more than one period in the system. Recent studies reported the detection of orbiting hot Jupiters around weak-line T Tauri stars \citep[e.g.,][]{2017MNRAS.465.3343D,2017MNRAS.467.1342Y,2020MNRAS.tmp.3555K} and around the accreting T Tauri star CI Tau \citep{2016ApJ...830...15J}. Short orbit planetary mass companions could contribute to the flux in the emission lines of accreting systems, either directly through accretion onto the protoplanet or indirectly via the gravitational perturbation they may induce on the star-disk interaction. The longer period detected in V2129 Oph, which is significantly longer than the stellar rotation period, could be attributed to such a disturbance. However, we have so far found no independent evidence supporting this interpretation. We do measure radial velocity variations with an amplitude of a few km/s in the system, but, as shown in Sect. \ref{sec:optical} from a bisector analysis, these radial velocity variations are best accounted for by stellar spots. Indeed, the dynamical accretion-ejection process in CTTSs makes it quite challenging to detect a possible contribution from an inner embedded planet to the variability of the system.

\section{Conclusions} \label{sec:concl}
In this work we revisited the variability pattern of the CTTS V2129 Oph on a timescale of a few rotational periods to probe the star-disk magnetospheric interaction region. Using high-resolution spectroscopic time series, we analyzed veiling variations, as well as the changing shape and strength of optical and infrared emission lines. 

From the photospheric spectrum, we measured the stellar radial velocity 
and got consistent results from optical ($v_{rad}=-7.1\pm 0.8\,\mathrm{km\ s^{-1}}$) and infrared ($v_{rad}=-6.8\pm0.6\,\mathrm{km\ s^{-1}}$) data. We find the radial velocity to be rotationally modulated by either two sets of cold spots located at nearly opposite longitudes at the stellar surface or an elongated polar spot with latitudinal extensions at opposite azimuths. We recover the stellar rotation period of 6.53 days. These spots also dominate the photometric variability of the system. 
Furthermore, we derived the wavelength-dependent veiling across the optical and near-infrared ranges. 
Our results support the idea that optical veiling arises from a hot spot and that infrared veiling originates from dust in the illuminated inner disk.

The optical emission lines are broad and variable on a night-to-night basis and support the idea of magnetically controlled accretion close to the star. High-velocity redshifted absorptions, most conspicuous in the $\mathrm{H}\beta$ and HeI 10830\AA\ lines and also seen in the H$\alpha$ and Pa$\beta$ profiles, appear below the continuum level at some phases, which is a result of the main accretion funnel flow crossing the line of sight. V2129 Oph presents only a narrow component in the $\mathrm{He}$\,I\ 5876$\mathring{\mathrm{A}}$ line, redshifted by about $v_{rad_{HeI}}=(3\pm2)\,\mathrm{km\ s^{-1}}$, which is commonly associated with the accretion post-shock region.  The $\mathrm{He}$\,I\ 10830$\mathring{\mathrm{A}}$ line additionally provides evidence for outflowing gas as it presents blueshifted absorptions below the continuum level at all rotational phases. The NaI D lines exhibit transient high-velocity blueshifted absorptions, indicative of episodic outflows. 

We applied a Lomb-Scargle periodogram analysis to all emission lines, and, surprisingly,
different lines exhibit different periodicities and sometimes even a varying period across the line profile. A period of 6.0 days, shorter than the 6.53-day rotational period, is measured in the $\mathrm{He}$\,I\ 5876$\mathring{\mathrm{A}}$ line profile.
Conversely, a longer period of 8.5 days is measured in the H$\alpha$ profile and in the red wing of the H$\beta$ profile. 
The longer period suggests that the emission arises from a structure located beyond the disk corotation radius.
While the origin of the multiple periods seen in the line profiles remains puzzling, we discuss various possibilities for the circumstellar configuration, including a trailing funnel flow, multiple accretion flows, and an external disturbance.

The combination of quasi-simultaneous optical and near-infrared spectroscopy allows for a detailed characterization of the system's properties and variability. Comparing our results to those obtained from previous campaigns performed on V2129 Oph over the last 12 years, we find that the mass accretion rate onto the star hardly varied and that both the shape and intensity of the emission line profiles remained fairly stable on this timescale. However, the variability pattern of the lines changed drastically, with new periodicities appearing that were either not seen or were not dominant during previous monitoring campaigns. Our data provide further evidence that the environment of CTTSs can be highly dynamic on a timescale of a few years and suggest that the topology of CTTS surface magnetic fields may evolve significantly over this period.

\begin{acknowledgements}
We thank the referee for the suggestions that helped to clarify this paper, and Nicola Astudillo and Jorge Martins for performing some of the HARPS observations at ESO La Silla. We would like to thank Carlo Felice Manara, Scott Gregory, Gaspard Duchene, Clement Baruteau, and Claire Davies for  carefully reading the manuscript and give suggestions to improve the paper. APS and SHPA acknowledge support from CNPq, CAPES and Fapemig. This project has received funding from the European Research Council (ERC) under the European Union’s Horizon 2020 research and innovation programme (grant agreement No 742095; SPIDI: Star-Planets-Inner Disk-Interactions; http://www.spidi-eu.org and grant  agreement  No.  740651 NewWorlds). This  work  was  conducted  during  a  scholarship  supported  by  the  International  Cooperation Program    CAPES/COFECUB Foundation at  the  University  of  Grenoble.  Financed  by  CAPES –Brazilian Federal Agency for Support and Evaluation of Graduate Education within the Ministry of Education of Brazil. JR acknowledges funding from the European Research Council (ERC) under the European Union’s Horizon 2020 research and innovation program (grant agreementNo. 682393 AWESoMeStars). A.C. acknowledge funding from ANR of France under contract number ANR-18-CE31-0019 (SPlaSH). This work is supported by the French National Research Agency in the framework of the Investissements d' Avenir program (ANR-15-IDEX-02), through the funding of the "Origin of Life" project of the University of Grenoble.
\end{acknowledgements}

\bibliographystyle{aa}   
\bibliography{ref}

\begin{appendix}
\section{Residual emission line profiles}\label{sec:residual}

We show here the residual profiles obtained after removing the photospheric lines, using the WTTS V819 Tau as a template (see Sect. \ref{sec:opticallines} for ESPaDOnS and HARPS data and Sect .\ref{sec:infraredlines} for SPIRou data).  Before computing the residual profiles, by subtracting the template spectra from the V2129 Oph profiles, we removed any chromospheric emission component seen in the template line profiles. This was done by simply putting the chromospheric emission component back to the continuum level. Indeed, the chromospheric emission level may differ between V2129 Oph and the template, and, comparing the ESPaDOnS and HARPS template spectra taken at different epochs, it even appears to be  variable in the template.

\begin{figure*}
 \begin{center}
\subfigure[]{\includegraphics[width=4.0cm]{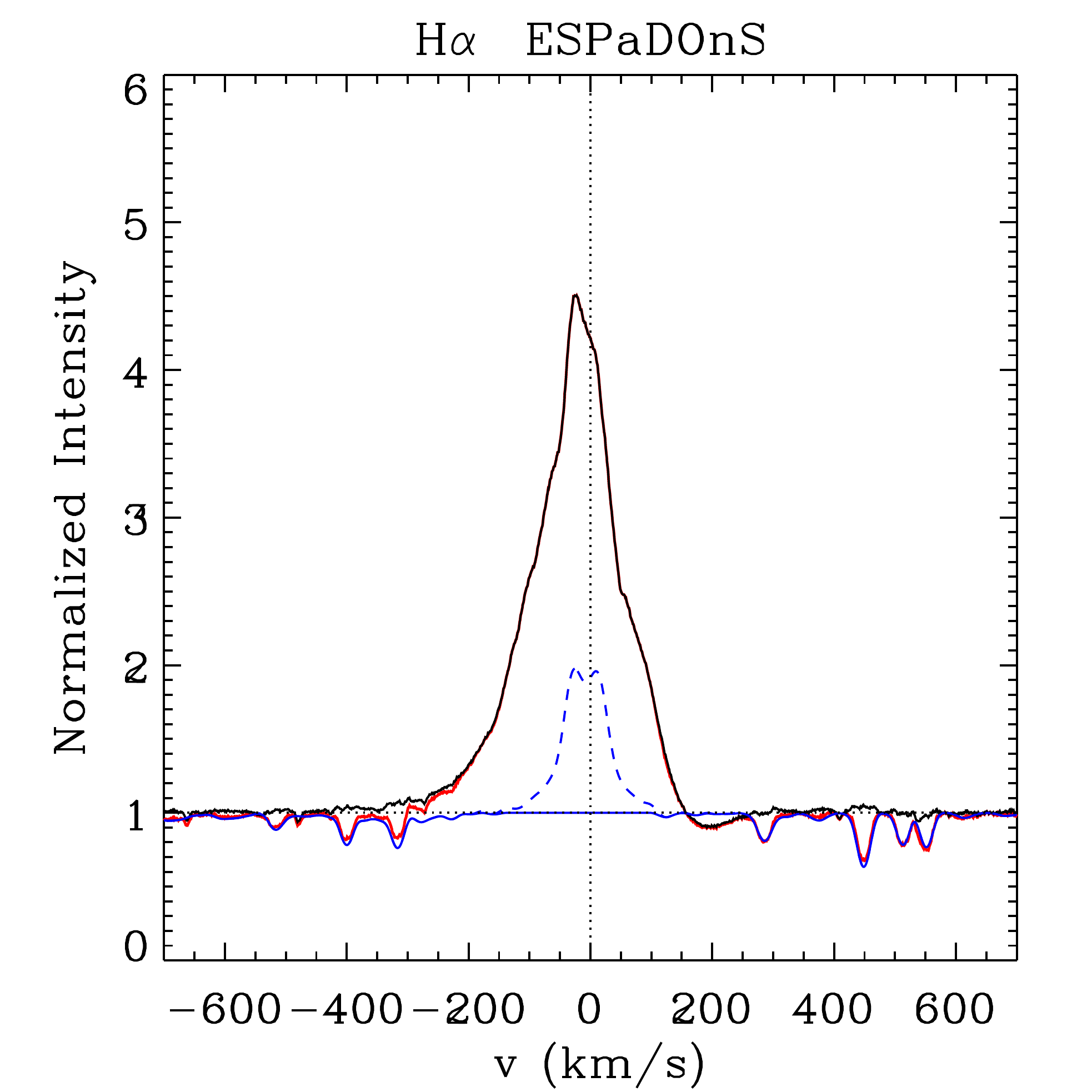}}
\subfigure[]{\includegraphics[width=4.0cm]{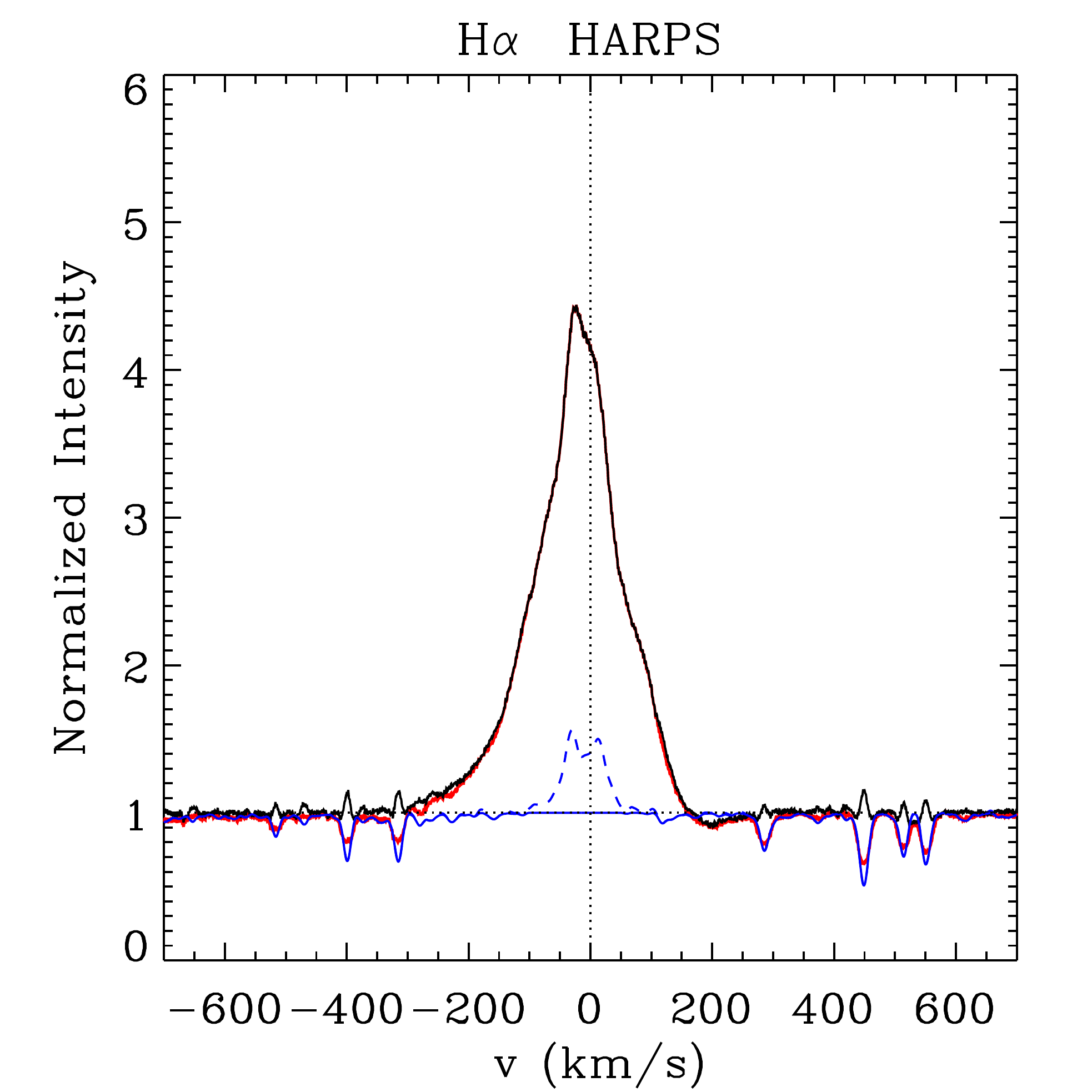}}
\subfigure[]{\includegraphics[width=4.0cm]{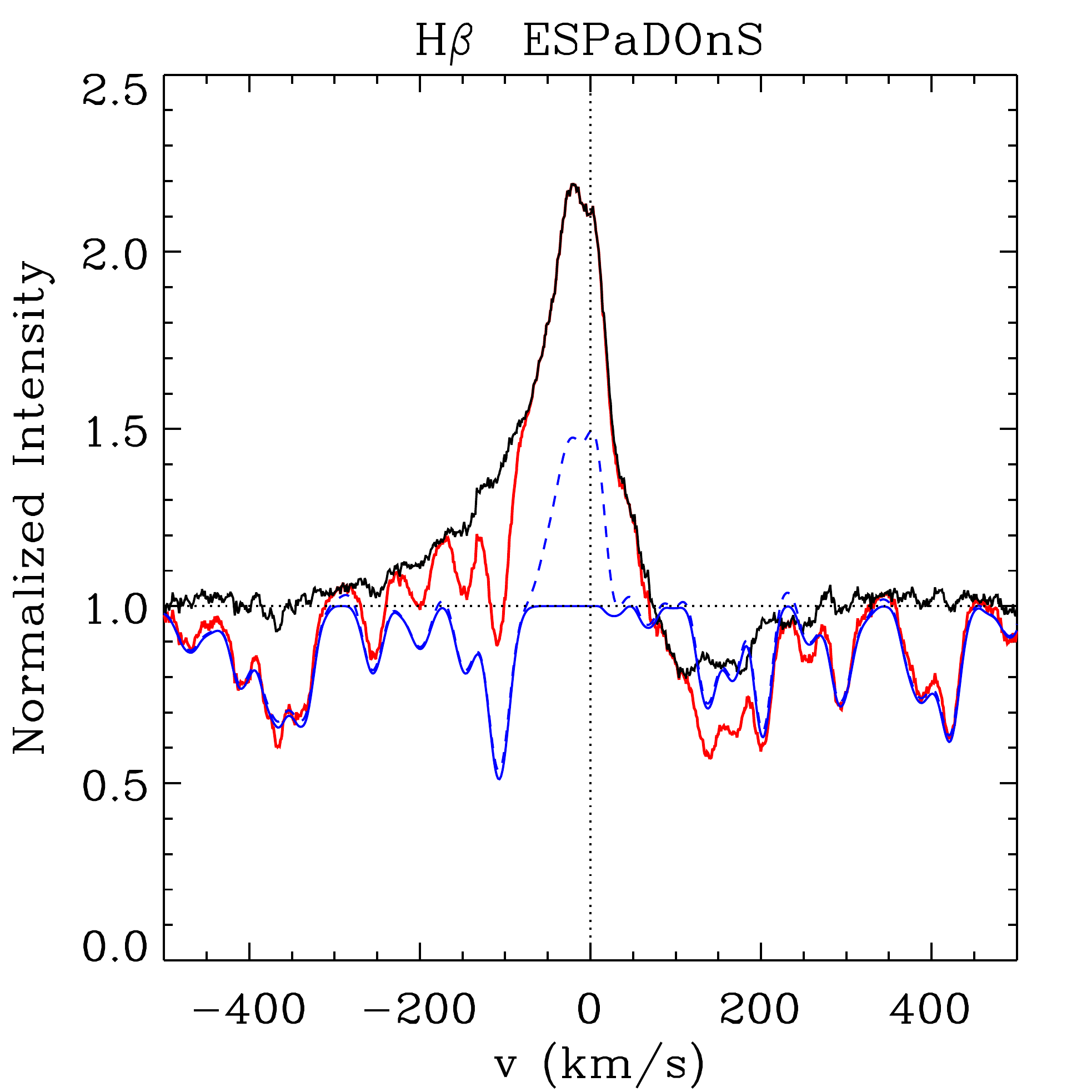}}
\subfigure[]{\includegraphics[width=4.0cm]{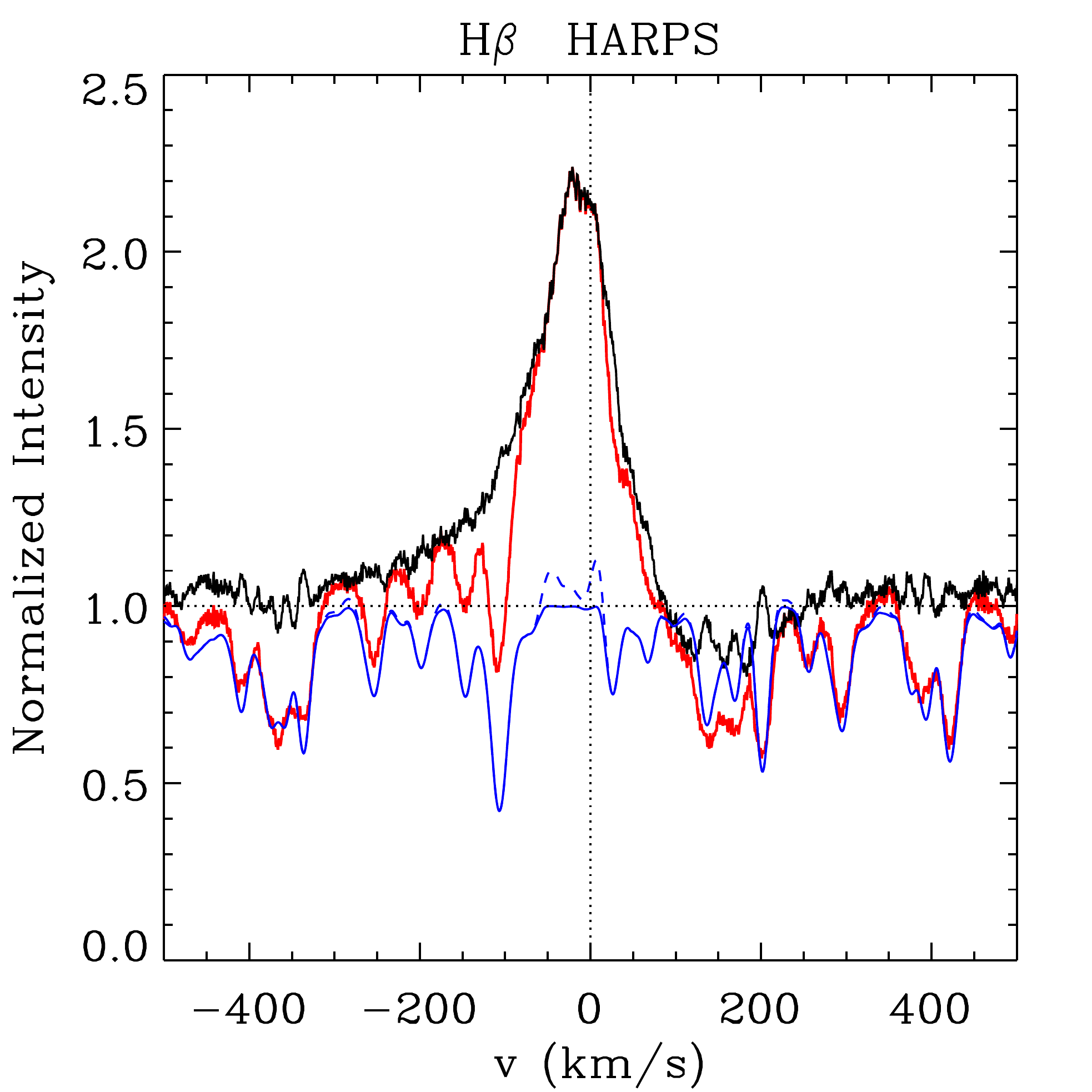}}
\subfigure[]{\includegraphics[width=4.0cm]{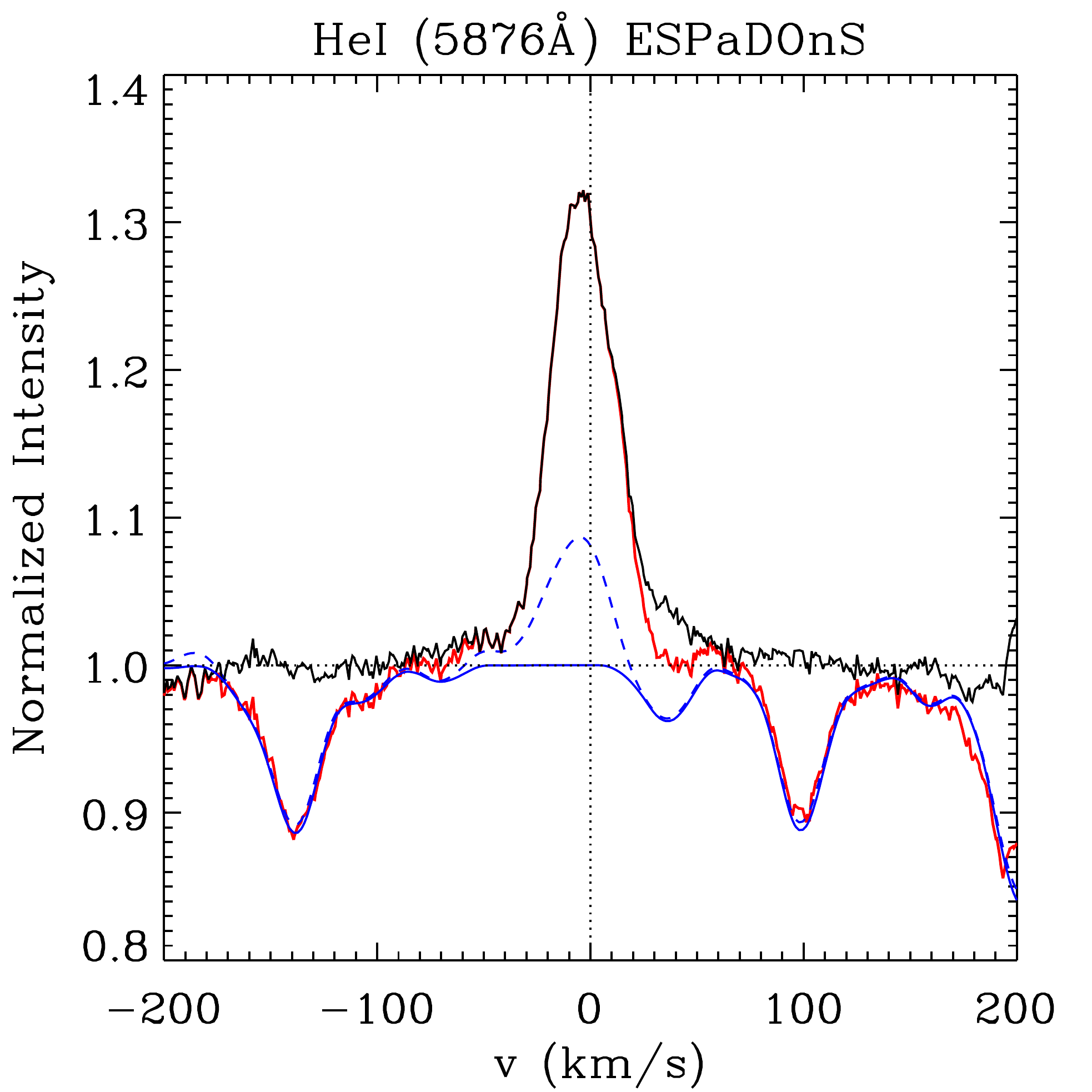}}
\subfigure[]{\includegraphics[width=4.0cm]{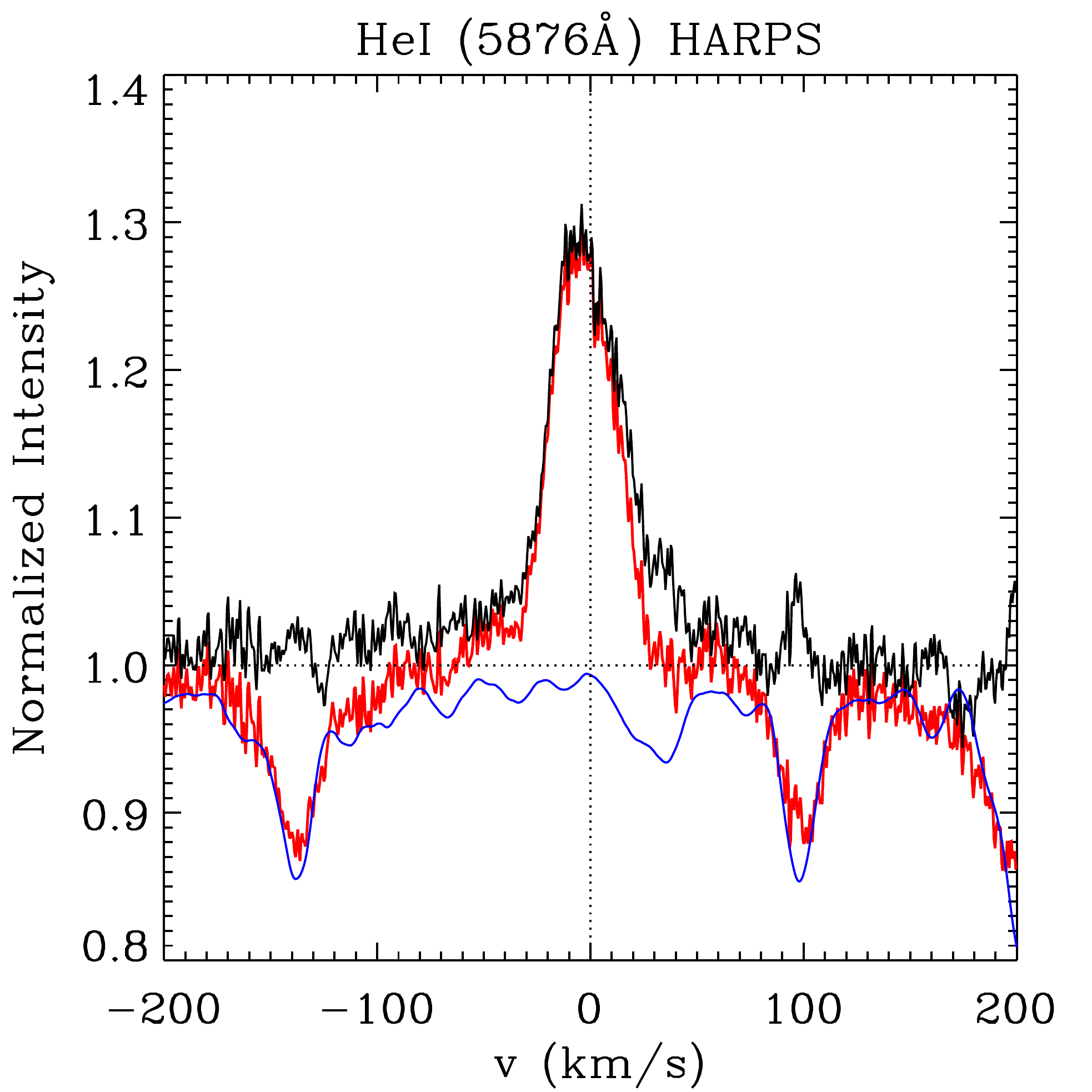}}
\subfigure[]{\includegraphics[width=4.0cm]{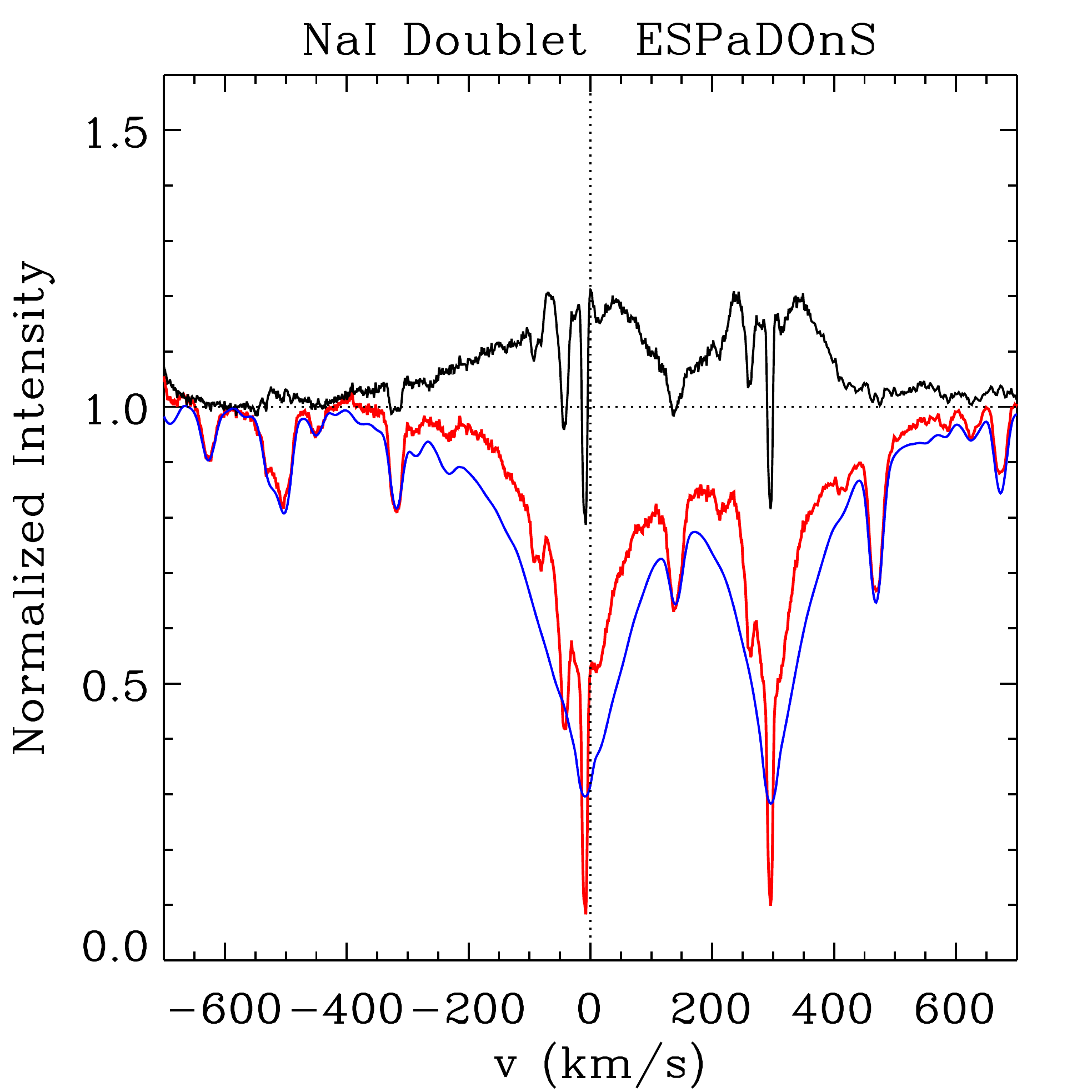}}
\subfigure[]{\includegraphics[width=4.0cm]{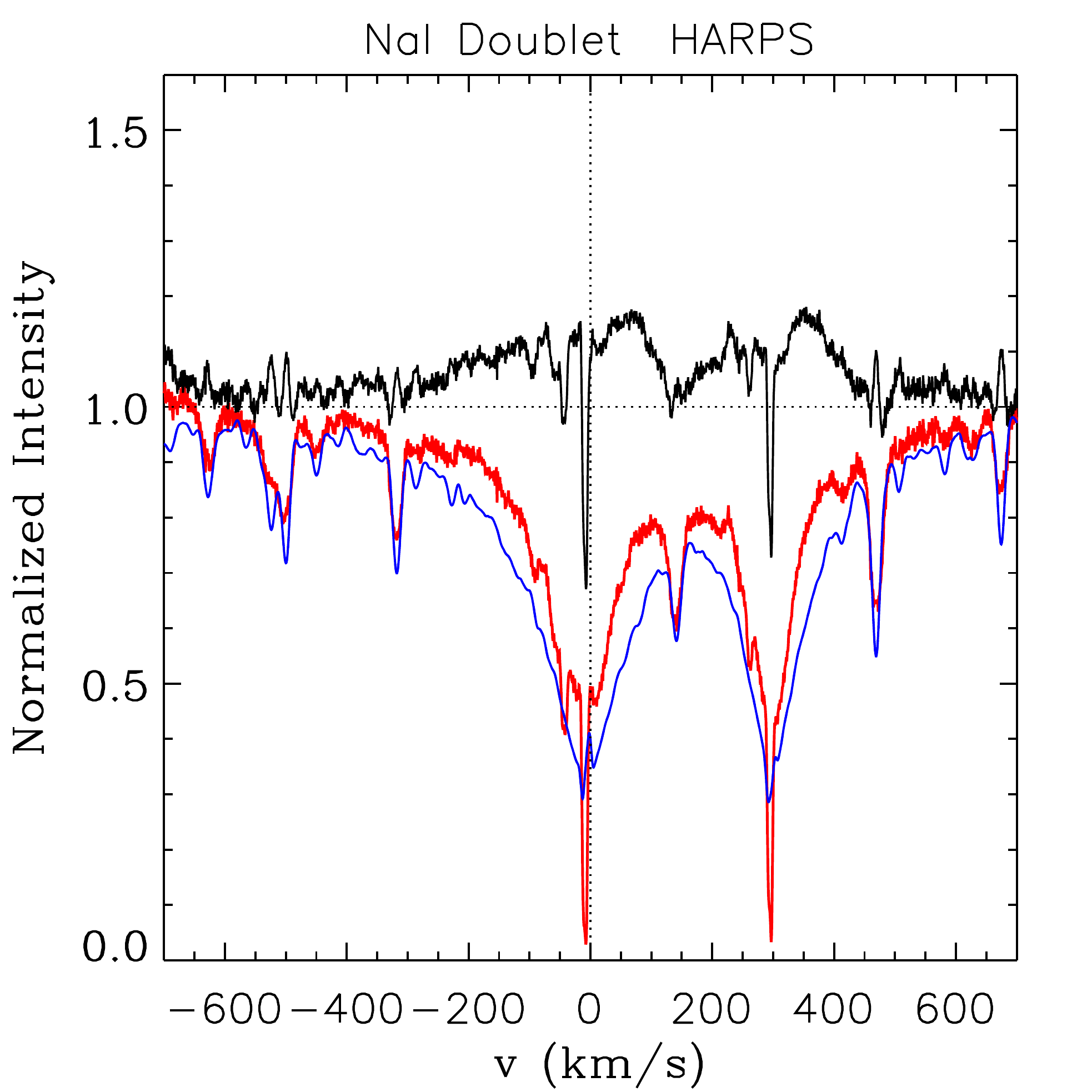}}
\subfigure[]{\includegraphics[width=4.0cm]{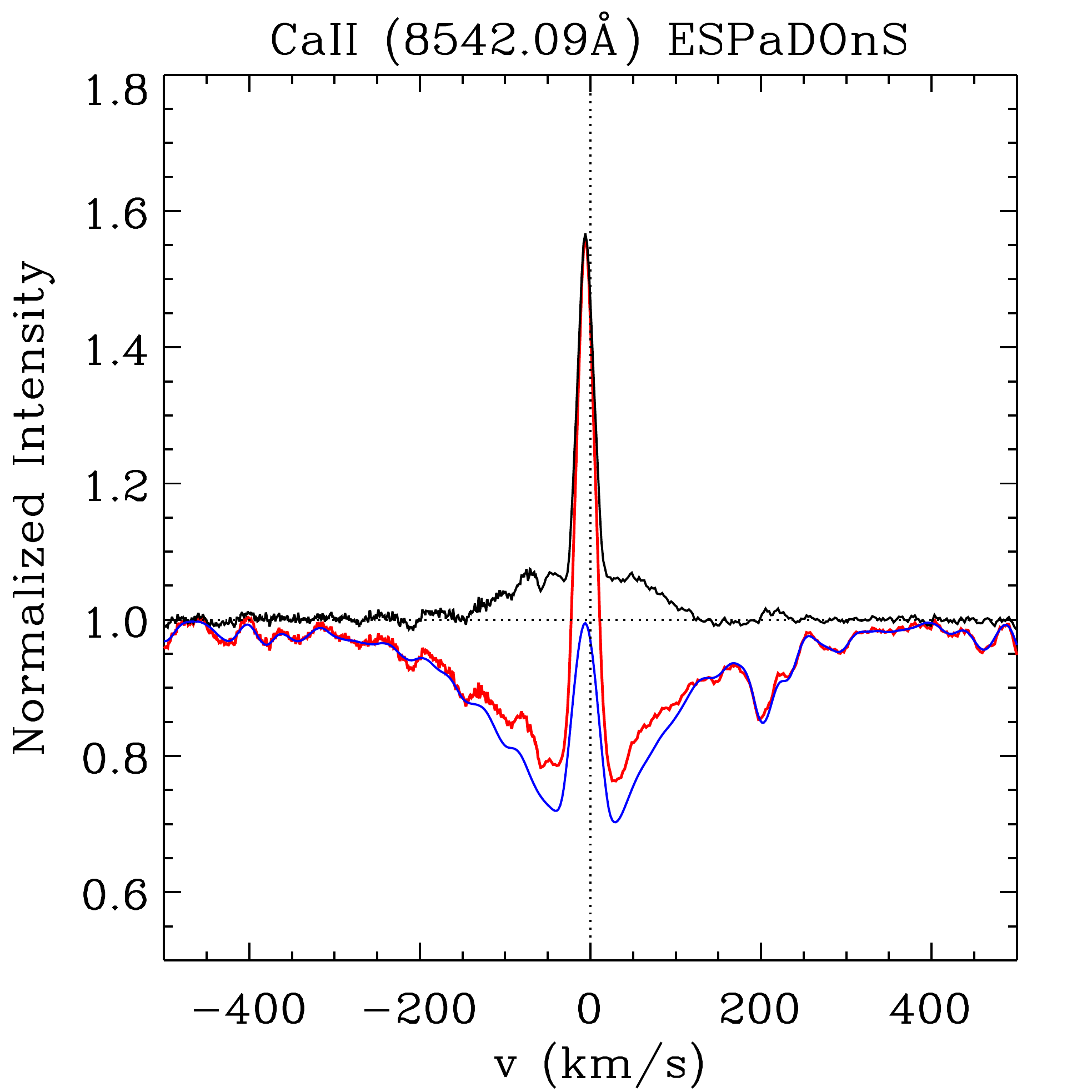}}
\subfigure[]{\includegraphics[width=4.0cm]{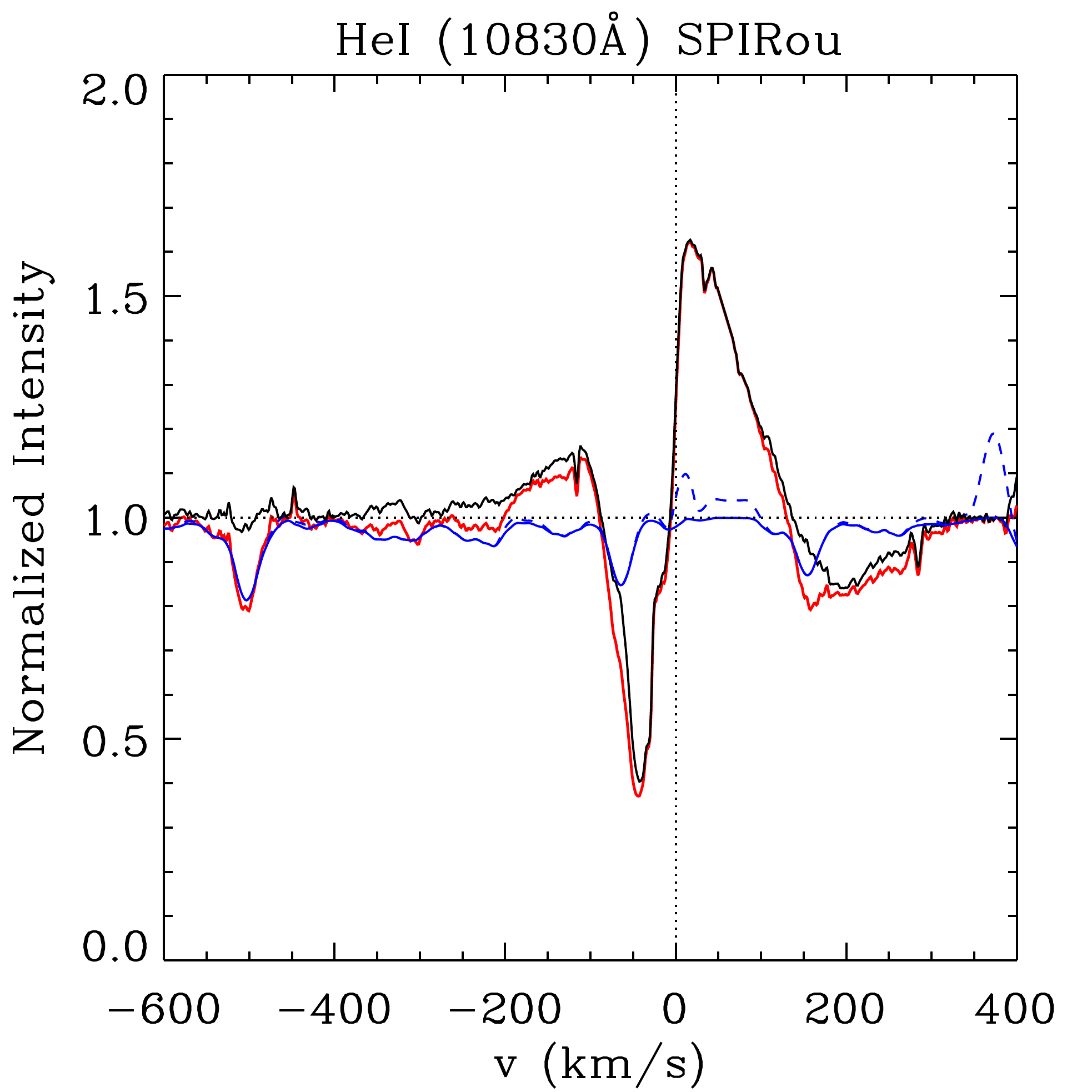}}
\subfigure[]{\includegraphics[width=4.0cm]{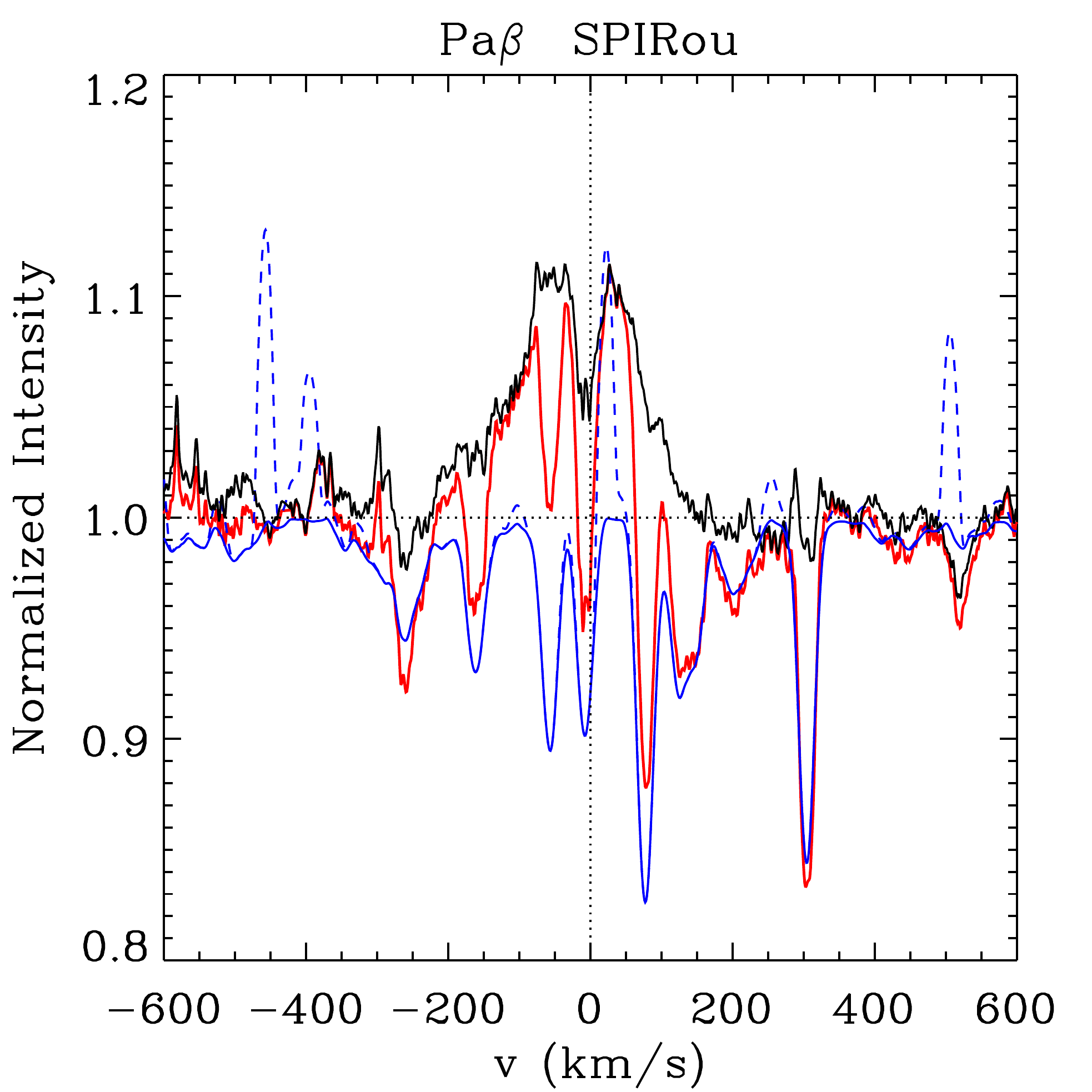}}
 \end{center}
\caption{\label{fig:resid} Mean profiles of optical and infrared circumstellar lines. Shown is a comparison of the lines before (red) and after (black) removing the photospheric contribution. The template used, V819 Tau, is shown in blue.  Before computing the residual profiles, we removed the chromospheric emission of the template by setting it to the continuum level (solid blue curve). The original template spectra are also shown (dashed blue line). The narrow emission lines in the Pa$\beta$ template are sky emission components.} 
\end{figure*}

\section{Optical and infrared radial velocities}
\begin{figure} 
\centering
\includegraphics[scale=0.40]{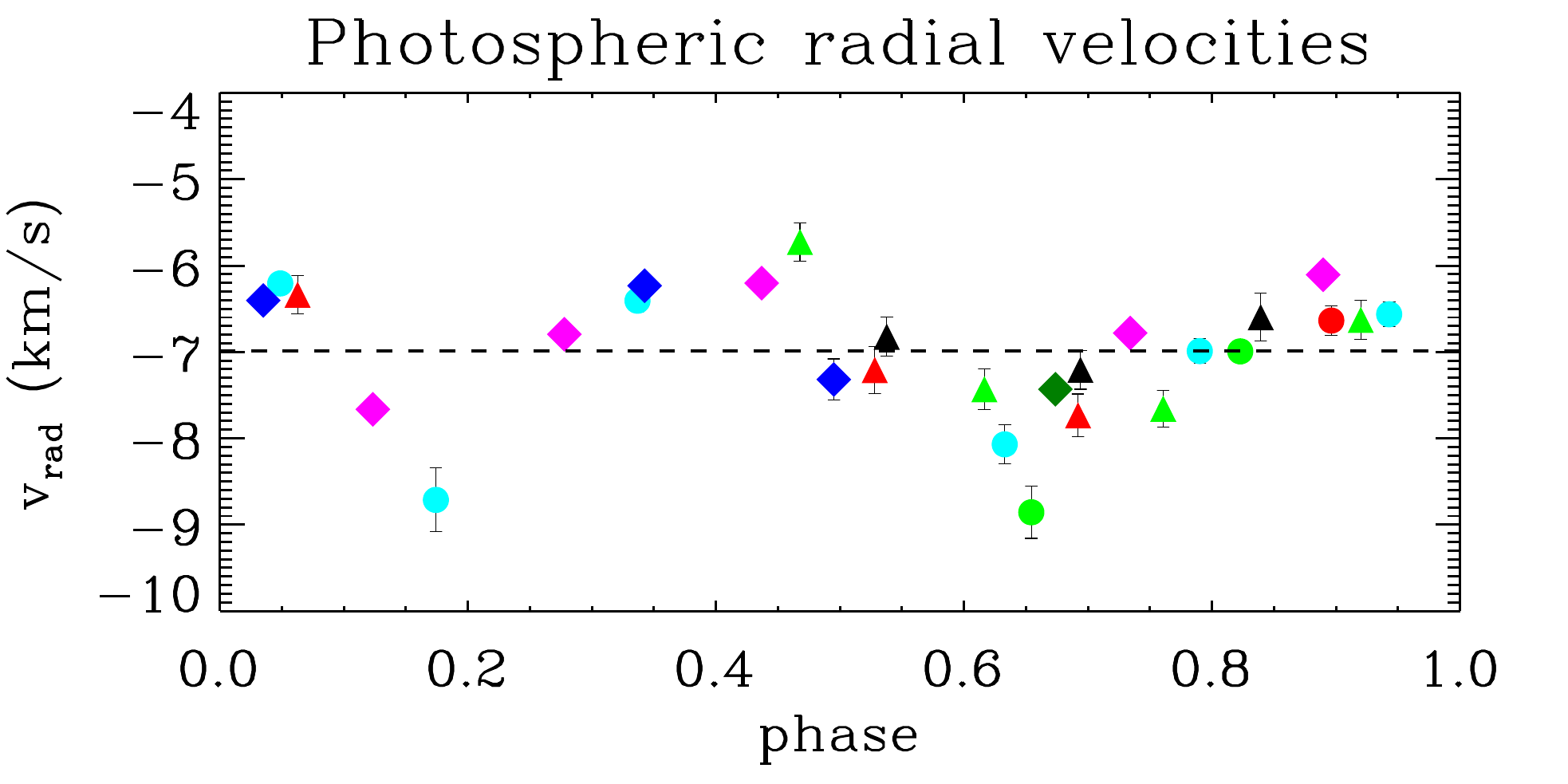}
\caption{\label{fig:opirvrad}  Radial  velocity  of  photospheric  lines  measured  with ESPaDOnS (triangles), HARPS (circles), and SPIRou (diamonds) spectra and plotted in phase with the ephemeris from \cite{2012A&A...541A.116A} and \cite{2007MNRAS.380.1297D}. The colors represent different cycles.} 
\end{figure}

In Fig. \ref{fig:opirvrad} we show, on the same plot, the optical and infrared radial velocities computed in Sects. \ref{sec:optveiling} and \ref{sec:irveling}, respectively. The shape of the optical and infrared radial velocity variations is similar. The amplitude of the infrared radial velocity curve appears to be about twice as small as the amplitude of the optical one, though SPIRou measurements are lacking toward the radial velocity minimum.

\section{Times series analysis of the data from \cite{2012A&A...541A.116A}}\label{sec:data2012}
We present the time series analysis of the V2129 Oph data used by \cite{2012A&A...541A.116A}.  The data set consists of 52 observations obtained in 2009: 24 from ESPaDOnS and 28 from HARPS (see their paper for more information). We measured the periodogram using these 52 nights together. In 2009, all the lines were more intense than during our observational campaign, and the periodicity in the redshifted absorptions and the $\mathrm{He}$\,I line corresponded to the rotational period of the star ($6.53\,\mathrm{days}$). The $\mathrm{H}\alpha$ line also presented a longer period ($\sim8.5\,\mathrm{days}$), similar to what was seen in our data (see Sect. \ref{sec:period}). 

\begin{figure}
 \begin{center}
{\includegraphics[width=4.3cm]{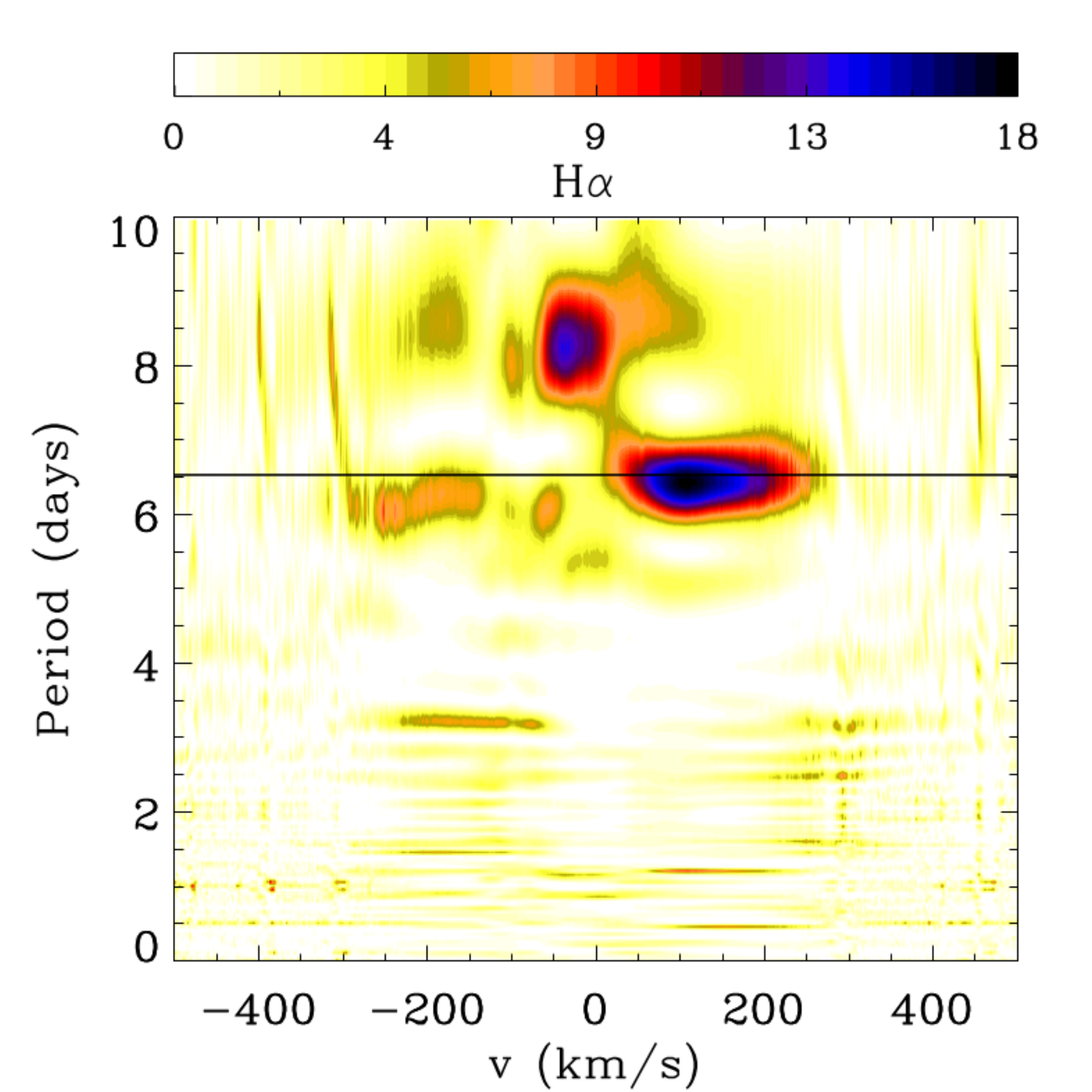}}
{\includegraphics[width=4.3cm]{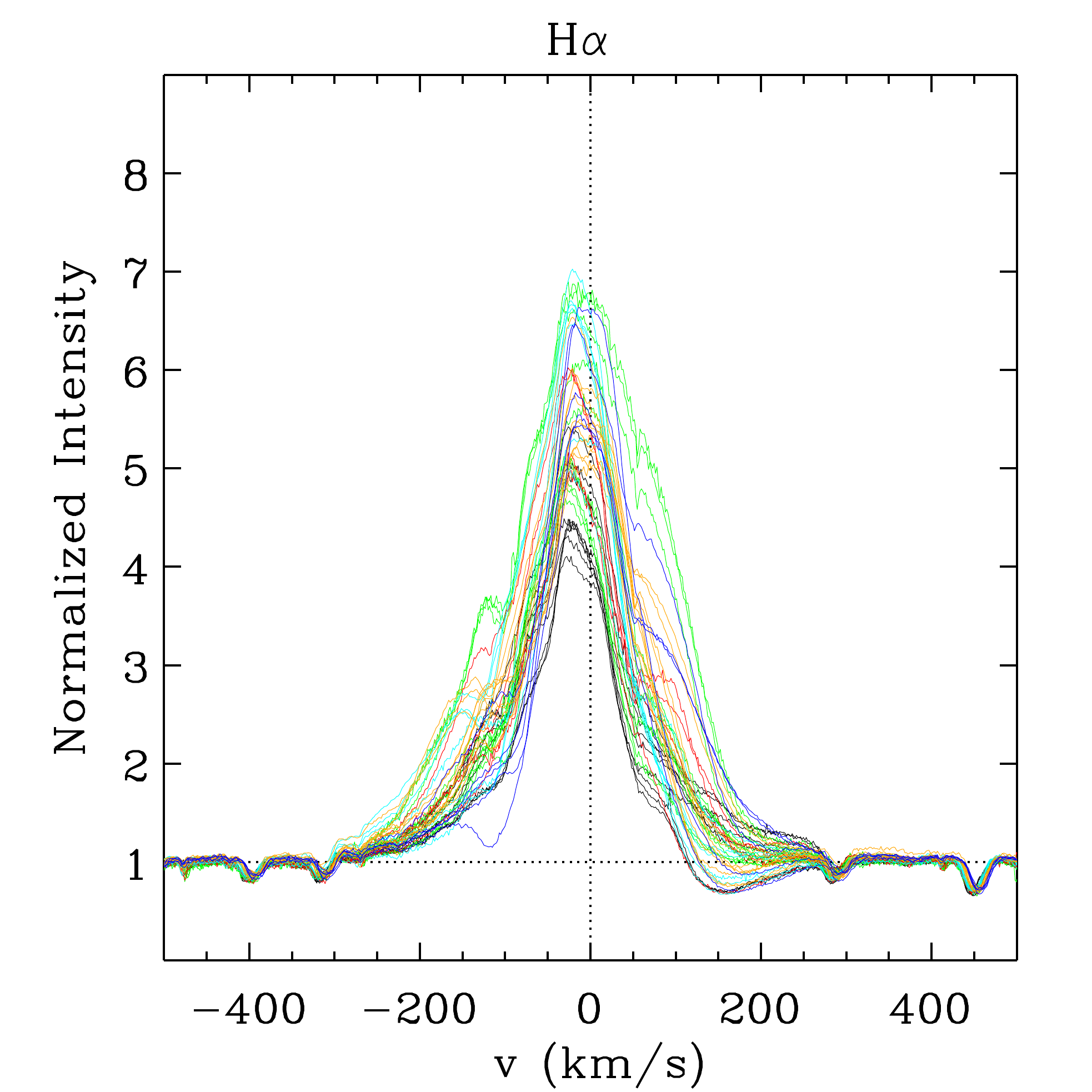}}

{\includegraphics[width=4.3cm]{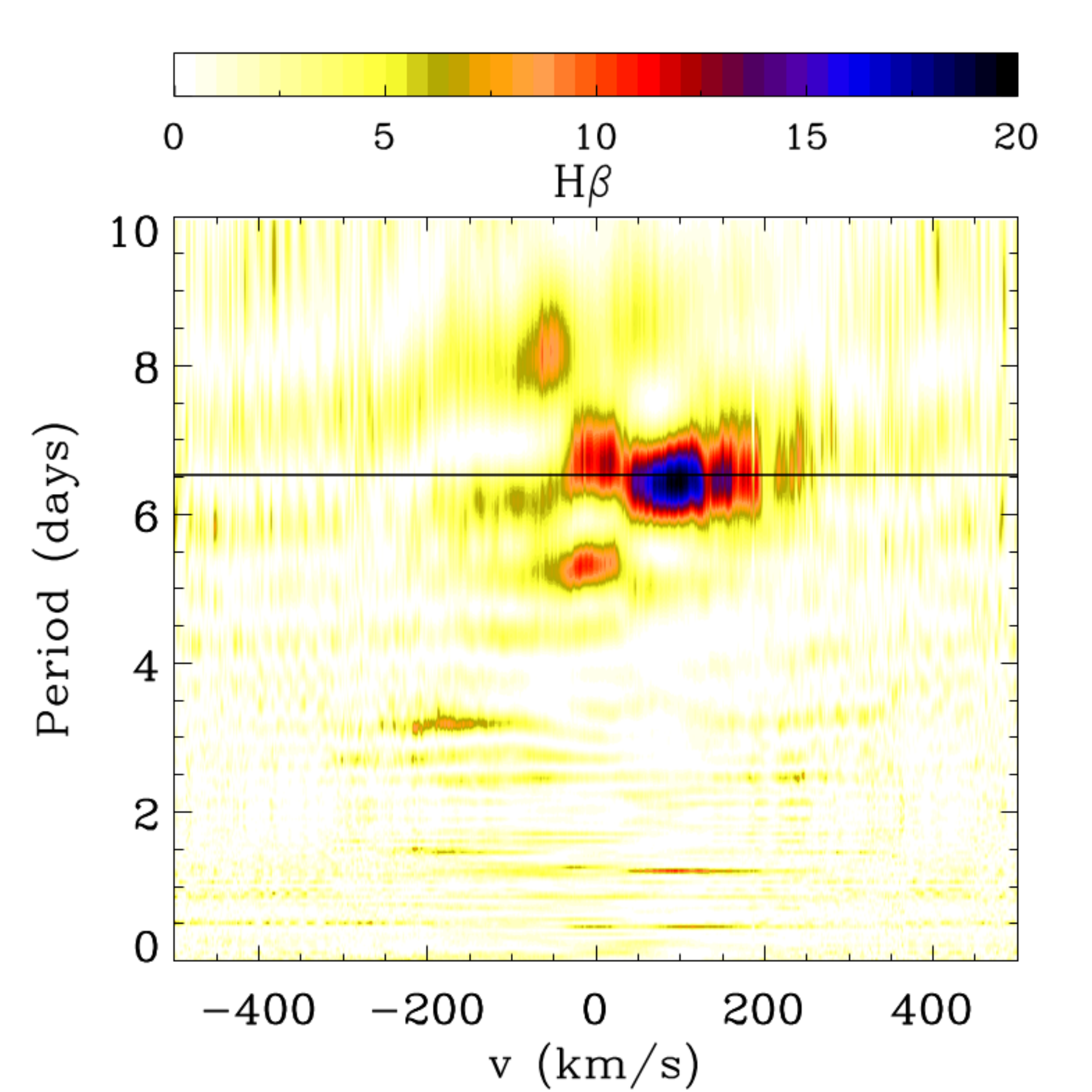}}
{\includegraphics[width=4.3cm]{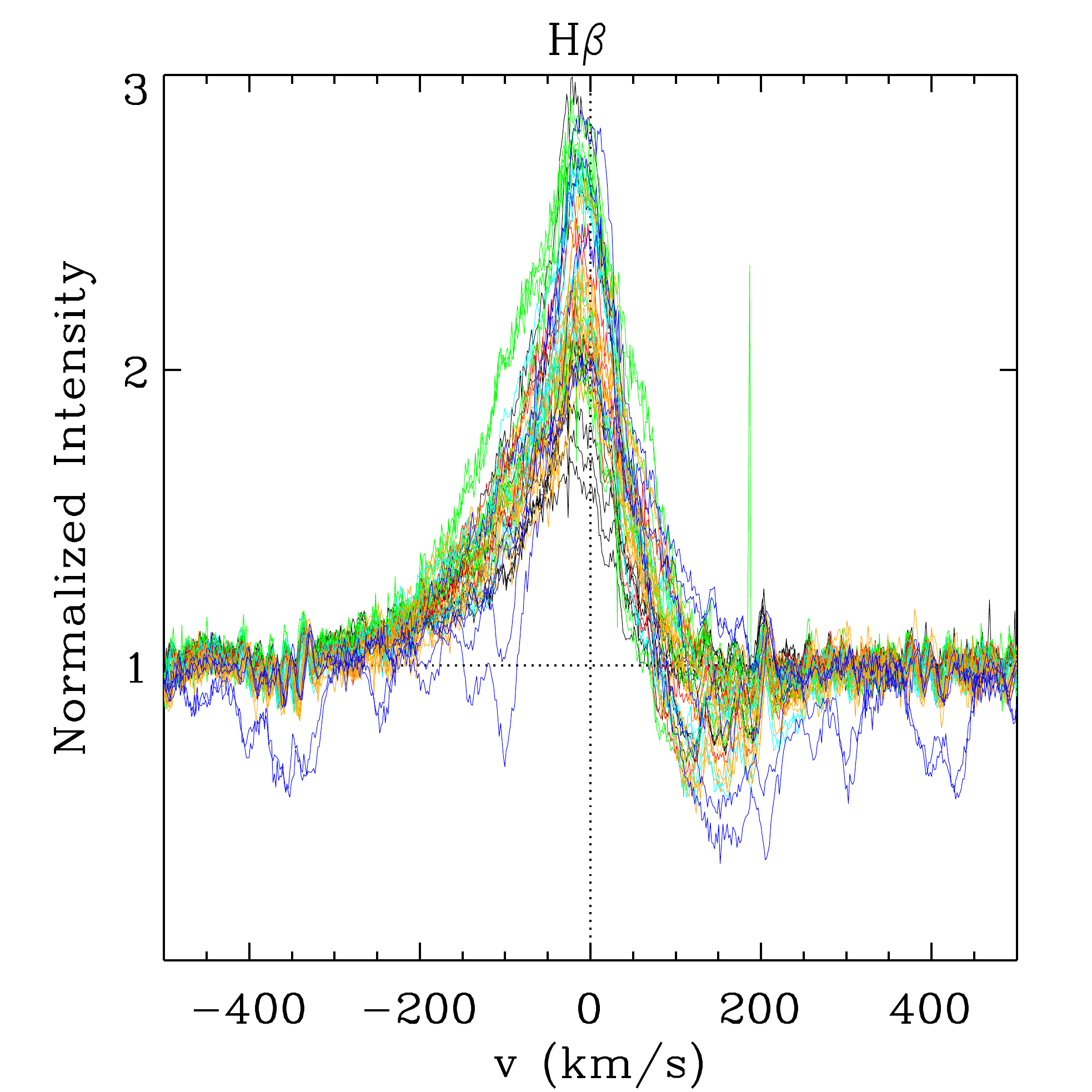}}

{\includegraphics[width=4.3cm]{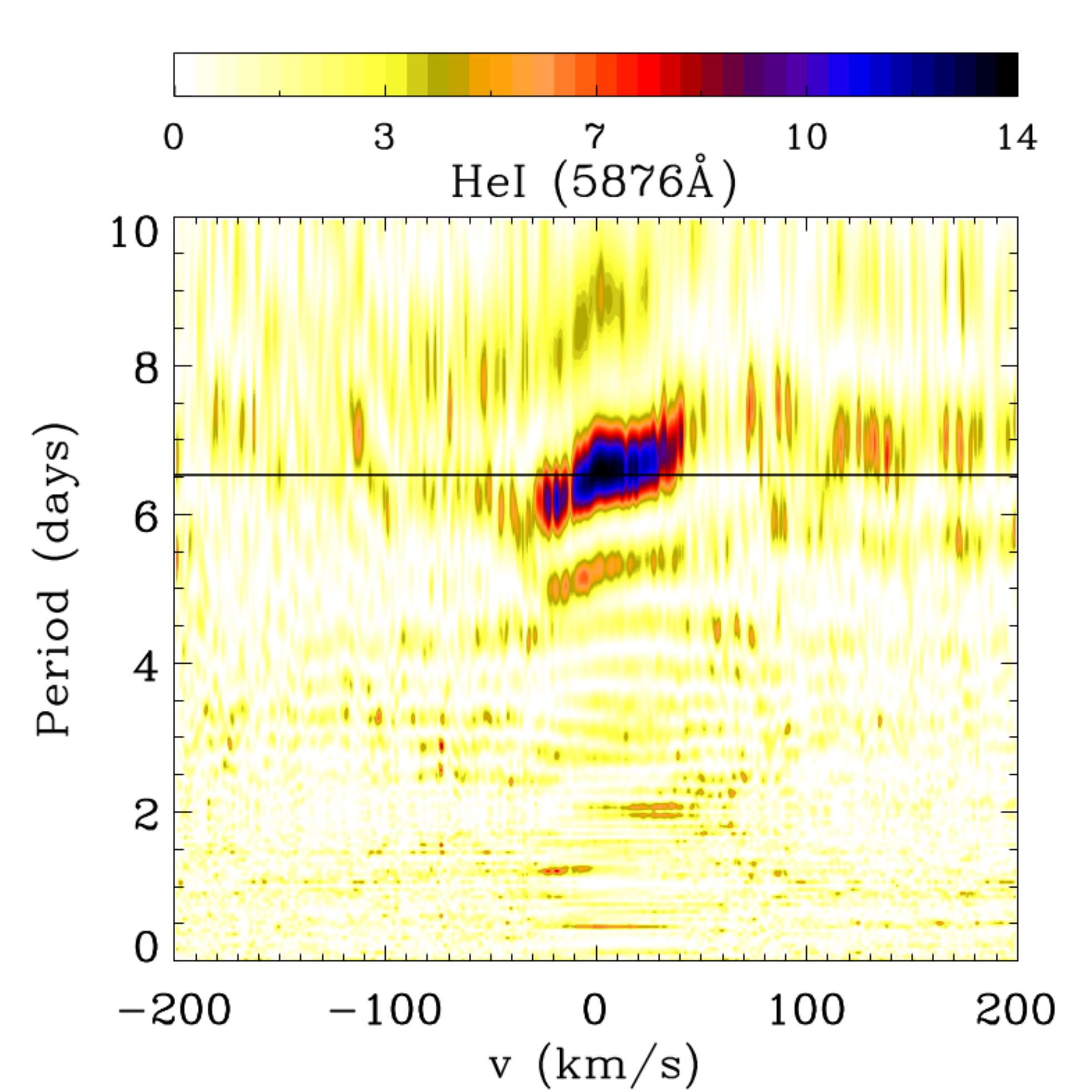}}
{\includegraphics[width=4.3cm]{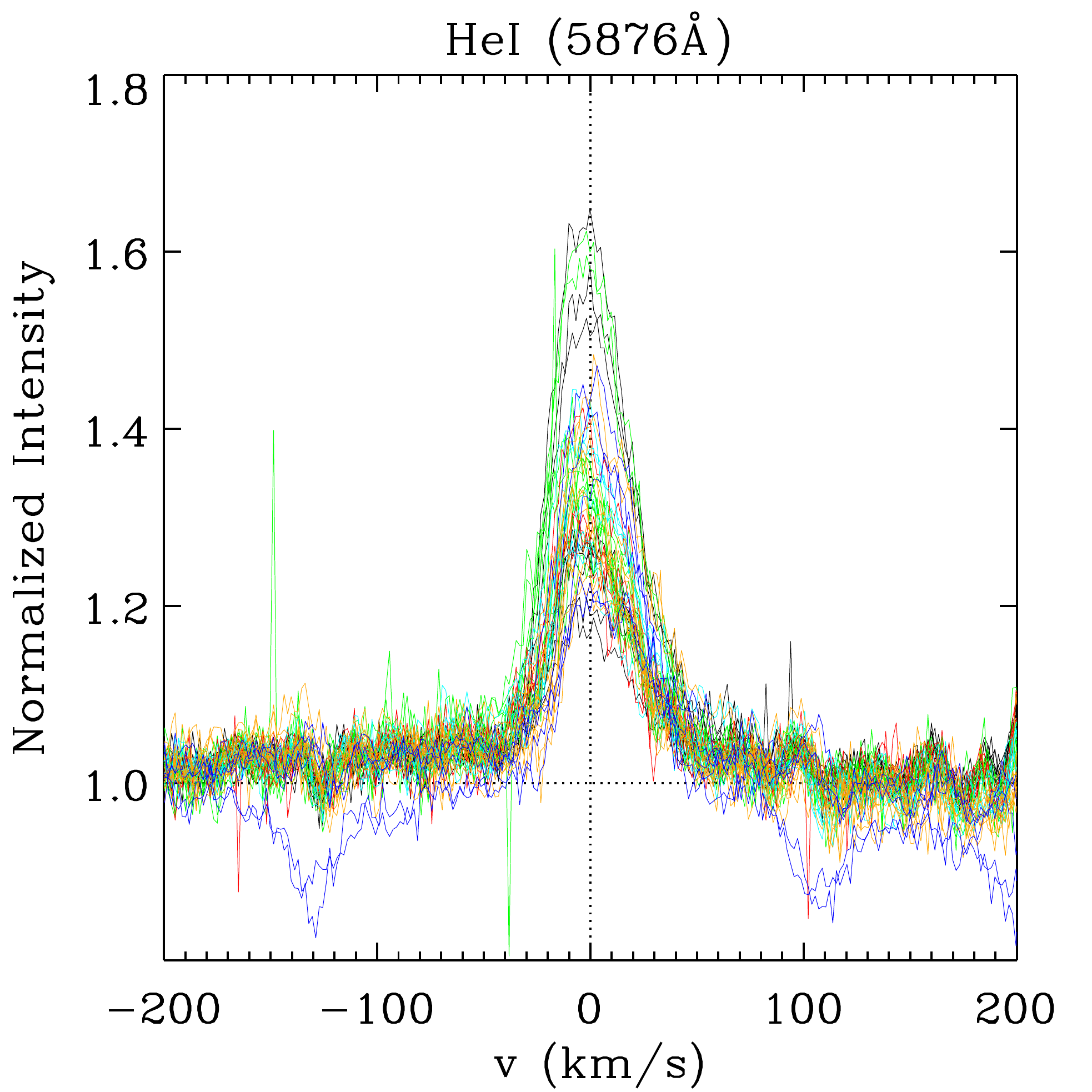}}
 \end{center}
\caption{Time series analysis of optical emission lines of the V2129 Oph data from \cite{2012A&A...541A.116A}. Shown are bidimensional periodograms ({\it left}) and the corresponding line profiles ({\it right}). In the {\it left} panels, the color range corresponds to the normalized power of the periodogram, and the horizontal solid line represents the rotation period of the star, $\mathrm{P}=6.53\,\mathrm{days}$. In the {\it right} panels, different colors correspond to different rotation cycles.}
\end{figure}

\end{appendix}

\end{document}